\documentclass[prd,superscriptaddress,amsfonts,amssymb,amsmath,twocolumn,floatfix,showkeys]{revtex4-2}

\usepackage{bm}
\usepackage{amsfonts}
\usepackage{latexsym}
\usepackage{graphicx}
\usepackage{amsmath}
\usepackage{palatino}
\usepackage{bigints}
\usepackage{mathpazo}
\usepackage{textcomp}
\usepackage{booktabs}
\usepackage{dcolumn}
\usepackage{booktabs}
\usepackage{multirow}
\usepackage{hyperref}
\hypersetup{colorlinks,citecolor=blue}
\usepackage{amsmath}
\usepackage{xcolor}
\usepackage{orcidlink}
\usepackage[caption=false]{subfig}
\usepackage{commath}
\def \nn  {\nonumber}
\graphicspath{{Graphics/}}

\def\be{\begin{equation}}
\def\ee{\end{equation}}
\def\bea{\begin{eqnarray}}
\def\eea{\end{eqnarray}}

\def\nn{\nonumber}

\def\prd{Phys. Rev. D}

\def\mnras{MNRAS}

\def\apss{Astr.Space Sci.}

\definecolor{vividviolet}{rgb}{0.62, 0.0, 1.0}
\definecolor{amaranth}{rgb}{0.9, 0.17, 0.31}
\definecolor{palatinateblue}{rgb}{0.15, 0.23, 0.89}
\definecolor{brightpink}{rgb}{1.0, 0.0, 0.5}
\definecolor{cornflowerblue}{rgb}{0.39, 0.58, 0.93}
\definecolor{deepcarminepink}{rgb}{0.94, 0.19, 0.22}
\definecolor{radicalred}{rgb}{1.0, 0.21, 0.37}
\hypersetup{ linktoc=all,
    colorlinks, linkcolor={palatinateblue},
    citecolor={brightpink}, urlcolor={amaranth}
}

\begin{document}

\title{Geodesic deviation in the $q$-metric}

\author{Anuar~\surname {Idrissov}}
\email{anuar.idrissov@gmail.com}
\affiliation{%
Department of Physics, Nazarbayev University, Kabanbay Batyr 53, 010000 Astana, Kazakhstan.}
\affiliation{%
Fesenkov Astrophysical Institute, Observatory 23, 050020 Almaty, Kazakhstan.}

\author{Kuantay~Boshkayev}
\email{kuantay@mail.ru}
\affiliation{Al-Farabi Kazakh National University, Al-Farabi av. 71, 050040 Almaty, Kazakhstan.}
\affiliation{Institute of Nuclear Physics, Ibragimova, 1, 050032 Almaty, Kazakhstan.}

\author{Konstantinos F. Dialektopoulos}
\email{kdialekt@gmail.com}
\affiliation{Department of Mathematics and Computer Science, Transilvania University of Brasov, 500091 Brasov, Romania}

\author{Ainur~\surname{Urazalina}}
\email{y.a.a.707@mail.ru}
\affiliation{Al-Farabi Kazakh National University, Al-Farabi av. 71, 050040 Almaty, Kazakhstan.}
\affiliation{Institute of Nuclear Physics, Ibragimova, 1, 050032 Almaty, Kazakhstan.}

\author{Daniya~\surname{Utepova}}
\email{utepova\_daniya@mail.ru}
\affiliation{Abai Kazakh National Pedagogical University, Dostyk av., 13, Almaty, 050010, Kazakhstan}
\affiliation{Al-Farabi Kazakh National University, Al-Farabi av. 71, 050040 Almaty, Kazakhstan.}

\date{\today}

\begin{abstract}

We consider the tidal forces between test particles falling along geodesics in the exterior spacetime generated by a static and axially symmetric compact matter source with non-vanishing mass quadrupole. Specifically, we analyze the radial and angular geodesic deviation, compare it with that of the Schwarzschild spacetime, and investigate the impact of the deformation parameter $q$, at different polar angles $\theta$ with respect to the vertical symmetry axis. Furthermore, we examine the geodesic deviation for the case of non-constant $\theta$ during the radial fall. It is shown that the presence of the deformation parameter affects the behavior of the geodesic deviation vectors, depending on its value.  In particular, we observe that for arbitrary values of $q$ and $\theta$ the behavior of the deviation vector differs as it approaches the singularity at $r = 2m$. Above all, we can witness either stretching or compressing of the deviation vector for various combinations of $q$ and $\theta$. These findings provide insight into the effects of quadrupole deformation on the motion of test particles in the vicinity of the central object.

\end{abstract}


\maketitle

\section{Introduction}

The study of geodesic deviation in terms of exact solutions of Einstein's equations is important as it allows to characterize the tidal forces produced by compact objects \cite{misner1973gravitation,poisson2004relativist,d1992introducing,hobson2006general,schutz2022first}.
Tidal effects are observed everywhere in the universe, from the solar system to binary star systems, neutron stars and black holes. In particular, it is the tidal acceleration between two arms of gravitational wave detectors that recently allowed LIGO and Virgo to detect the gravitational waves emitted by binary black hole mergers \cite{abbott2016observation,scientific2017gw170104}. In fact, it can be argued that tidal forces and gravitational waves are highly interrelated concepts \cite{goswami2021tidal}. 
Geodesic deviation in static and spherically symmetric spacetimes has been thoroughly studied starting from the Schwarzschild black hole \cite{Pirani:1956tn,1990AN....311..219F,2015arXiv150806457P,1989JMP....30.1794B,Vandeev:2022gbi}, 
to the Reissner-Nordstr\"om \cite{2016EPJC...76..168C,2005CSF....23..313A,2010Ap&SS.330..107G}
and some regular extensions of Schwarzschild \cite{sharif2018tidal} such as the charged Hayward black hole \cite{2020IJMPD..2941014L}.
Other theoretical studies on static spherical symmetry include \cite{shahzad2017tidal,nandi2001tidal,cardoso2013tidal,hong2020tidal,sharif2018tidal,toshmatov2023tidal}.  Also in \cite{li2021tidal} tidal forces were investigated in the 4D-Einstein-Gauss-Bonnet black hole spacetime. In classical general relativity, the analysis of tidal forces in the Kerr spacetime was done in \cite{2020EPJP..135..334L, Marck1983, 1993MNRAS.262..881M, Chicone:2004ic, Chicone:2006rm}.

However, there are not many works investigating geodesic deviation in static axially symmetric spacetimes \cite{Bini:2012zze}. The topic is interesting and important for realistic physical systems as one can expect the presence of a non-vanishing mass quadrupole moment to bear significant consequences on the tidal forces exerted on test particles in the vicinity of the source. The study of tidal forces in static axially symmetric spacetimes is also interesting as it may provide clues at the behavior of gravitational collapse of a non-spherical massive object. It is generally believed that the higher multipole moments are radiated away through gravitational waves, but as of now no analytical mechanism showing how this occurs has been found.

In the present article we study the geodesic deviation in the Zipoy-Voorhees spacetime \cite{zipoy1966topology,voorhees1970static}, sometimes referred to as $\gamma$-metric, $\delta$-metric or $q$-metric,  which is a well known vacuum, static and axially symmetric solution of Einstein's equations. The $q$-metric is interesting because it is continuously linked to the Schwarzschild solution via the value of only one parameter i.e. $q$, which is related to the mass quadrupole and characterizes the departure of the source from spherical symmetry \cite{Papadopoulos:1981wr, quevedo2011exterior}. This allows one to study the effects of the presence of mass quadrupole and compare with the black hole case. 

It must be noted that the line element presents a curvature singularity at the null surface where the black hole horizon occurs, thus making the spacetime considerably different from Schwarzschild \cite{Quevedo:2010mn}. The $q$-metric has been widely studied over the years including motion of test particles \cite{Herrera:1998rj, Chowdhury:2011aa, Boshkayev:2015jaa, Capistrano:2019qdv}, interior solutions \cite{Stewart:1982, Herrera:2004ra}, photon sphere and shadow \cite{Abdikamalov:2019ztb, Arrieta-Villamizar:2020brc, Shaikh:2022ivr, Turimov:2023lbn}, motion of spinning particles \cite{Toshmatov:2019bda}, charged particles \cite{Benavides-Gallego:2018htf, Faraji:2021vid}, particle collisions \cite{Malafarina:2020kmk}, quasiperiodic oscillations \cite{Toshmatov:2019qih, Benavides-Gallego:2020fri,2024MNRAS.531.3876B}, lensing of photons and neutrinos \cite{Boshkayev:2020igc, Chakrabarty:2021bpr, Chakrabarty:2022fbd} and accretion disks \cite{Boshkayev:2021chc, Shaikh:2021cvl}. Extensions of the $q$-metric have also been considered, including the rotating case \cite{Quevedo:1991zz, Toktarbay:2014yru, Allahyari:2019umx, Li:2022eue}, the charged case \cite{Gurtug:2021noy}, the case with the inclusion of the Newman-Unti-Tamburino (NUT) charge \cite{Halilsoy:1992zz, Narzilloev:2020qdc} and even wormholes \cite{Gibbons:2017jzk, Narzilloev:2021ygl}.  
It should be noted that solutions involving quadrupole moments can be rather complicated, making the pursuit of exact analytical results challenging. Consequently, most computations have been performed using numerical methods.

In the present article, in order to gain some insight into the effect of deviations from spherical symmetry on the tidal forces experienced by falling test particles, we examine two situations, each involving two particles descending radially and with a fixed azimuthal angle $\phi$: (i) particles following trajectories with the same polar angle $\theta$, initially separated by a radial deviation vector $\eta^r$; (ii) particles following trajectories with the same $r$, initially positioned apart by an angular deviation vector $\eta^{\theta}$. 

The paper is organized as follows: In Sect.~\ref{sec:geod_dev}, we introduce the $q$-metric, highlighting the  main physical features, geodesics, and geodesic deviation equations. In Sect.~\ref{sec:radial fall}, we analyze the radial fall for different scenarios, while in Sect.~\ref{sec:interpretation} we discuss the physical interpretation and their theoretical implications. In Sect.~\ref{sec:conclusions}, we conclude and discuss future perspectives. Throughout this paper, we use natural units $G = c = 1$, and the Lorentzian signature $(+,-,-,-)$.

\section{Geodesic deviation in the {\it q}-metric}
\label{sec:geod_dev}

The $q$-metric is an exact solution of the vacuum field equations, which extends the Schwarzschild spacetime from spherical symmetry to describe the field outside prolate or oblate spheroids. It belongs to the Weyl class of spacetimes, which encompasses static, axially symmetric, and vacuum solutions of the Einstein equations that approach flatness at infinity \cite{stephani2009exact}. It is well known that all metrics within the Weyl class are obtained from solutions of the Laplace equation in the two-dimensional flat space in cylindrical coordinates \cite{weyl1917theory}. The unique feature of the $q$-metric is that it is continuously connected to the Schwarzschild metric through the value of one parameter, $q$ which is then interpreted as related to the source's deformation \cite{quevedo2011mass}. 
In Schwarzschild-like coordinates $\{t,r,\theta,\phi\}$ the line element of the $q$-metric is
\begin{widetext}
\be\label{eq:metric}
ds^2= \left(1 - \frac{2 m}{r}\right)^{1+q} d t^2 - \left(1 - \frac{2 m}{r}\right)^{-q} \left[ \left(1 + \frac{m^2 \sin^2{\theta} }{r^2 - 2 m r} \right)^{-q(2+q)} \left( \frac{d r^2}{1-\frac{2 m}{r}} + r^2 d \theta^2 \right)+ r^2 \sin^2{\theta} d \phi^2\right]\,.
\ee 
\end{widetext}
The other parameter appearing in the line element, $m$ can be interpreted in analogy to Schwarzschild as a mass parameter. 
Notice that the ADM mass of the $q$-metric, as measured by an observer at infinity, is $M_0 = m(1+q)$, which reduces to $m$ in the Schwarzschild limit $q=0$. In addition, the deformation parameter can only take values $q>-1$, while the values of $q > 0$ correspond to an oblate geometry, while the values of $q < 0$  correspond to a prolate geometry.
In the following, we consider the behavior of radial geodesics in the $q$-spacetime. For a detailed overview of geodesics in the $q$-metric one can refer to \cite{Herrera:1998rj}.

Due to the metric being static and axially symmetric, we have two conserved quantities related to time translations and rotations about the symmetry axis. These conserved quantities are the (specific) energy $E$ and angular momentum $L$ per unit mass of the test particles and are obtained from $g_{tt}\dot{t} = E$ and $g_{\phi \phi}\dot{\phi} = - L$. These give
\bea \label{E}
    \dot{t}&=& E \left(1 - \frac{2m}{r}\right)^{-(1+q)}\,, \\ \label{L}
    \dot{\phi}&=& -\frac{L}{r^2\sin^2\theta} \left(1 - \frac{2 m}{r}\right)^{q}\,.
\eea
In the case of infalling geodesics confined in the $\{r,\theta\}$ plane we must have $L=0$, which allows to write the equation of motion for the infalling particles as a first order equation as
\begin{eqnarray} \label{eq:rad}
    \dot{r}^2=  \left(E^2-\Phi^2 \right)  \left(1+\frac{m^2 \sin^2{\theta}}{r^2-2 m r}\right)^{q (2+q)}\,,
\end{eqnarray}
where
\begin{align}
    \Phi^2 & = \left(1-\frac{2 m}{r}\right)^{q+1} \times\\ \nonumber
    & \times \left[
    r^2 \left(1 - \frac{2 m}{r} \right)^{-q} \left(1 + \frac{m^2 \sin^2{\theta}  }{r^2 - 2 m r} \right)^{-q(q+2)} \dot{\theta}^2 -1\right],   
\end{align}
where the dot defines the differentiation with respect to an affine parameter $s$, which for timelike geodesics may coincide with the proper time. The simpler case of purely radial geodesics is then obtained for $\dot{\theta}=0$. It is worth noticing that while in the spherically symmetric case the choice of the angle $\theta$ is arbitrary and thus we can always confine the analysis to $\theta={\rm const}.$ the same is not true in the axially symmetric case.

The most general geodesic equations can be derived from the Lagrangian per unit mass
\begin{equation}
    {\cal L} = \frac{1}{2} g_{\alpha\beta} \dot{x}^{\alpha} \dot{x}^{\beta}\,.
\end{equation}
by evaluating Euler-Lagrange equations \cite{Herrera:1998rj},
\begin{equation}
    \frac{d}{d s} \left(\frac{\partial {\cal L}}{\partial \dot{x}^{\alpha}} \right) - \frac{\partial {\cal L}}{\partial x^{\alpha}} = 0\,.
\end{equation}
For the metric ~\eqref{eq:metric} it follows that geodesics in the $\{r,\theta\}$ plane become \cite{Boshkayev:2015jaa} 
\begin{widetext}
\bea
    \ddot{t}& = & - \frac{2 m \left(1+ q\right) \dot{r} \dot{t}}{r \left( r - 2 m \right)} ,\\
    \ddot{r}& = & \frac{m \dot{r}^2 \left(m \sin ^2{\theta} (m q (q+3)+m-q (q+2) r)+(q+1) r (r-2 m)\right)}{r (r - 2m) \left(m^2 \sin^2{\theta} +r (r-2 m)\right)} -\\ \nonumber
    && - \frac{1}{r^3} m \dot{t}^2 (q+1) (r-2 m) \left(1-\frac{2 m}{r}\right)^{2 q} \left(\frac{m^2 \sin^2{\theta }}{r^2-2 m r}+1\right)^{q (q+2)},\\ \label{ddot-theta}
    \ddot{\theta} & = & - \frac{m^2 \dot{r}^2 q (q+2) \sin {2 \theta} }{r (r - 2 m) \left(m^2 -m^2 \cos {2 \theta} -4 m r+2 r^2\right)}.
\eea
\end{widetext}
with the remaining equation obviously being $\ddot{\phi} = 0$. Notice that the first is just the derivative of the conservation equation \eqref{E}. On the other hand, for purely radial in-fall, i.e. $\dot{\theta}= \dot{\phi}=0$, we see that equation \eqref{ddot-theta} also vanishes.

The geodesic deviation equation describes the relative acceleration between nearby geodesics in a curved spacetime. It is an important tool for investigating the physical properties of spacetime geometry, since it provides information about the tidal forces produced by the source. In this respect, the geodesic deviation plays a crucial role in understanding the effects of strong gravity in general relativity and has numerous applications in astrophysics, including the study of gravitational waves and the tidal forces experienced by test particles in the vicinity of extreme compact objects such as black holes and neutron stars.
The geodesic equation is written as \cite{d1992introducing}
\begin{equation}\label{eq:GDE}
    \frac{D^2 \eta^{\alpha}}{d s^{2}} - R^\alpha_{\mu\nu\rho} u^{\mu} u^{\nu} \eta^{\rho} = 0,
\end{equation}
where  $D^2/d s^{2}$ is the covariant derivative, $ R^\alpha_{\mu\nu\rho}$ is the Riemann tensor, $u^{\mu}=d x^{\mu}/d s$ is the 4-velocity tangential to the geodesic and the geodesic deviation vector is given by $\eta^{\rho}$.
Following Shirokov~\cite{1973GReGr...4..131S}, one can write geodesic deviation equations in terms of connection coefficients. Thus, the first term of Eq.~\eqref{eq:GDE} can be written as
\bea \nn
        \frac{D^2 \eta^{\alpha}}{d s^{2}}&=& \frac{d^2 \eta^{\alpha}}{ d s^2}   + \Gamma^{\alpha}_{\mu\nu} u^{\mu} \frac{d \eta^{\nu}}{d s}  + \frac{\partial \Gamma^{\alpha}_{\mu\nu}  }{\partial x^\delta} u^{\mu} u^{\delta} \eta^{\nu} + \\ \label{eq:covdiv}
        && - \Gamma^{\alpha}_{\mu\nu} \Gamma^{\alpha}_{\delta\lambda} u^{\delta} u^{\lambda} \eta^{\nu} + \Gamma^{\alpha}_{\mu\nu} \Gamma^{\nu}_{\delta\lambda} u^{\mu} u^{\delta} \eta^{\lambda}\,,
\eea
while the second term will be
\bea \nn
    R^\alpha_{\mu \nu \rho} u^{\mu} u^{\nu} \eta^{\rho} &=& \left( \frac{}{} \Gamma^{\alpha}_{\mu\nu,\rho}  -\Gamma^{\alpha}_{\mu\rho,\nu}  + \Gamma^{\lambda}_{\mu\nu} \Gamma^{\alpha}_{\lambda\rho}+\right. \\ \label{eq:riemtens}
    && \left. \frac{}{} - \Gamma^{\lambda}_{\mu\rho} \Gamma^{\alpha}_{\lambda\nu}    \right) u^{\mu} u^{\nu} \eta^{\rho}\,.
\eea
After substituting ~\eqref{eq:covdiv} and ~\eqref{eq:riemtens} into ~\eqref{eq:GDE} we end up with
\begin{align} \label{eq:explgde}
    \frac{d^2 \eta^{\alpha}}{d s^2} + 2 \Gamma^{\alpha}_{\mu\rho} u^{\mu} \frac{d \eta^{\rho}}{d s} + \frac{\partial \Gamma^{\alpha}_{\mu\rho}}{\partial x^{\nu}} u^{\mu} u^{\rho} \eta^{\nu} = 0\,.
\end{align}
To keep the notation as light as possible, we shall write the line element ~\eqref{eq:metric} in terms of the metric functions $g_{\mu\nu}(r,\theta)$ 
so that the complete set of geodesic deviation equations for the case of radial fall $\dot{\theta}=0$, $ \phi = 0$, $\dot{\phi}=0$ can be written in the following form: 
\begin{widetext}
\bea
\label{eq:dev_t}
\ddot{\eta}^{t} &=& \frac{\eta^r \dot{r} \dot{t} \left( ( g_{tt,r})^2 - g_{tt} g_{tt,rr} \right)}{g_{tt}^2} - \frac{ g_{tt,r} \left(\dot{r} \dot{\eta}^t+\dot{\eta}^r \dot{t} \right)}{g_{tt}}\,,\\
\label{eq:dev_r}
\ddot{\eta}^{r}  &=& \frac{ g_{tt,r} \dot{t} \dot{\eta}^t}{g_{rr}}  -\frac{\dot{r}}{2 g_{rr}}  \left( 2 \dot{\eta}^{\theta}  g_{rr,\theta} +2 \dot{\eta}^r  g_{rr,r} + \eta^{\theta} \dot{r} g_{rr,r \theta} \right)
-\frac{\left(\eta^{\theta} g_{rr,\theta} + \eta^r g_{rr,r} \right)\left( g_{tt,r} \dot{t}^2 - \dot{r}^2 g_{rr,r}  \right) }{2 g_{rr}^2}+ \\ \nn 
&&  -  \frac{\eta^r}{2 g_{rr}}
\left(\dot{r}^2 g_{rr,rr} - \dot{t}^2  g_{tt,rr} \right) \,,\\
\label{eq:devtheta}
\ddot{\eta}^{\theta} &=& \frac{\dot{r}}{2 g_{\theta \theta}} \left( 2 \dot{\eta}^r g_{rr,\theta} + \eta^{\theta} \dot{r}  g_{rr,\theta \theta} - 2 \dot{\eta}^{\theta}  g_{\theta \theta,r} + \eta^r \dot{r} g_{rr,r \theta}   \right) - \frac{\dot{r}^2  g_{rr,\theta} \left(\eta^{\theta}  g_{\theta\theta,\theta} + \eta^r g_{\theta \theta,r}\right) }{2 g_{\theta \theta}^2}\,,\\
\label{eq:devphi}
\ddot{\eta}^{\phi} &=& \frac{ g_{\phi \phi,r}}{g_{\phi \phi}} \dot{r} \dot{\eta}^{\phi} \,.
\eea
\end{widetext}

In the following, we numerically solve the geodesic deviation equations to find the deviation vector $\eta^{\alpha}$
in terms of the radial coordinate $r$. From Eqs. ~\eqref{eq:dev_r} -~\eqref{eq:devphi} ,
we obtain a differential equation with respect to the radial coordinate (as an independent variable), using
\begin{eqnarray} \label{eq:transform}
    \frac{ d^2 \eta^{\alpha}}{d s^2} =\frac{d r}{d s} \frac{d}{d r} \Bigg( \dot{r} \frac{d \eta^{\alpha}}{d r} \Bigg) = \dot{r}^2 \frac{d ^2 \eta^{\alpha}}{d r^2}  + \frac{1}{2} \frac{d (\dot{r})^2}{d r} \frac{d \eta^{\alpha}}{d r} 
\end{eqnarray}

\begin{figure*}

{\hfill
\includegraphics[width=0.45\hsize,clip]{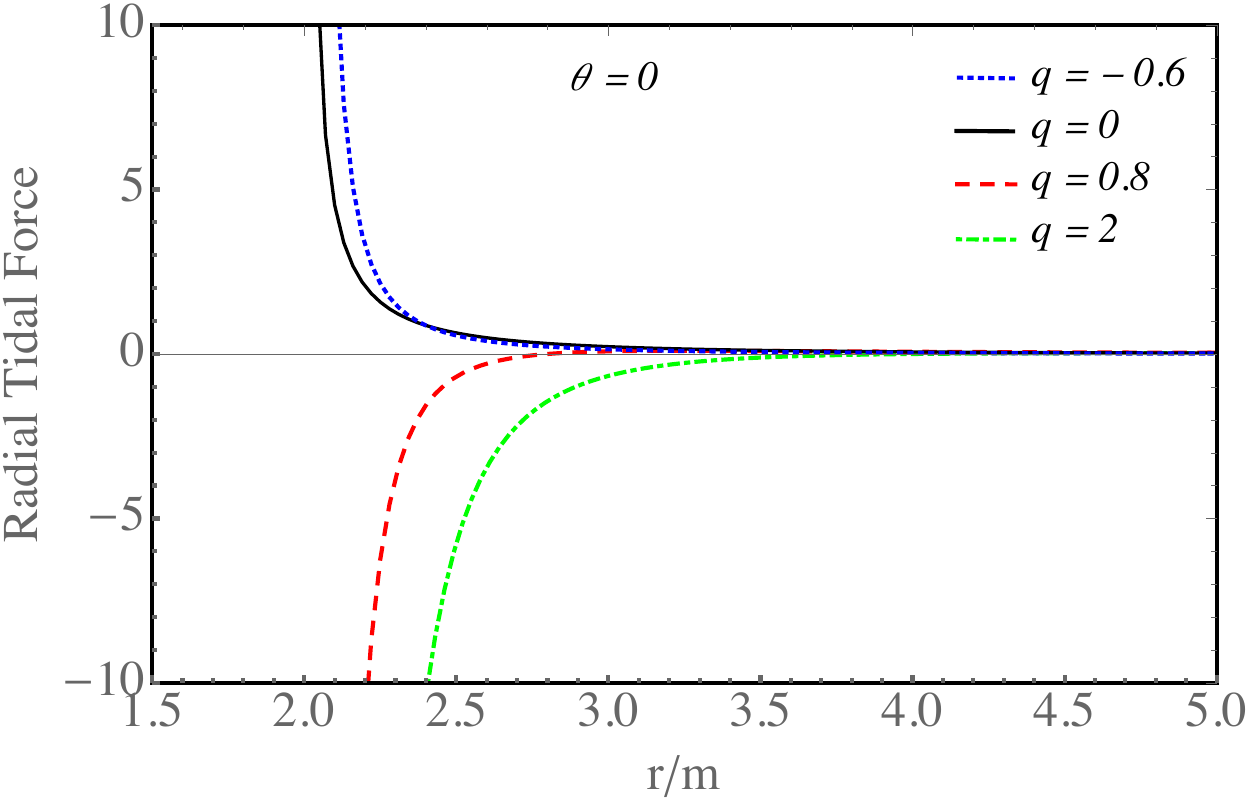}\hfill
\includegraphics[width=0.45\hsize,clip]{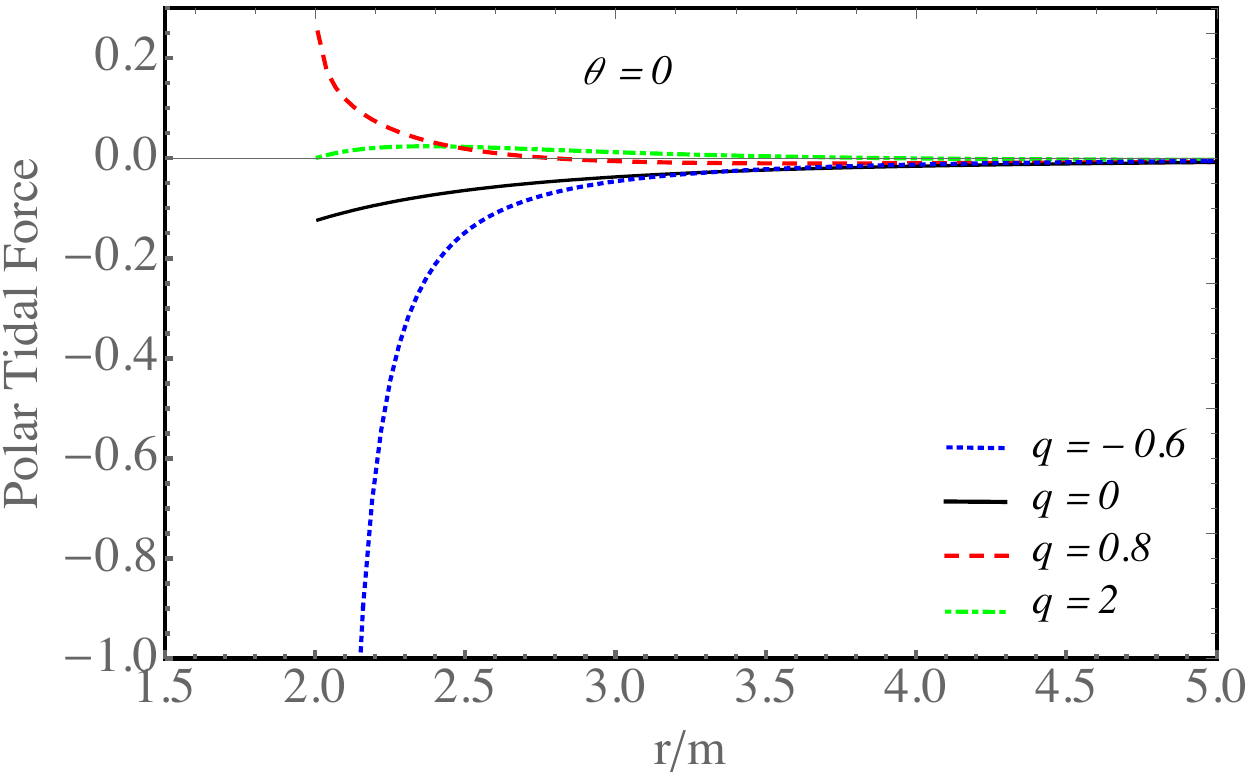}\hfill}

{\hfill
\includegraphics[width=0.45\hsize,clip]{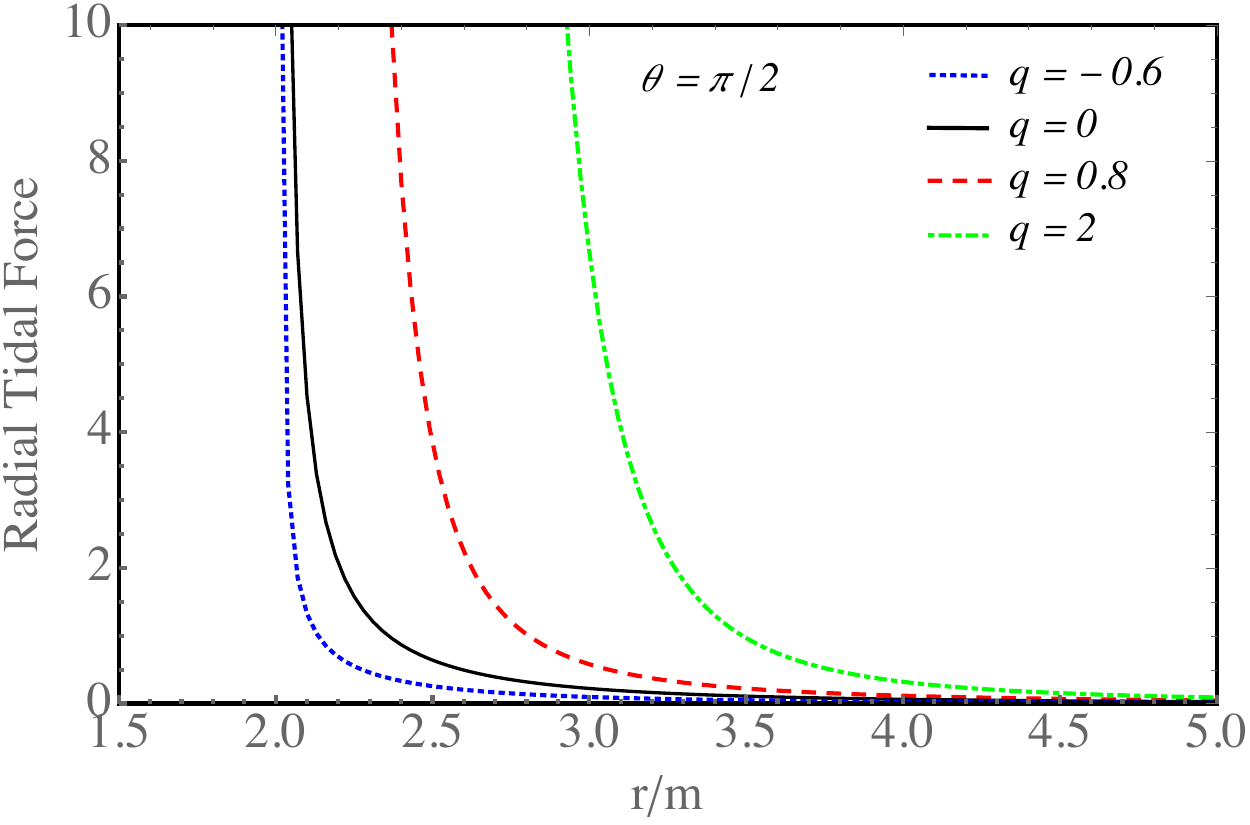}\hfill
\includegraphics[width=0.45\hsize,clip]{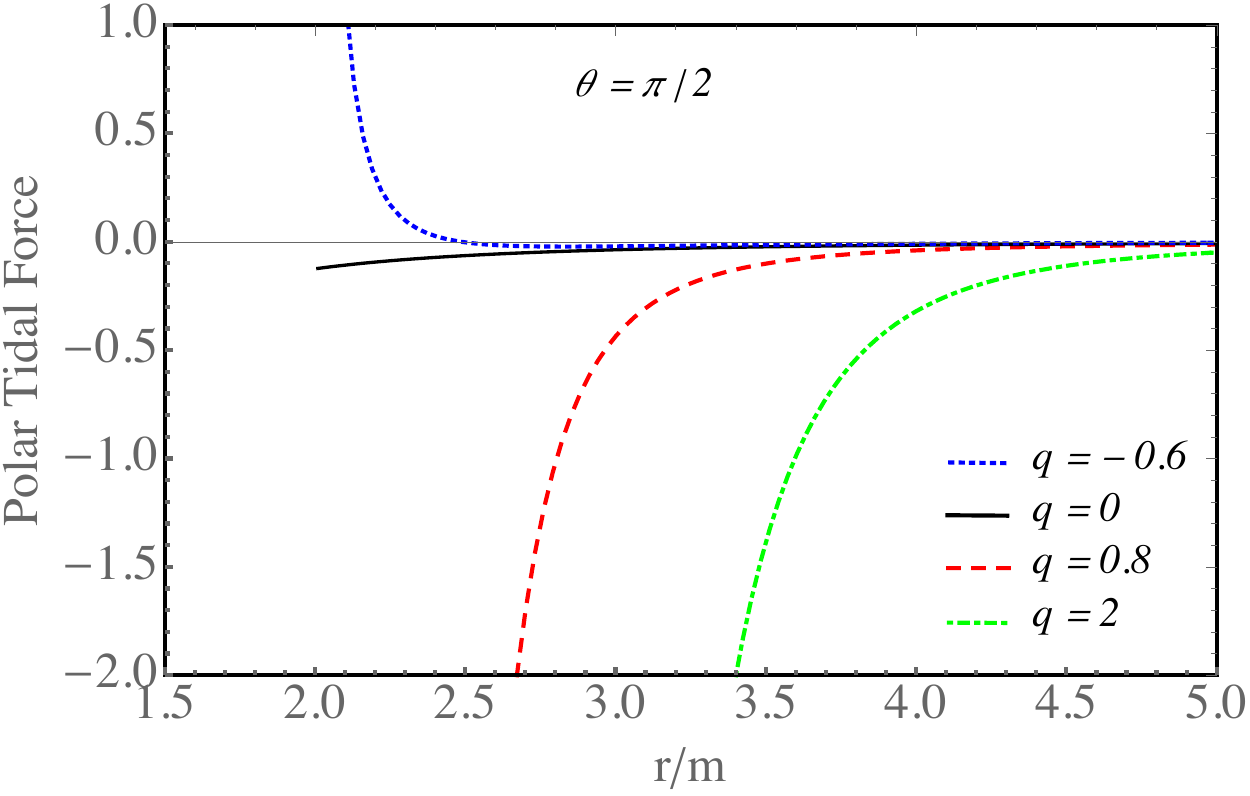}\hfill}

\caption{Dependence of tidal forces on $r/m$ for different values of quadrupole parameter $q$. Left panel. Radial tidal force vs $r/m$: (top) along symmetry axis $\theta=0$, (bottom) equatorial plane $\theta=\pi/2$. 
Right panel. Polar tidal force vs $r/m$: (top) along symmetry axis $\theta=0$, (bottom) equatorial plane $\theta=\pi/2$.}
\label{fig:rtf_atf}
\end{figure*}

\begin{figure*}

{\hfill
\includegraphics[width=0.45\hsize,clip]{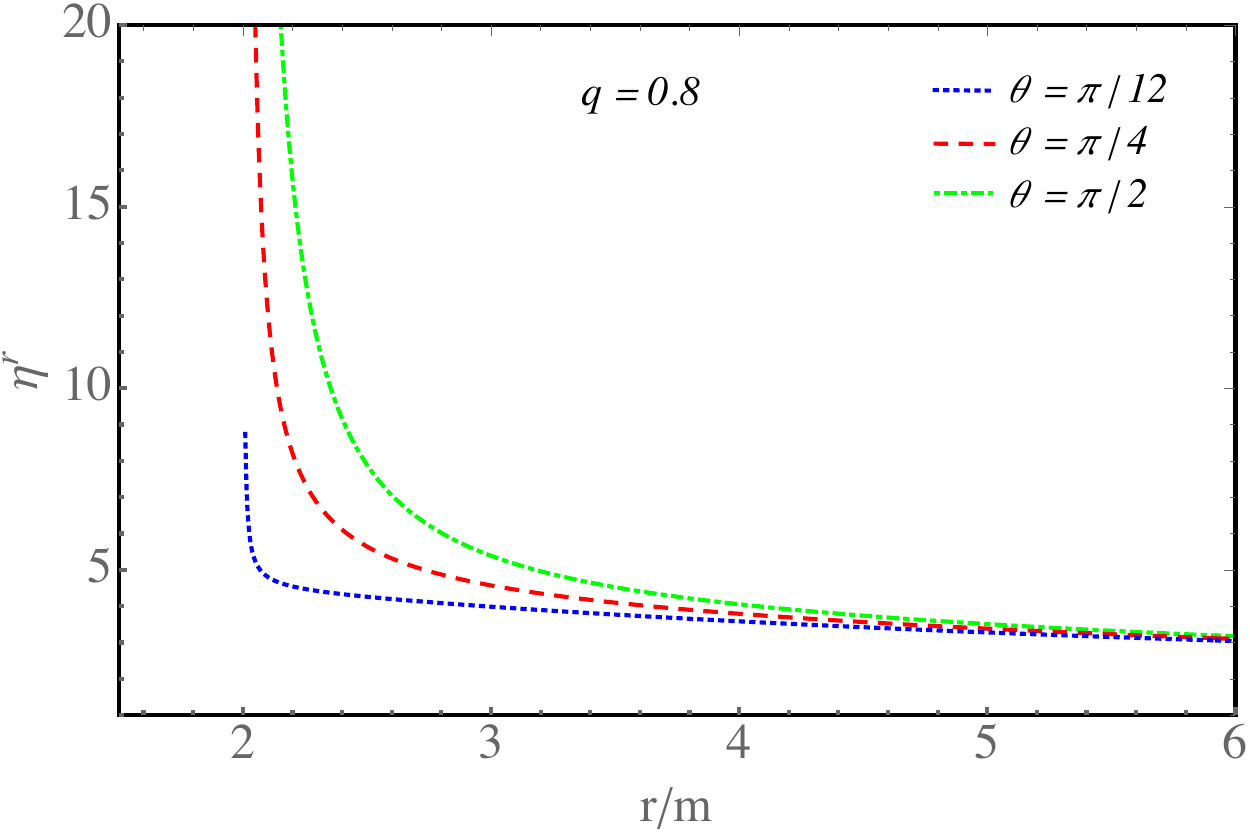}\hfill
\includegraphics[width=0.45\hsize,clip]{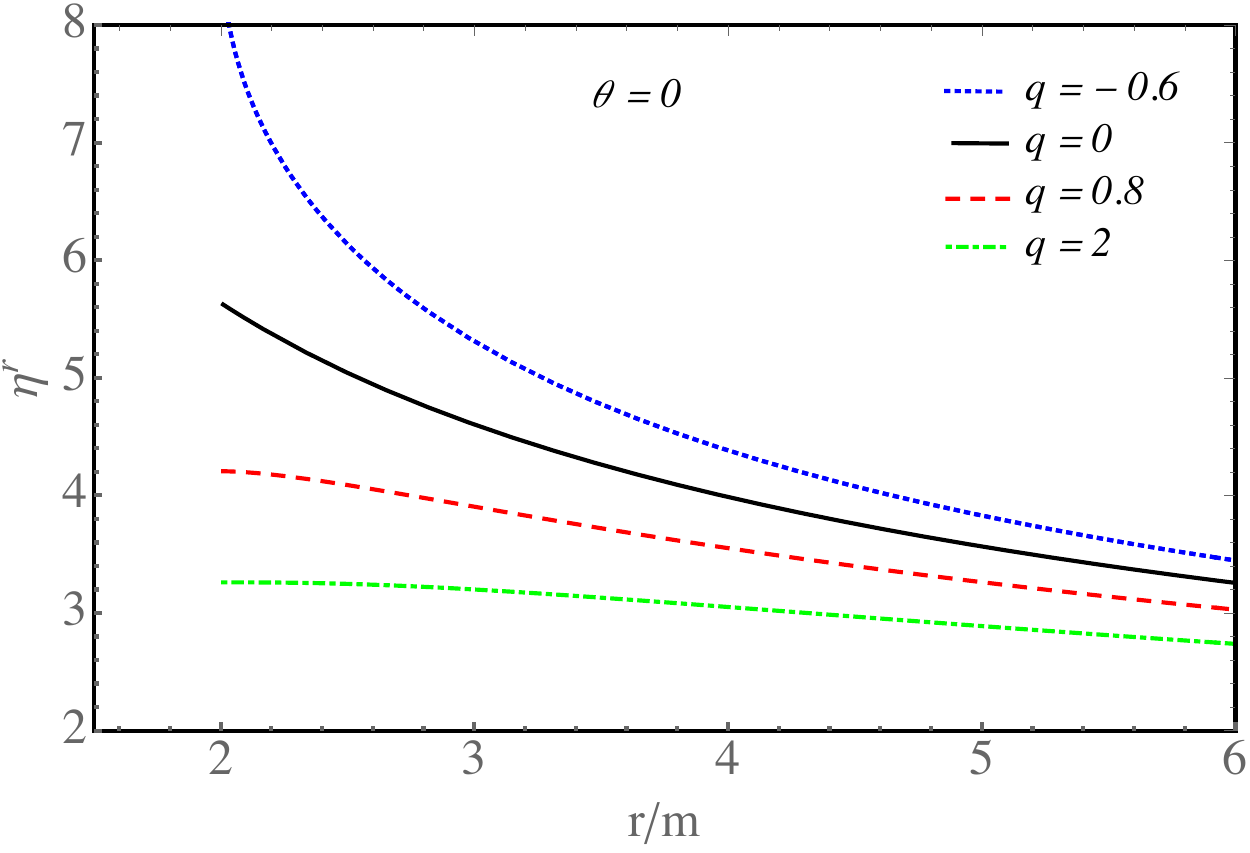}\hfill}

{\hfill
\includegraphics[width=0.45\hsize,clip]{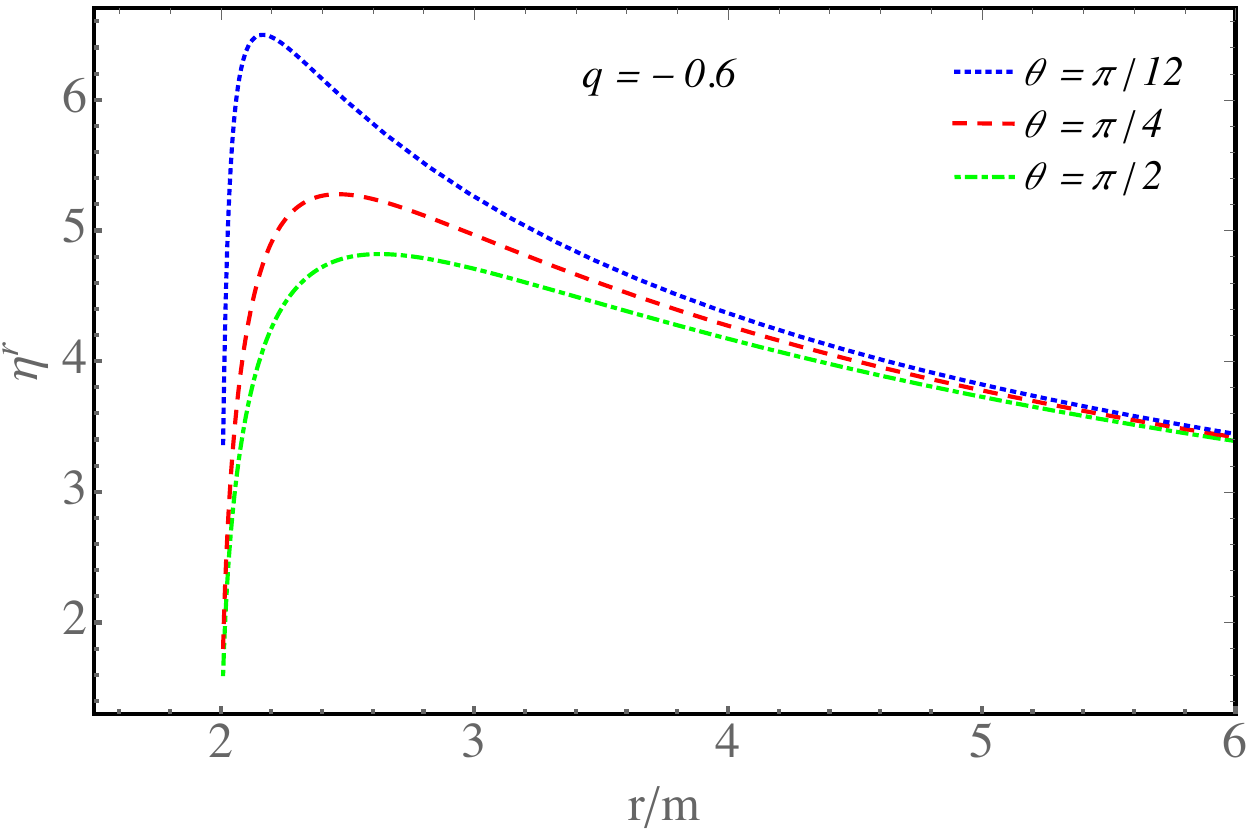}\hfill
\includegraphics[width=0.45\hsize,clip]{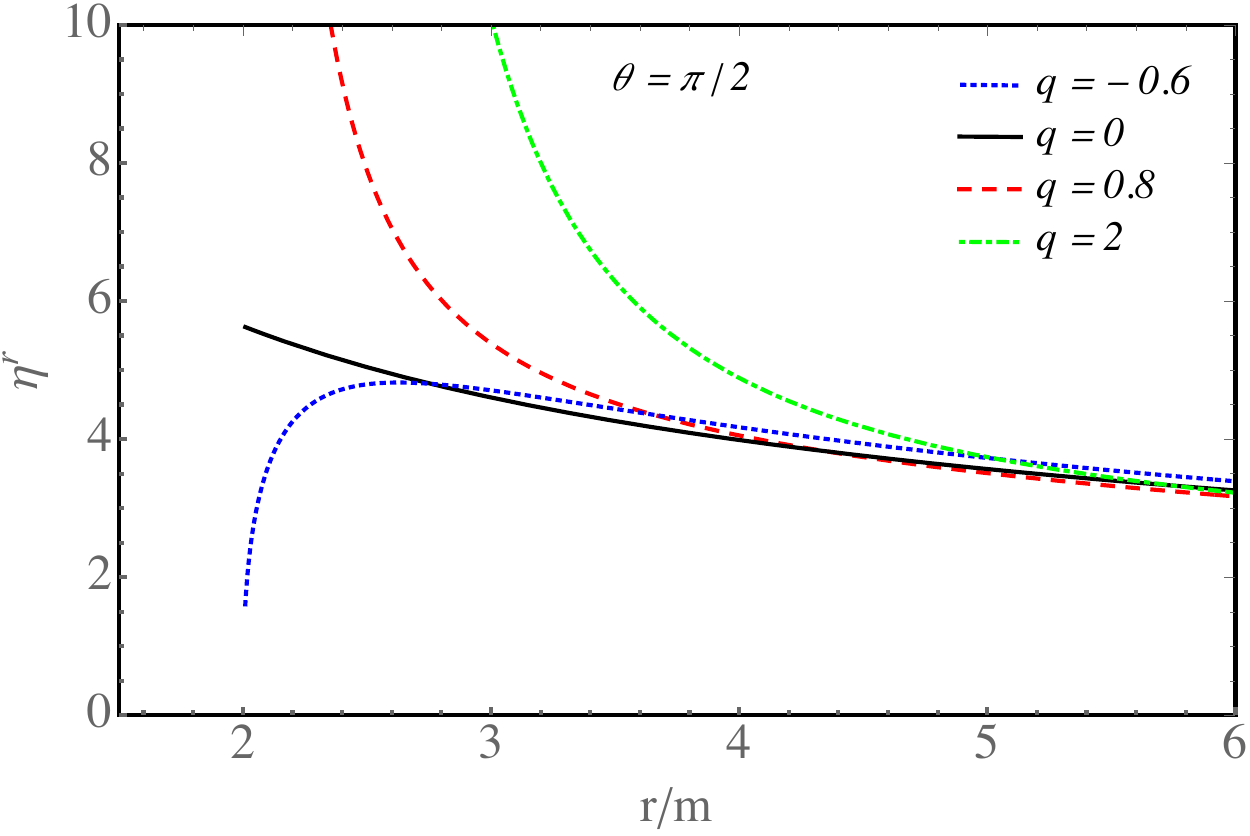}\hfill}

\caption{Radial deviation vector ($\eta^{r}$) on $r/m$. Left panel: for fixed $q$ and arbitrary polar angle.  Right panel: for fixed polar angle ($\theta=\pi/2$ and $\theta=0$) and arbitrary $q$. IC:   $\eta^{r} \left[ r_b\right]=1$, $\dot{\eta}^{r} \left[ r_b\right]=0$, $r_b = 100m$, $m  = 1$.}
\label{fig:rdv}
\end{figure*}
By substituting ~\eqref{eq:transform} into Eqs. ~\eqref{eq:dev_r} -~\eqref{eq:devphi} one can get the full set of ordinary differential equations describing the evolution of the separation between geodesics. We set the initial position for one of the test particles at $r = r_b =100m$ and impose that the particles are initially at rest. Then $\eta^{r}(r_b)$ and $\eta^{\theta}(r_b)$ represent the initial distance of the second particle in the radial and angular directions, respectively. We then choose the other initial conditions
\begin{align}
\eta^{\alpha} (r_b) & = 1 & \dot{\eta}^{\alpha} (r_b) & = 0 \label{eq:IC1}
\end{align}
with $\alpha=r,\theta$. Notice that $\dot{\eta}^{r}$ and $\dot{\eta}^{\theta}$ represent the evolution of the deviation vector in the radial and angular directions, respectively and therefore show that the particles are getting separated when $\dot{\eta}^{\alpha}>0$, while conversely the particles are approaching each other when $\dot{\eta}^{\alpha}<0$.

\section{Radial fall}
\label{sec:radial fall}

Let us consider a neutral test particle radially falling in the geometry given by the metric Eq.~\eqref{eq:metric}. From Eq.~\eqref{eq:GDE} one can calculate the tidal forces for both the radial and the angular components. Since the spacetime is axially symmetric in Fig.~\ref{fig:rtf_atf} we plot the radial profile of radial and angular tidal force for different values of $q$\footnote{For clarity, under angular tidal force we consider polar tidal component, $\theta$, of axially symmetric spacetime, since contrary to spherical symmetric spacetime azimuthal, $\phi$, and polar, $\theta$, angle are not the same values.}. It can be seen that from far away distances at $r=10m$ the deformation parameter does not heavily influence the trajectory of the test particle. But as it approaches the singularity at $r=2m$ the impact of the deformation parameter $q$ plays a role. For example, for the negative value of $q$ the radial tidal force has a lower magnitude compared to the positive ones. Also, at $q=0$ one can recover the Schwarzschild solution. For $q>0$ there are maximum values of the radial tidal force, which implies that the deformation parameter has a critical value. 

From the right panel of Fig.~\ref{fig:rtf_atf}, one can see that for negative values of the quadruple parameter $q$, as a test particle approaches $r=2m$, the angular tidal force has greater influence on the particle, compared to positive values of $q$. As expected, for $q=0$ the angular tidal force corresponds to the Schwarzschild metric.

\begin{figure*}

{\hfill
\includegraphics[width=0.45\hsize,clip]{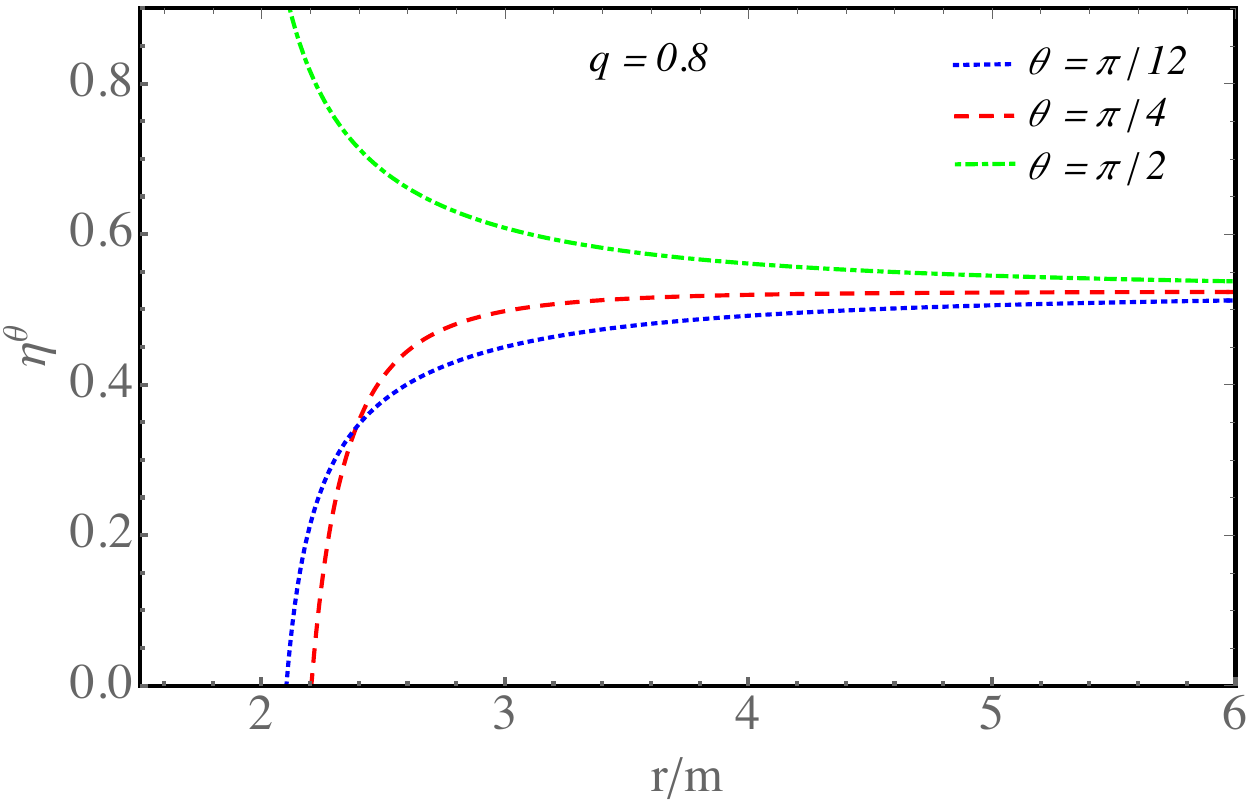}\hfill
\includegraphics[width=0.45\hsize,clip]{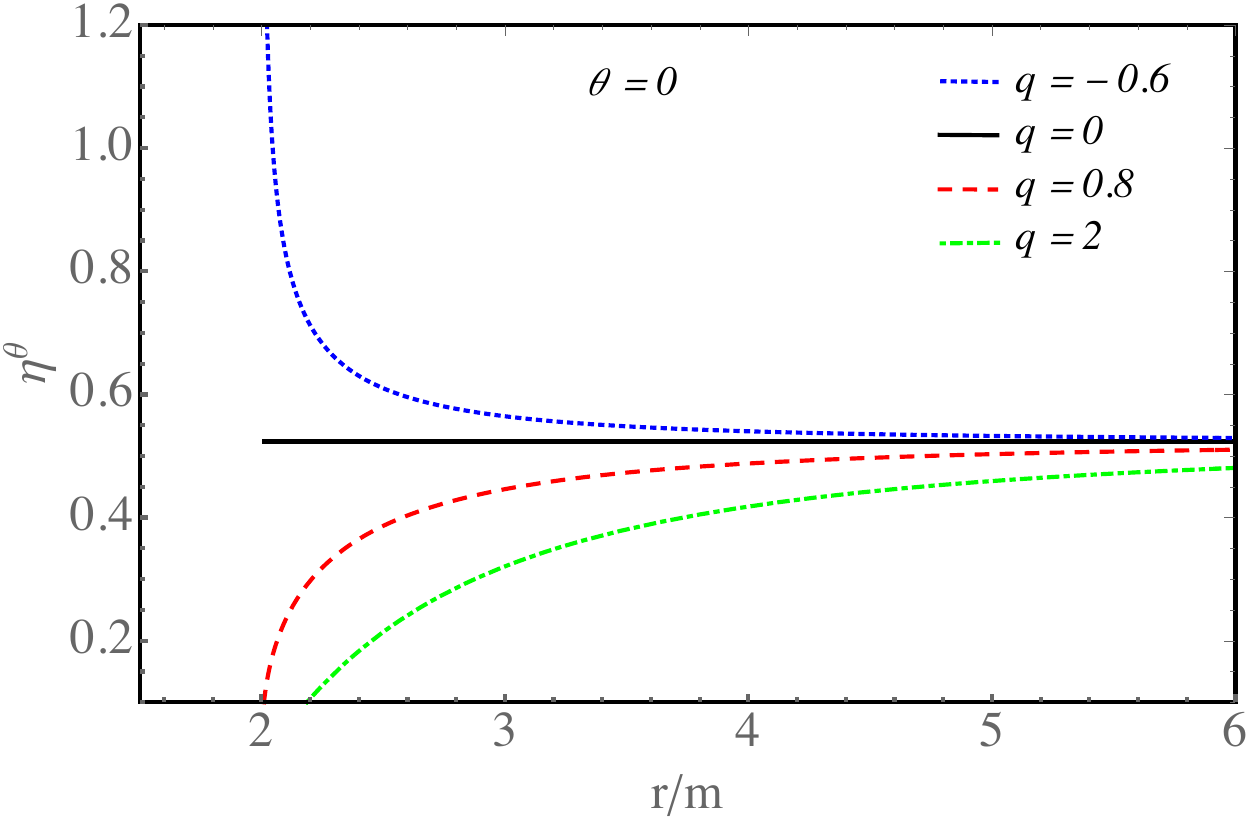}\hfill}

{\hfill
\includegraphics[width=0.45\hsize,clip]{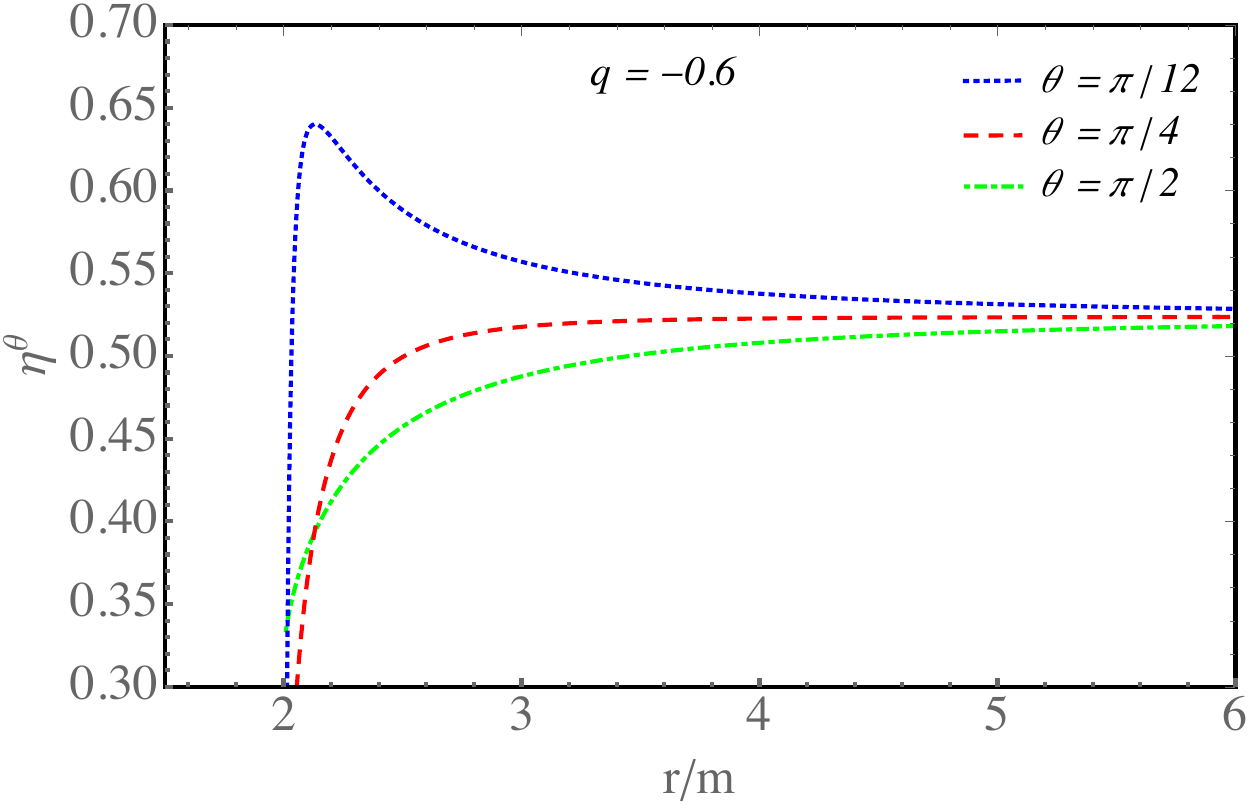}\hfill
\includegraphics[width=0.45\hsize,clip]{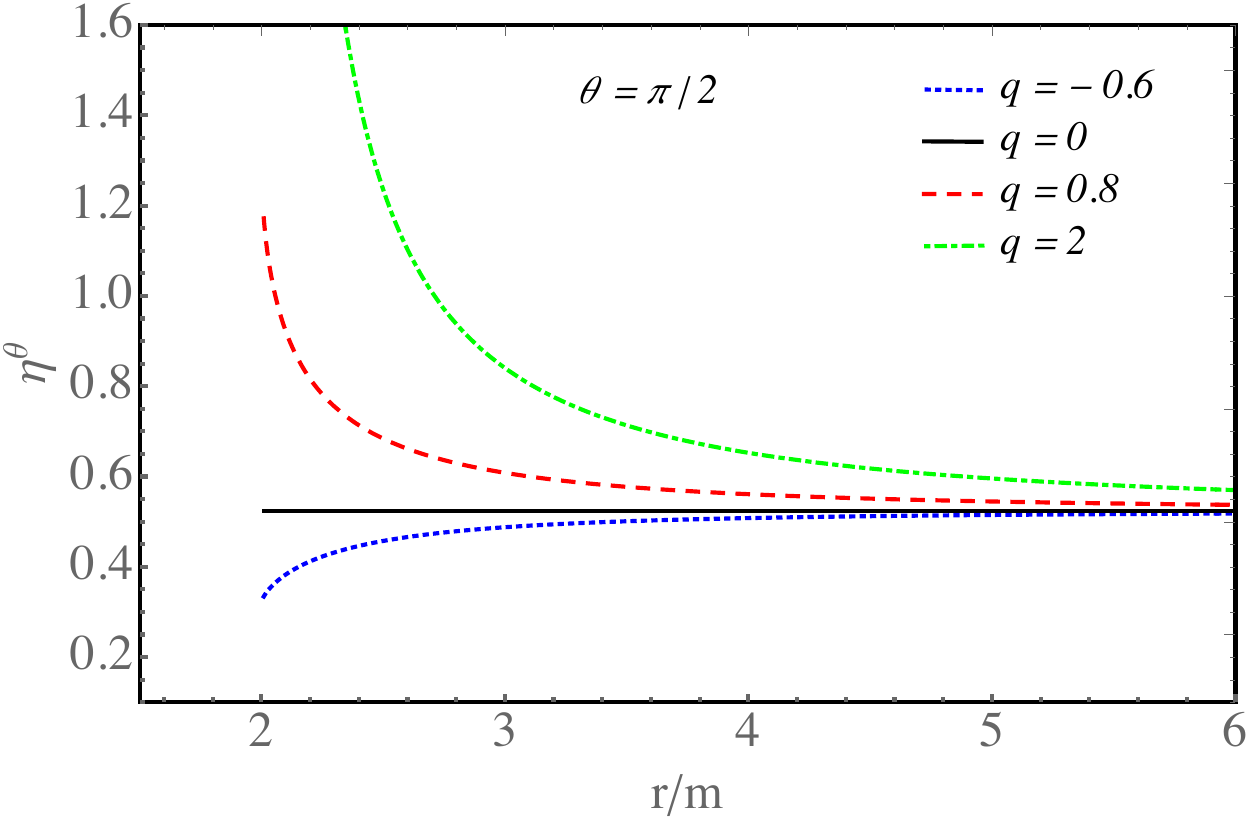}\hfill}

\caption{ Angular deviation vector ($\eta^{\theta}$) on $r/m$. Left panel: for fixed $q$ and arbitrary polar angle. Right panel: for fixed polar angle ($\theta=\pi/2$ and $\theta=0$) and arbitrary $q$. IC:  $\eta^{\theta} \left[ r_b\right]=\pi/6$, $\dot{\eta}^{\theta} \left[ r_b\right]=0$, $r_b = 100m$, $m = 1$.}
\label{fig:adv}
\end{figure*}

\subsection{Geodesic deviation vector vs {\it r}}

The analysis of radial fall of two test particles separated by a radial deviation vector $\eta^{r}$ with constant $\phi$ and $\theta$ and different values of $q$ are presented in Fig.~\ref{fig:rdv}. In the top-left panel, we plot the radial deviation vector for $q = 0.8$ and different $\theta$. We see that $\eta ^r$ increases closer to the equatorial plane as we approach the singularity at $2m$. For negative $q$ (bottom-left panel), particles that move farther from the equatorial plane deviate more strongly compared to particles closer to the equatorial plane. Respectively, in the top-right panel one can see that the radial deviation vector gradually increases as it approaches the surface singularity, and for negative values of the deformation parameter, the increase becomes steeper for test particles that fall perpendicularly to the central object. In the case of the equatorial plane, for positive $q$ we notice the same behavior as before, but for negative $q$ the radial deviation vector decreases, which means that the test body is squeezed in the direction of the fall. 

In Fig.~\ref{fig:adv} we plot how the angular deviation vector changes with the radial distance for different values of $q$ and $\theta$. In the top-left panel, one can see that for $q = 0.8$ the angular deviation vector decreases as it approaches the singularity, except for $\theta = \pi/2$ where it increases. In the bottom-left panel, we see that for a negative value of the deformation parameter in the equatorial plane $\theta=\pi/2$ the angular deviation vector decreases, while at $\theta=\pi/12$ it stretches as it approaches the singularity. For larger values of $\theta$ it gets compressed immediately. Respectively, in the right panel of Fig.~\ref{fig:adv} we see that the behavior of the angular deviation vector is the same for $\theta = 0$ and $\theta = \pi/2$ but with opposite influence from the values of $q$. Specifically, for $\theta=\pi/2$ the compression of the angular deviation vector near $r=2m$ occurs when $q<0$, while for $\theta=0$ when $q>0$. 

\subsection{Radial and angular deviation vs proper time}

In order to consider a more realistic scenario, where the angular orientation, $\theta$, is not constant throughout the radial fall, we should consider the deviation vector as a function of proper time instead of distance. In this case, in addition to the influence of the quadruple parameter $q$, we can see the impact of deviation from the straight radial trajectory, induced by the alteration of $\theta(s)$.

\begin{figure*}[ht]

{\hfill
\includegraphics[width=0.42\hsize,clip]{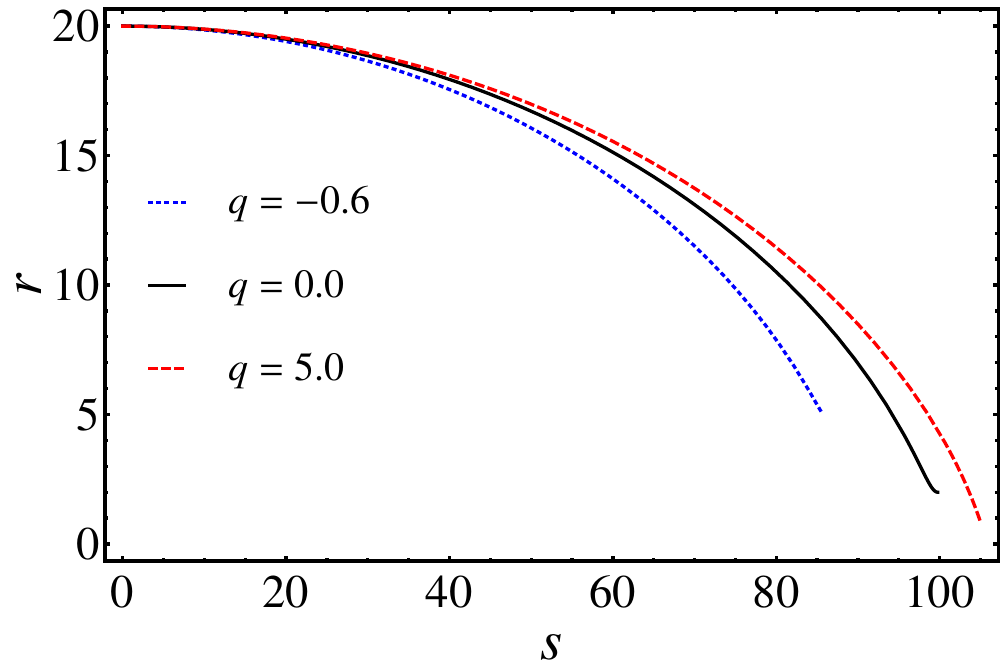}\hfill
\includegraphics[width=0.42\hsize,clip]{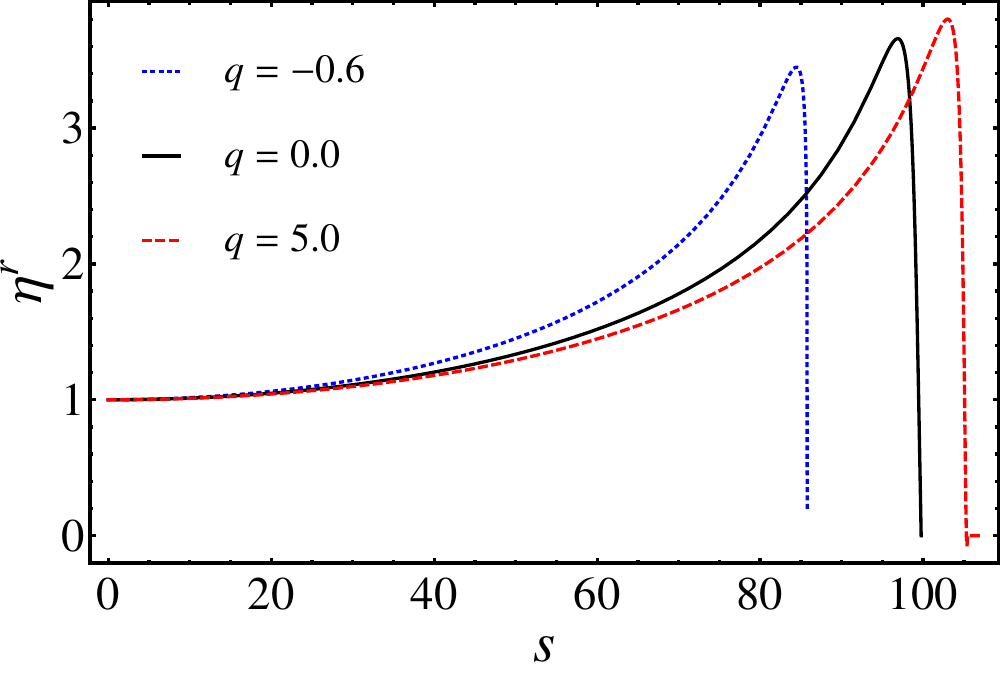}\hfill}

\caption{Radial free fall of test particles at the pole $\theta=0$ of a deformed object with different quadrupole parameter $q$. Left: radial coordinate $r$ versus proper time $s$. Right: radial deviation $\eta^r$ versus $s$. The black solid curve indicate the free fall in the Schwarzschild spacetime. Initial conditions (ICs) along with the parameters of the source are as follows: $M_0=1$, $r_b=20M_0$, $E=0.95$, $r(0)=r_b$, $\theta(0)=0$, $\dot{r}(0)=0$, $\dot{\theta}(0)=0$, $\eta^r(0)=1$, $\eta^{\theta}(0)=0$, $\dot{\eta}^r(0)=0$, $\dot{\eta}^{\theta}(0)=0$.}
\label{fig:theta0}
\end{figure*}

In Fig.~\ref{fig:theta0}, we depict the evolution of the radial coordinates (left panel) with respect to proper time, for a test particle in the gravitational field of a static deformed object with constant total mass and different values of the deformation parameter. The test particle undergoes a free fall starting from the pole. The dependence of the free fall time on $q$ is monotonically increasing, meaning that for larger values of $q$, the longer the free fall time is. In addition to this, no changes are expected in the polar coordinate.

In the right panels of Fig.~\ref{fig:theta0}, we present the evolution of the radial components of the deviation vector as a function of the proper time. Notably, the radial component exhibits spaghettification as test particles approach the central body. As anticipated, there are no changes in the angular component of the deviation vector and therefore they are not presented.

\begin{figure*}

{\hfill
\includegraphics[width=0.42\hsize,clip]{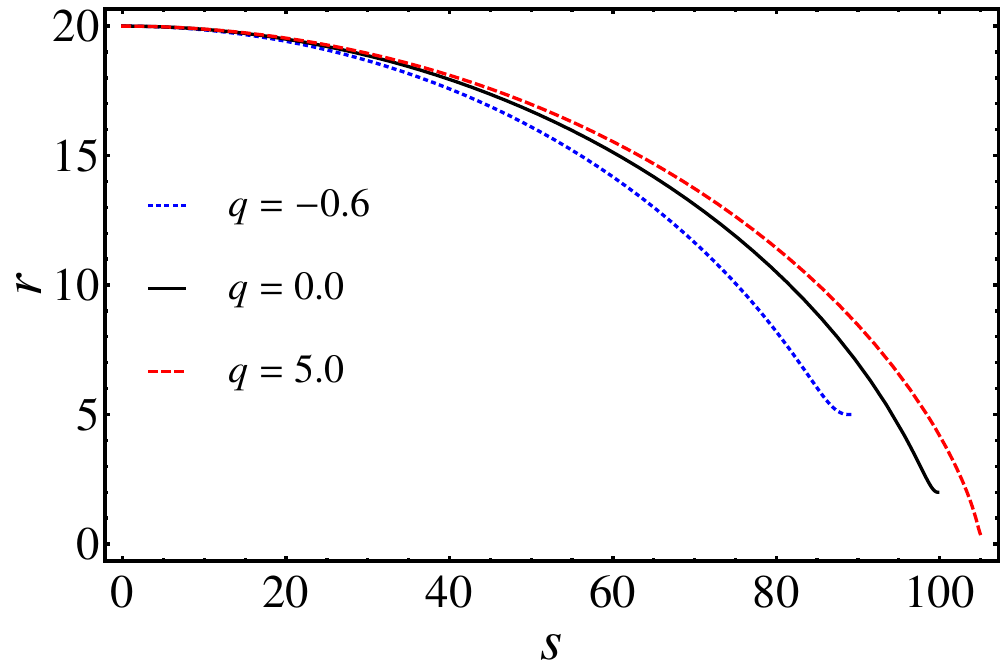}\hfill
\includegraphics[width=0.42\hsize,clip]{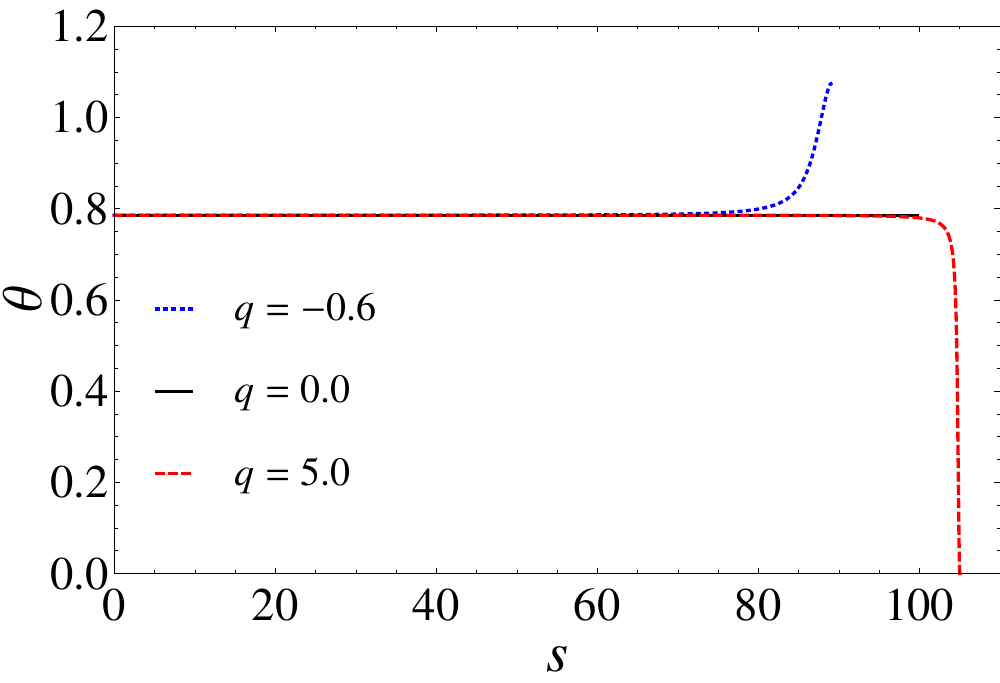}\hfill}

{\hfill
\includegraphics[width=0.42\hsize,clip]{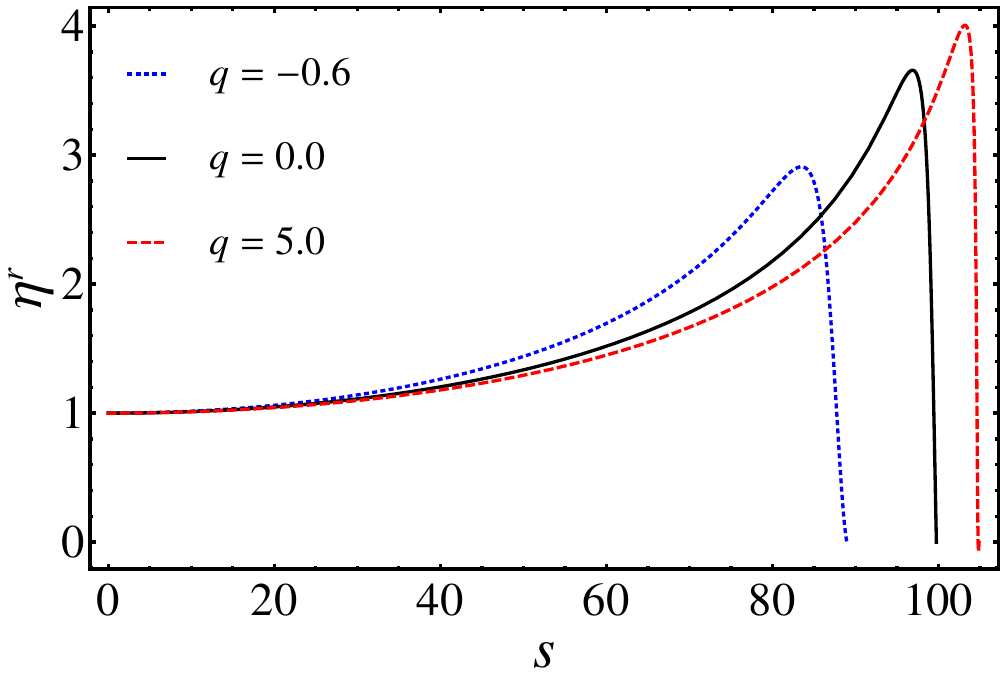}\hfill
\includegraphics[width=0.42\hsize,clip]{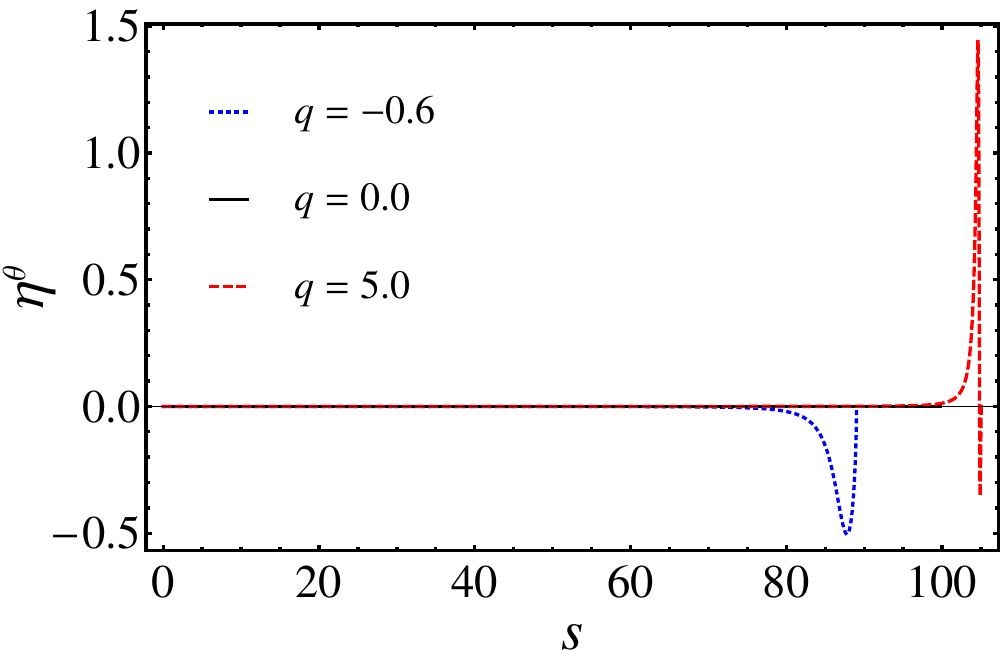}\hfill}

\caption{Radial free fall of test particles at $\theta=\pi/4$ of a deformed object with different quadrupole parameter $q$. Top panel. Left: radial coordinate $r$ versus proper time $s$. Right: polar coordinate $\theta$ versus $s$. Bottom panel. Left: radial deviation $\eta^r$ versus $s$. Right: angular deviation $\eta^{\theta}$ versus $s$. ICs are the same as in Fig.~\ref{fig:theta0} with the only exception: $\theta(0)=\pi/4$. }
\label{fig:theta45}
\end{figure*}

Similarly, in the top panel of Fig.~\ref{fig:theta45} the test particle undergoes a free fall, starting from the off-equatorial plane at $\theta=\pi/4$. The shape of the radial coordinate undergoes slight changes, and the free fall time changes for different values of $q$, increasing for prolate sources and decreasing for oblate ones. In contrast to the previous case, the polar coordinate varies with the deformation parameter. Specifically, for prolate sources ($q<0$), $\theta$ increases, whereas for oblate sources ($q>0$) it decreases.

In the bottom panel of Fig.~\ref{fig:theta45}, we depict the evolution of the radial and angular components of the deviation vector versus the proper time. Once again, the radial component demonstrates the spaghettification effect. Notably, in the case of the angular component of the deviation vector, the situation is intriguing: for prolate sources, $\eta^{\theta}$ decreases, while for oblate ones it increases as the two test particles approach the source of gravity.


\begin{figure*}

{\hfill
\includegraphics[width=0.42\hsize,clip]{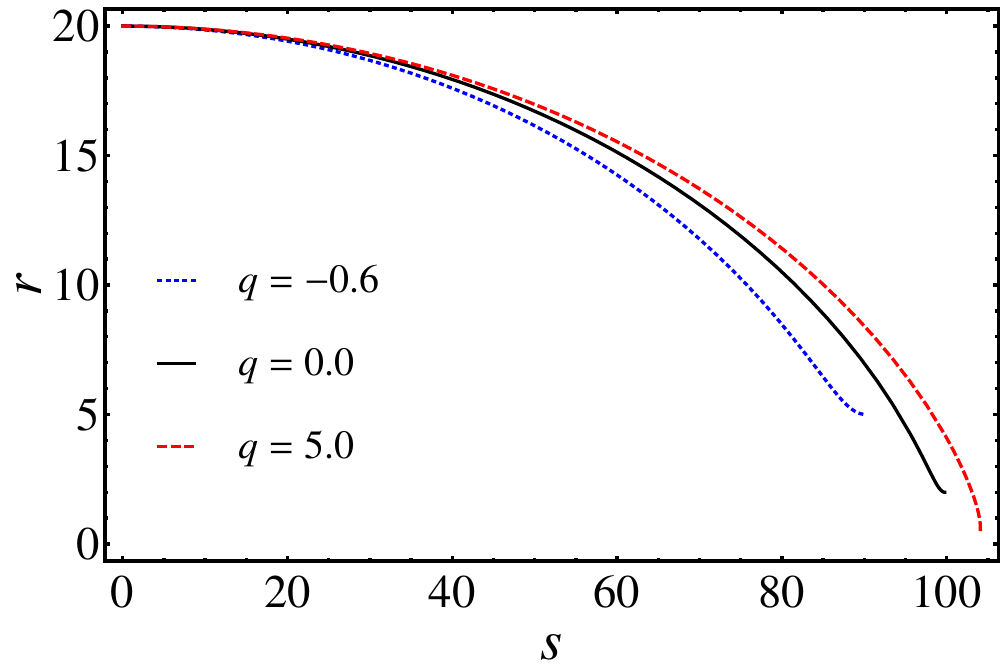}\hfill
\includegraphics[width=0.42\hsize,clip]{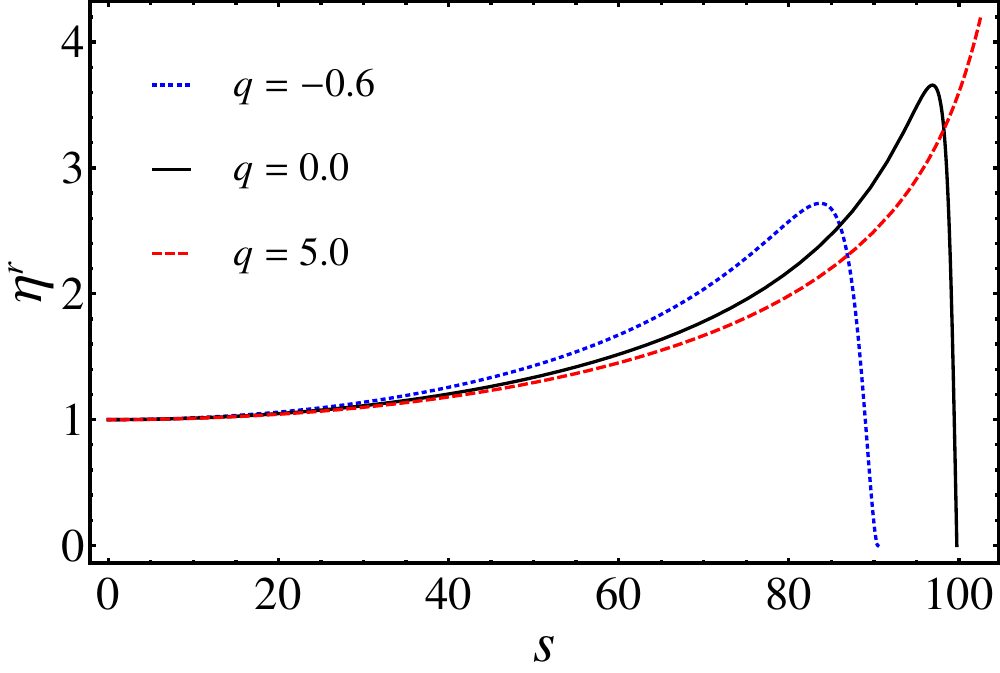}\hfill}

\caption{Radial free fall of test particles at the equatorial plane $\theta=\pi/2$ of a deformed object with different quadrupole parameter $q$. Left: radial coordinate $r$ versus proper time $s$. Right: radial deviation $\eta^r$ versus $s$. ICs are the same as in Fig.~\ref{fig:theta0} with the only exception $\theta(0)=\pi/2$. }
\label{fig:theta90}
\end{figure*}

In the same way, in Fig.~\ref{fig:theta90} we plot the same quantities, but in the equatorial plane, where $\theta=\pi/2$. We notice that the free fall time increases (decreases) for oblate (prolate) sources (left panel). Moreover, in the right panel, it is evident that for prolate sources, $\eta^r$ is smaller than in the field of oblate sources, indicating that the spaghettification effect is more pronounced in the field of oblate sources in the equatorial plane. As expected, $\theta$ and $\eta ^\theta$ are constant throughout the fall, therefore they are not presented here.

\begin{figure*}
{\hfill
\includegraphics[width=0.44\hsize,height=0.275\hsize,clip]{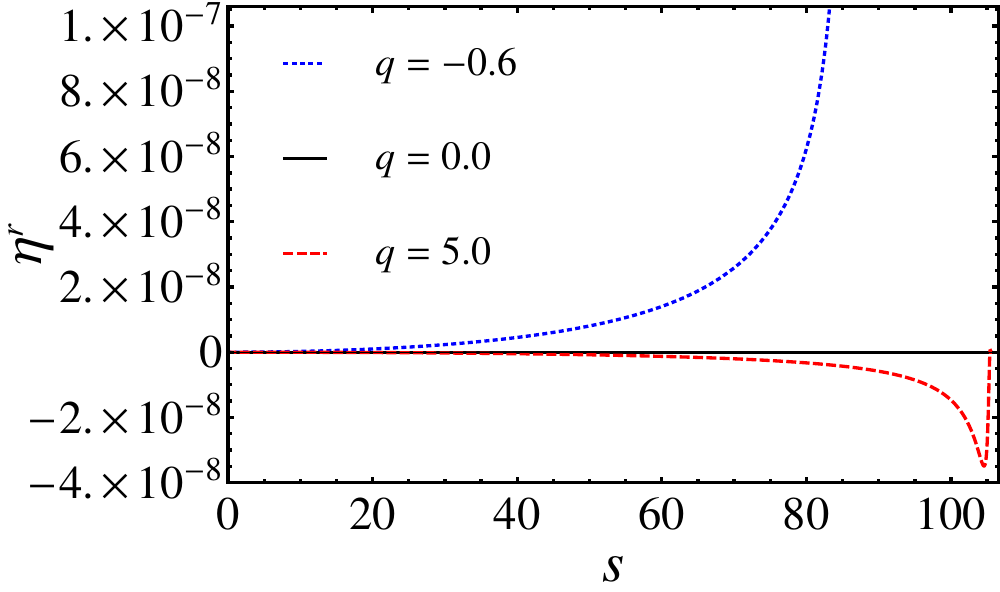}\hfill
\includegraphics[width=0.42\hsize,clip]{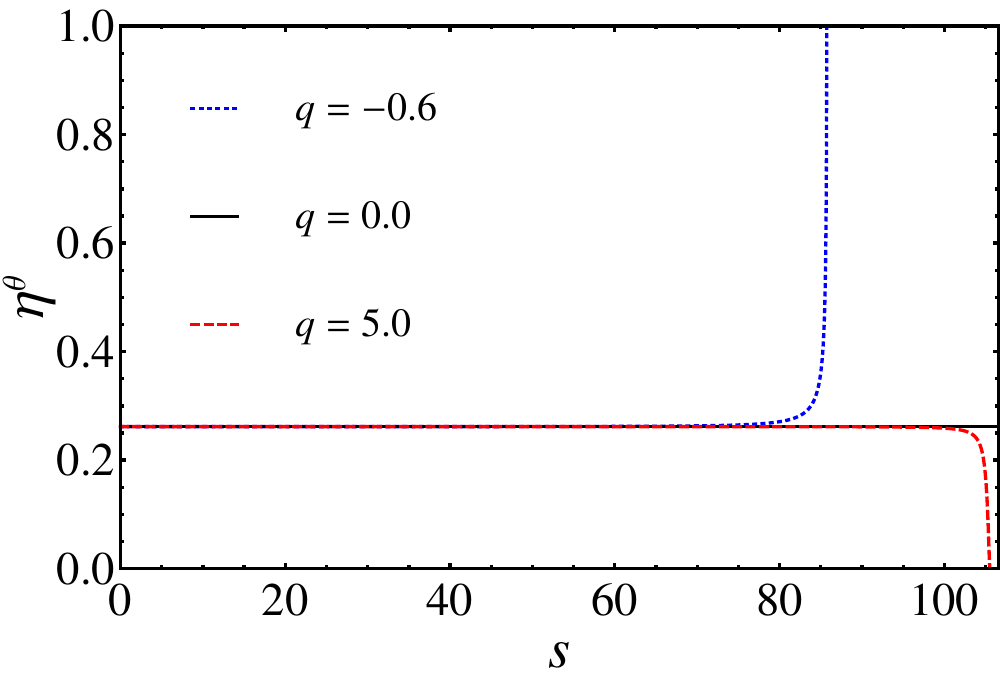}\hfill}
\caption{Radial free fall of test particles at a pole $\theta=0$ and $\eta^{\theta}=\pi/12$ in the field of a deformed object with different quadrupole parameter $q$.  Left: radial deviation $\eta^r$ versus $s$. Right: angular deviation $\eta^{\theta}$ versus $s$. ICs are $M_0=1$, $r_b=20M_0$, $E=0.95$, $r(0)=rb$,  $\dot{r(0)}=0$, $\dot{\theta(0)}=0$, $\eta^r(0)=0$,  $\dot{\eta}^r(0)=0$, $\dot{\eta}^{\theta}(0)=0$. }
\label{fig:1theta0}
\end{figure*}

\begin{figure*}
{\hfill
\includegraphics[width=0.42\hsize,clip]{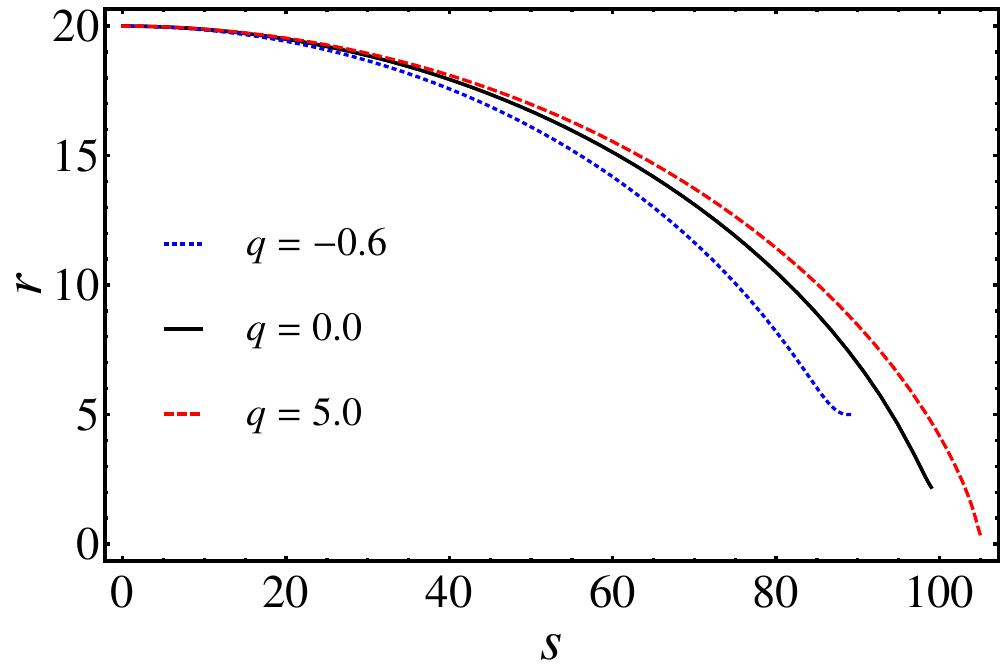}\hfill
\includegraphics[width=0.42\hsize,clip]{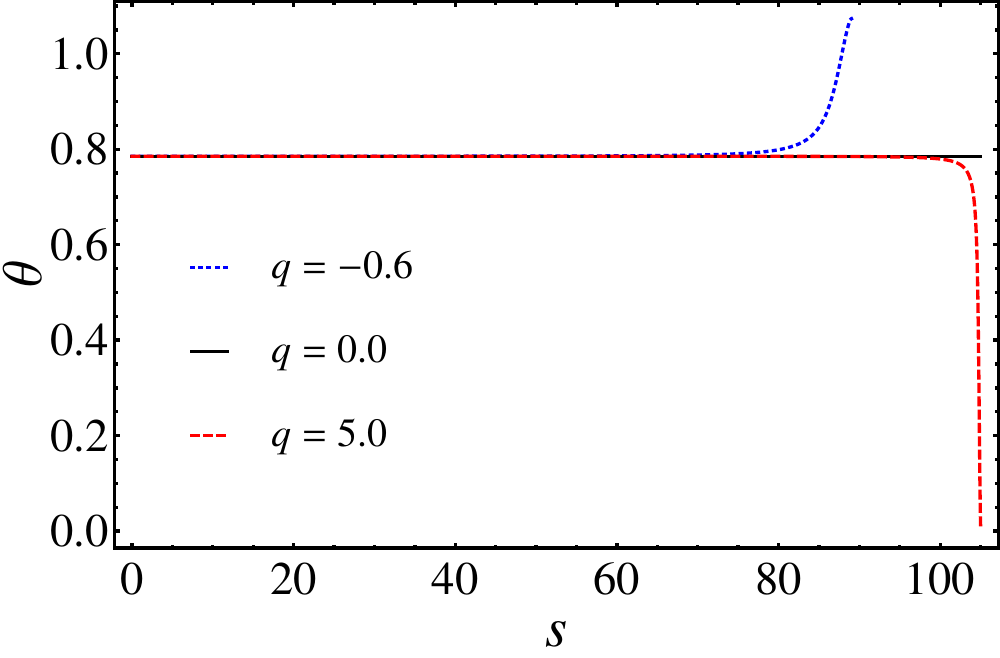}\hfill}

{\hfill
\includegraphics[width=0.42\hsize,clip]{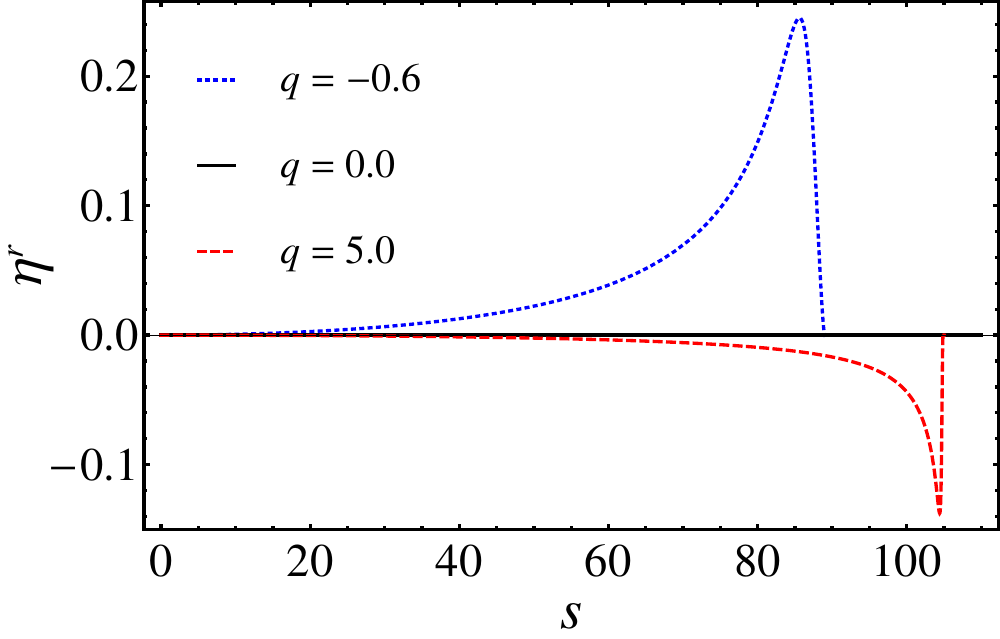}\hfill
\includegraphics[width=0.42\hsize,clip]{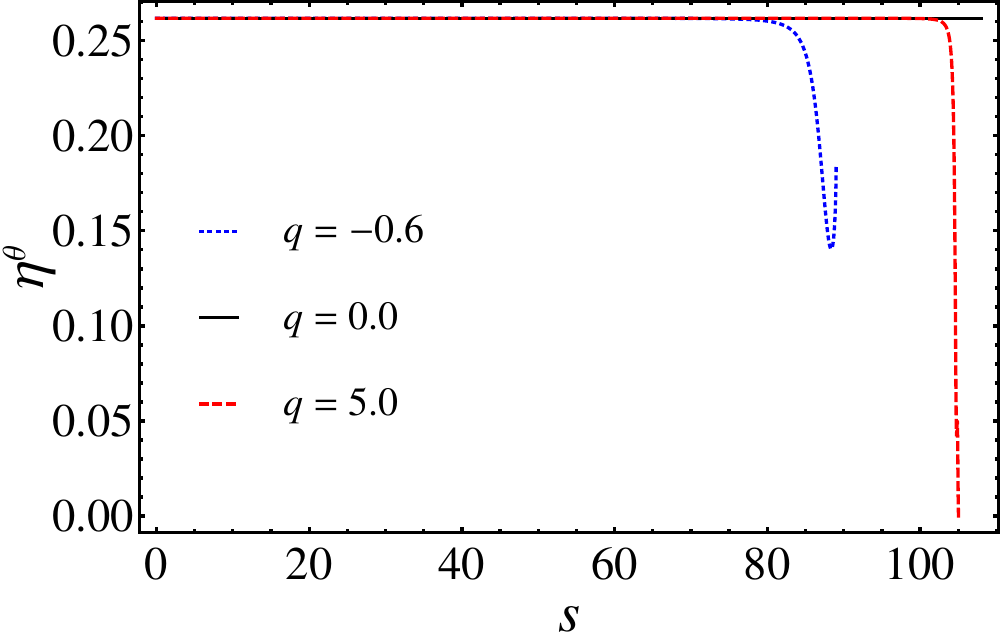}\hfill}
\caption{Radial free fall of test particles $\theta=\pi/4$ and $\eta^{\theta}=\pi/12$ of a deformed object with different quadrupole parameter $q$. Top panel. Left: radial coordinate $r$ versus proper time $s$. Right: polar coordinate $\theta$ versus $s$. Bottom panel. Left: radial deviation $\eta^r$ versus $s$. Right: angular deviation $\eta^{\theta}$ versus $s$. ICs are the same as in Fig.~\ref{fig:1theta0}. }
\label{fig:2theta45}
\end{figure*}

\begin{figure*}

{\hfill
\includegraphics[width=0.42\hsize,clip]{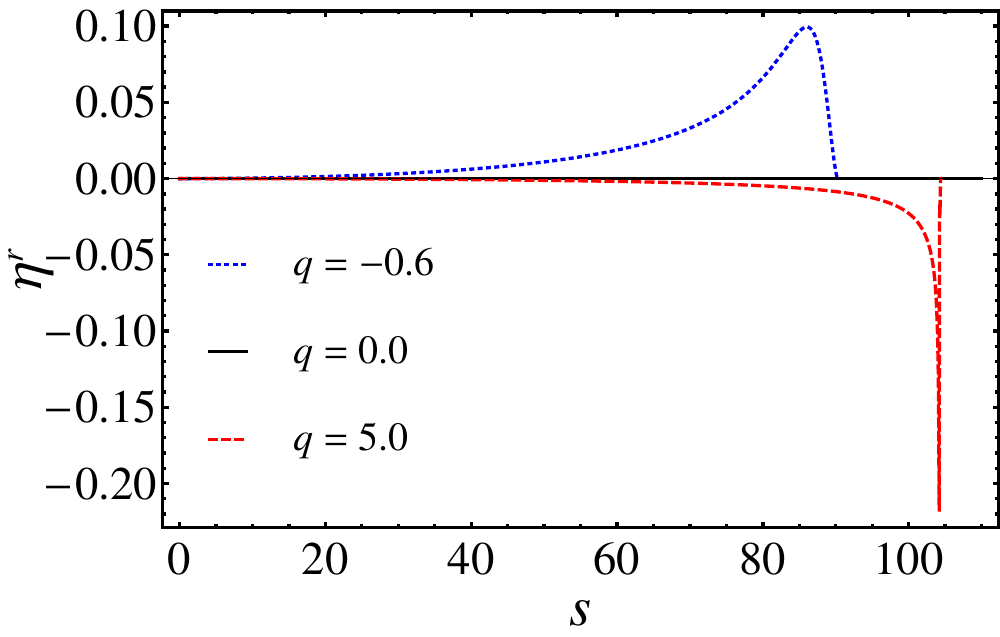}\hfill
\includegraphics[width=0.42\hsize,clip]{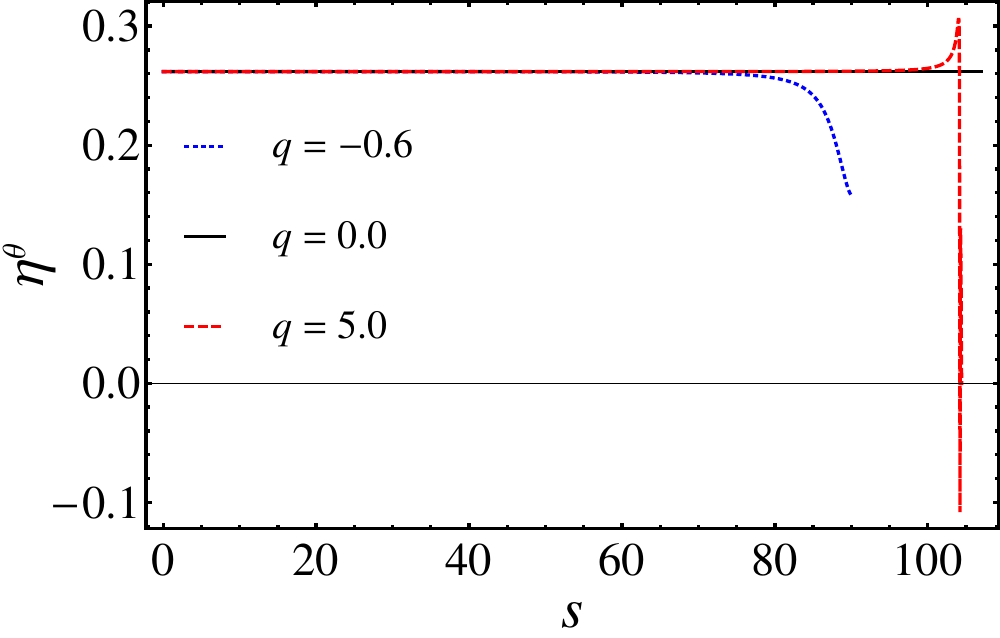}\hfill}
\caption{Radial free fall of test particles at $\theta=5\pi/12$ and $\eta^{\theta}=\pi/12$ of a deformed object with different quadrupole parameter $q$. Left: radial deviation $\eta^r$ versus $s$. Right: angular deviation $\eta^{\theta}$ versus $s$. ICs are the same as in Fig.~\ref{fig:1theta0}. }
\label{fig:3theta75}
\end{figure*}

In Figs.~\ref{fig:1theta0}-\ref{fig:3theta75} we illustrate $r$ and $\theta$ coordinates together with $\eta^r$ and $\eta^{\theta}$ components of the deviation vector, but instead of having $\eta ^\theta (0) = 0$, as before, we set the two particles apart with $\eta^\theta(0) = \pi/12$, and we study what happens for $\theta \in \{0,5\pi/12,\pi/2\}$ respectively. Their radial separation is kept vanishing, $\eta ^r = 0$. 

In particular, in Fig.~\ref{fig:1theta0}, the first particle is dropped from a pole and its angular coordinate remains unchanged throughout the free fall, regardless of the value of $q$. However, its radial coordinate decreases as shown in the left panel of Fig.~\ref{fig:theta0} . The parameter $q$ influences both the shape and size of the source with a fixed total mass $M_0$, thereby affecting the free fall time. Hence, in Fig.~\ref{fig:1theta0}, we depict the radial component (left) and the angular component (right) of the separation vector between the two particles. It is evident that for oblate (prolate) sources with $q>0$ ($q<0$), the separation decreases (increases), while for the $q=0$ case, no changes are observed.

In Fig.~\ref{fig:2theta45}, following the pattern of Fig.~\ref{fig:1theta0}, we depict the geodesic deviation of two particles with identical radial and angular separations. Here, the first particle is released at $\theta=\pi/4$ (top panel). We observe that its angular coordinate varies: it decreases (increases) for oblate (prolate) sources (right panel), while its radial coordinate behaves similarly to Fig.~\ref{fig:theta45} (top left panel). The behavior of the radial component of the separation vector reflects this trend: it decreases (increases) for oblate (prolate) sources (bottom left panel). In particular, the angular component of the separation vector consistently decreases (bottom right panel). This implies that regardless of $q$, excluding the cases where $q=0$, the two particles approach each other during their descent.

In Fig.~\ref{fig:3theta75}, the first particle is dropped at $\theta=5\pi/12$, while the second particle is released at the equatorial plane. Similarly to previous scenarios, the radial coordinate decreases. However, the angular coordinate now increases (decreases) for prolate (oblate) sources similar to the top right panel of Fig.~\ref{fig:2theta45}. Correspondingly, the radial component of the separation vector increases (decreases) for prolate (oblate) sources, while the opposite behavior is observed in the angular component of the separation vector.


\section{Interpretation of the results}
\label{sec:interpretation}

\begin{figure*}

{\hfill
\includegraphics[width=0.28\hsize,clip]{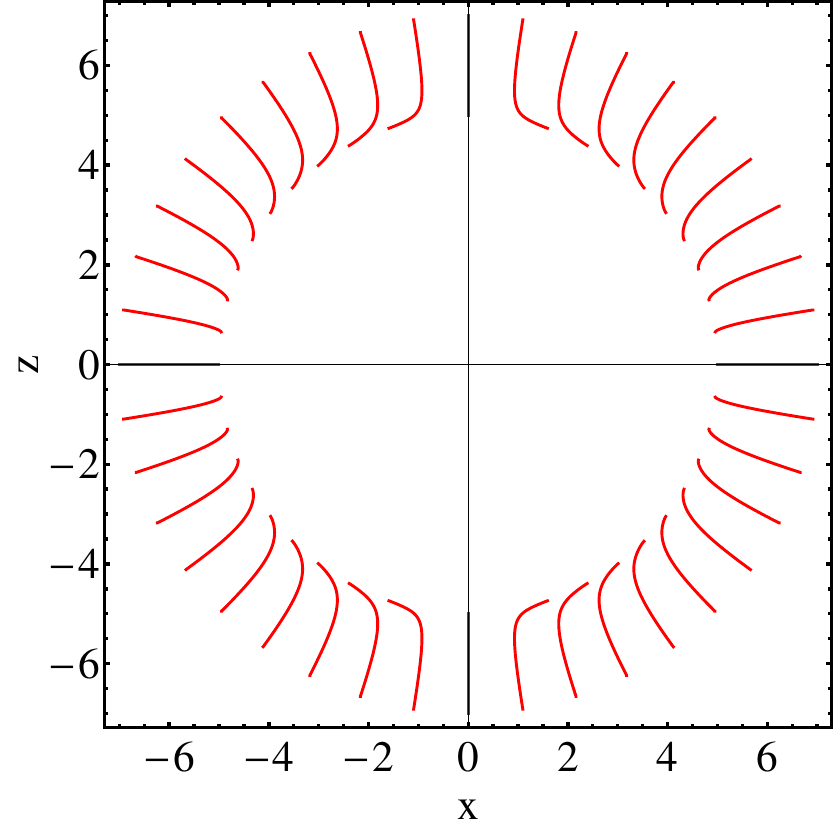}\hfill
\includegraphics[width=0.28\hsize,clip]{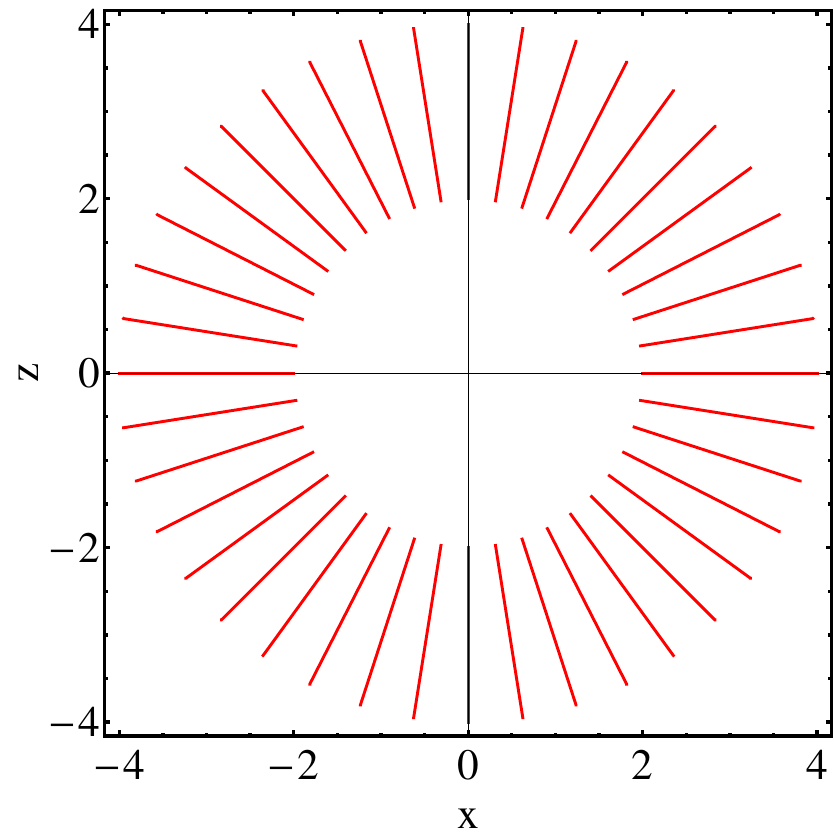}\hfill
\includegraphics[width=0.28\hsize,clip]{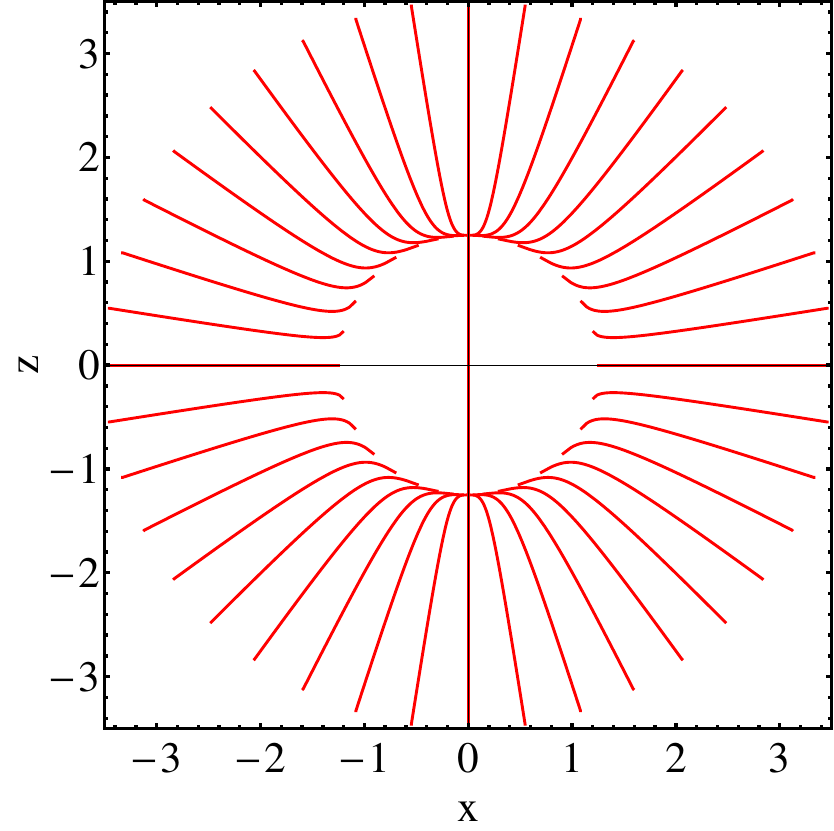}\hfill}

{\hfill
\includegraphics[width=0.28\hsize,clip]{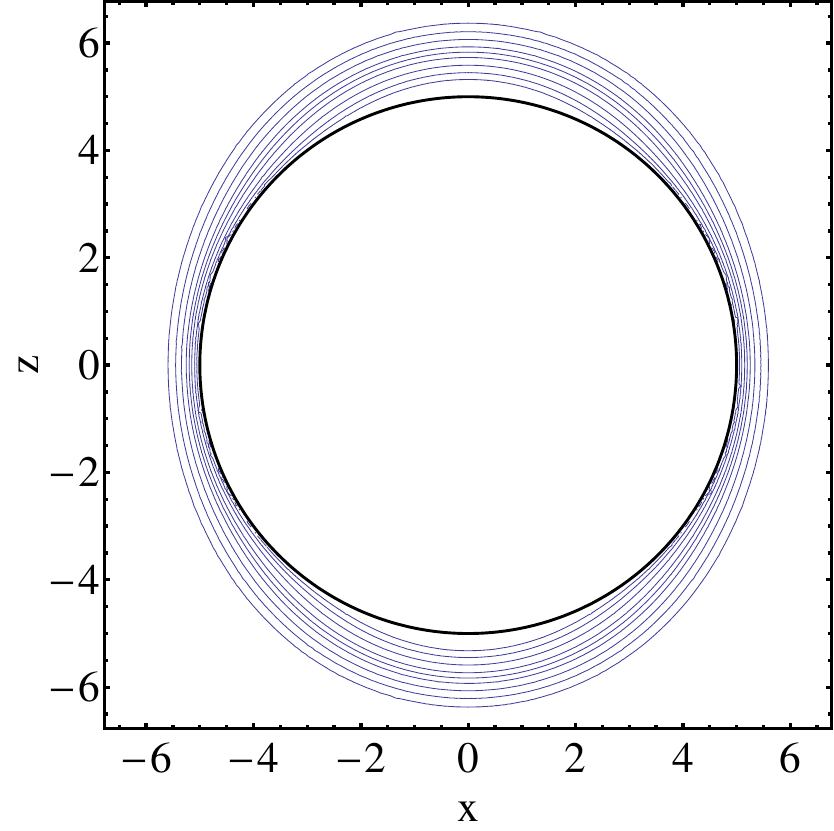}\hfill
\includegraphics[width=0.28\hsize,clip]{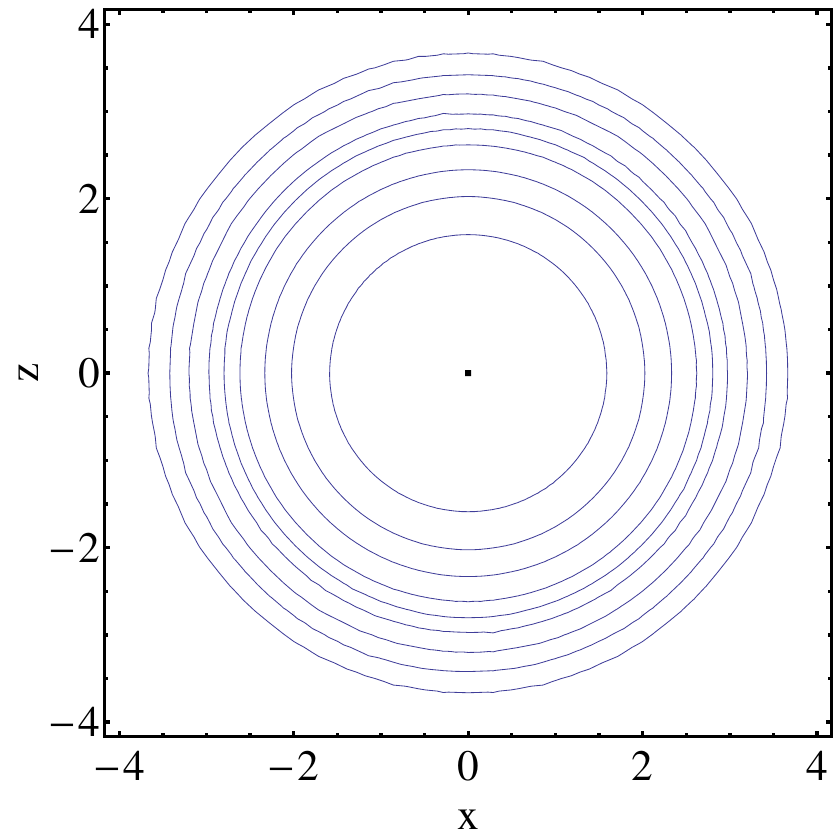}\hfill
\includegraphics[width=0.28\hsize,clip]{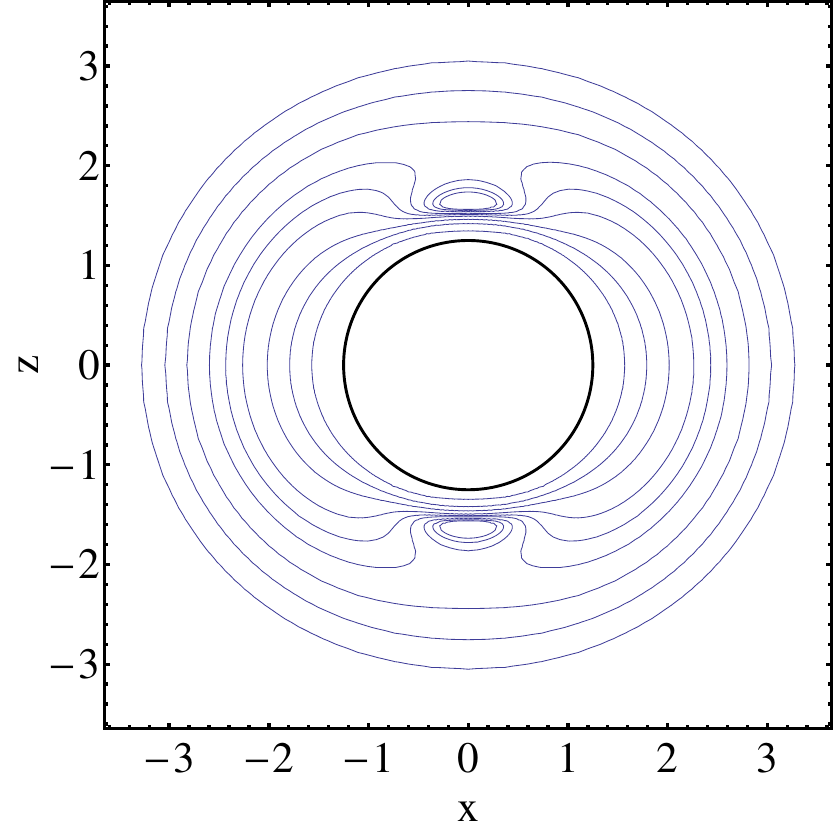}\hfill}

\caption{Top panel: Radial free fall geodesics of test particles on the $xz$-plane at different $\theta$ with a step of $\pi/20$ in the field of prolate ($q=-0.6$, left), spherical ($q=0$, middle) and oblate ($q=0.6$, right) sources. The initial radial coordinates of test particles are $\dot{r(0)}=7M_0$ (left),  $\dot{r(0)}=4M_0$ (middle) and $\dot{r(0)}=3.5M_0$ (right). The total mass of the source is $M_0=1$ km,  $x=r\sin\theta$ and $z=r\cos\theta$. Bottom panel: The Kretschmann scalar on the $xz$-plane. The values of $M_0$ and $q$ are the same as in the top panel. For the Kretschmann scalar we adopted values of [0.02, 0.03, 0.045, 0.07, 0.1, 0.15, 0.3, 0.7, 3.0] from outer to inner contours. Only the central circles of radius $r=2m=2M_0/(1+q)$ for the left and right panels, and the dot at the center of the middle panel show the Kretschmann scalar tending infinity, corresponding to a singularity.}
\label{fig:rfall_kretsch}
\end{figure*}

In order to understand the results presented in the previous section, we examine the radial free fall geodesics of test particles in the gravitational field of a deformed object described by the $q$-metric, on the $xz-$plane for different angles $\theta$, as depicted in Fig.~\ref{fig:rfall_kretsch} (top panel). It is obvious that in Schwarzschild spacetime, the geodesics of particles in radial fall are straight lines, connecting the initial points with the singularity at the center. However, the situation undergoes a significant change when considering the deformation of the source.

For prolate sources, the geodesics deviate from the initial angle $\theta$, towards the equatorial plane ($z=0$). Particles that initiate their motion in the equatorial plane and on the symmetry axis ($z$) will not experience any deviation. In contrast, for oblate sources, the deviation occurs towards the $z$ axis. These findings seem contradicting with our expectations, as one anticipates completely opposite behavior. We show in the following that this is the problem of the coordinates of the $q$-metric itself. In any case, Fig.~\ref{fig:rfall_kretsch} (top panel) summarizes the results presented in Figs.~\ref{fig:theta0} to ~\ref{fig:3theta75}.

It is worth mentioning that in Newtonian gravity, analogous to electrodynamics, one can construct gravitational force lines, which, in principle, could be utilized to explain geodesic deviations. Although in GR, gravity is not considered to be a force, the geodesics of test particles in radial free fall can be formally regarded as an analogy to force lines. Another crucial concept borrowed from electrodynamics is an equipotential surface in space or an equipotential contour on a plane. Once the potential of the sources is defined by assigning constant values to it, one can easily construct equipotential surfaces or contours. In the context of the $q$-metric, the gravitational potential can be readily determined from the condition $\Phi=(1/2)\log (g_{tt})$. Using the $g_{tt}$ component of the $q$-metric, the equipotential contours of deformed objects manifest themselves as concentric circles. This is attributed to the fact that $g_{tt}$ does not depend on the angle $\theta$. Initially, this may appear contradictory to common sense, as one might expect the equipotential contours of the deformed source to also be deformed, resembling spheroids or ellipsoids. However, this contradiction arises due to the specific choice of coordinates in the $q$-metric. Additionally, this choice makes it nontrivial to find the Newtonian limit of the $q$-metric at these coordinates. The weak field regime of the $q$-metric has been explored in \cite{2019PhRvD..99d4005A} using new approximate coordinate transformations that allow $g_{tt}$ to be a function of radial and angular coordinates $(r,\theta)$. However, exact coordinate transformations that enable the derivation of $g_{tt}$ in terms of both $(r,\theta)$ have not yet been discovered, a task that falls outside of the scope of the current work.

\subsection{The Kretschmann scalar}
As an alternative approach, rather than constructing equipotential surfaces (contours), we turn our attention to the Kretschmann scalar, which provides a measure of the curvature strength at a given point in spacetime. The higher the value of the Kretschmann scalar, the stronger the curvature and gravitational effects, and vice versa. In most general cases, the Kretschmann scalar is used in GR to search for real singularities of metrics. It is important to note that the Kretschmann scalar and equipotential surfaces are conceptually distinct; however, both offer insights into the strength of a gravitational field.

The Kretschmann scalar is defined as $K=R_{\alpha\beta\gamma\delta}R^{\alpha\beta\gamma\delta}$. A straightforward computation of it from the line element Eq.~\eqref{eq:metric} yields
\be
K=F_1F_2\left(F_3+\left[F_4+(1+q)^2 r^2\right] m^2 \sin ^2\theta\right) ,\ 
\ee
with
\bea
F_1&=&\frac{48m^2(1 + q)^2}{r^6(r-2m)^2(r(r-2m)+ m^2\sin^2\theta)} ,\ \\
F_2&=&\left(1-\frac{2m}{r}\right)^{2q}\left(1+\frac{m^2\sin^2\theta}{r^2-2mr}\right)^{2q(2+q)} ,\ \\
F_3&=&r(r-2m)(r-(2+q)m)^2 ,\ \\
F_4&=&(2+q)^2m(m(1+q+q^2/3)-(1+q)r) .\ \quad 
\eea
For vanishing $q=0$, we recover the Schwarzschild value $K=48m^2/r^6$. Unlike in the Schwarzschild metric, the hyper-surface $r=2m$ in the $q$-metric indicates a naked singularity, and the surface defined by the equation $r(r-2m)+ m^2\sin^2\theta=0$ shows a curvature singularity (for more details see \cite{Boshkayev:2015jaa}).

In Fig.~\ref{fig:rfall_kretsch} (bottom panel), we illustrate the contours of the Kretschmann scalar ranging from 0.02 to infinity for prolate, spherical, and oblate sources from left to right, respectively. For prolate and spherical sources, the contours of the Kretschmann scalar resemble the shape of a prolate spheroid and a sphere, respectively, as expected. Conversely, for an oblate source, the contours exhibit a resemblance to an ellipsoid, particularly near or at the surface singularity. Moreover, we observe additional contours along the $z$-axis, which bear a similar shape to the Kretschmann scalar in the Kerr metric \cite{2016PhRvD..94d4006S}.

\subsection{Post-Newtonian limit of the {\it q}-metric}

\begin{figure*}

{\hfill
\includegraphics[width=0.28\hsize,clip]{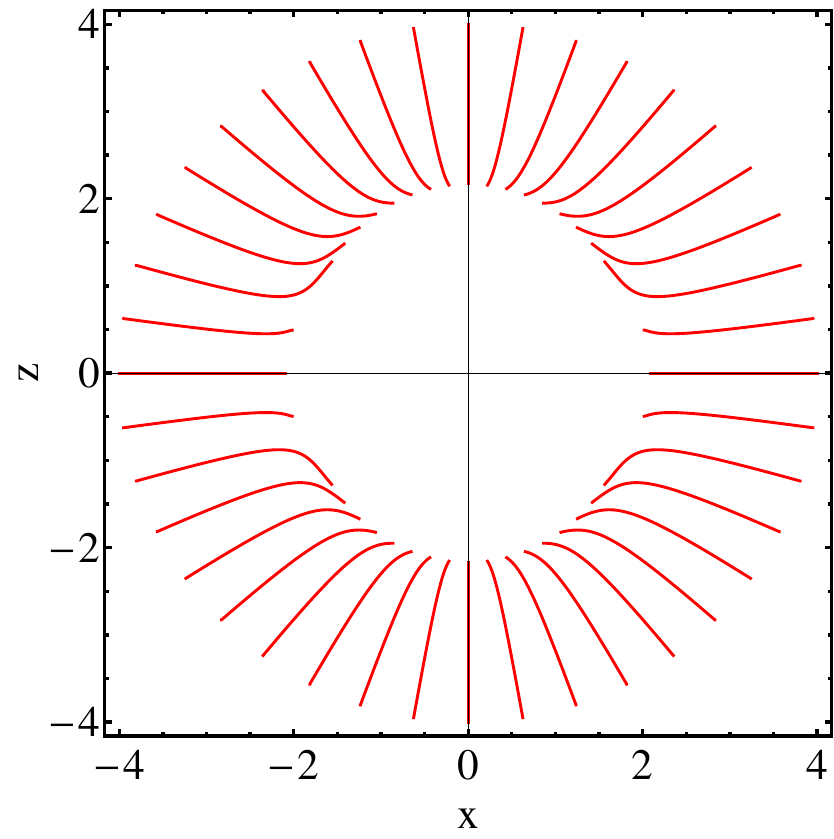}\hfill
\includegraphics[width=0.28\hsize,clip]{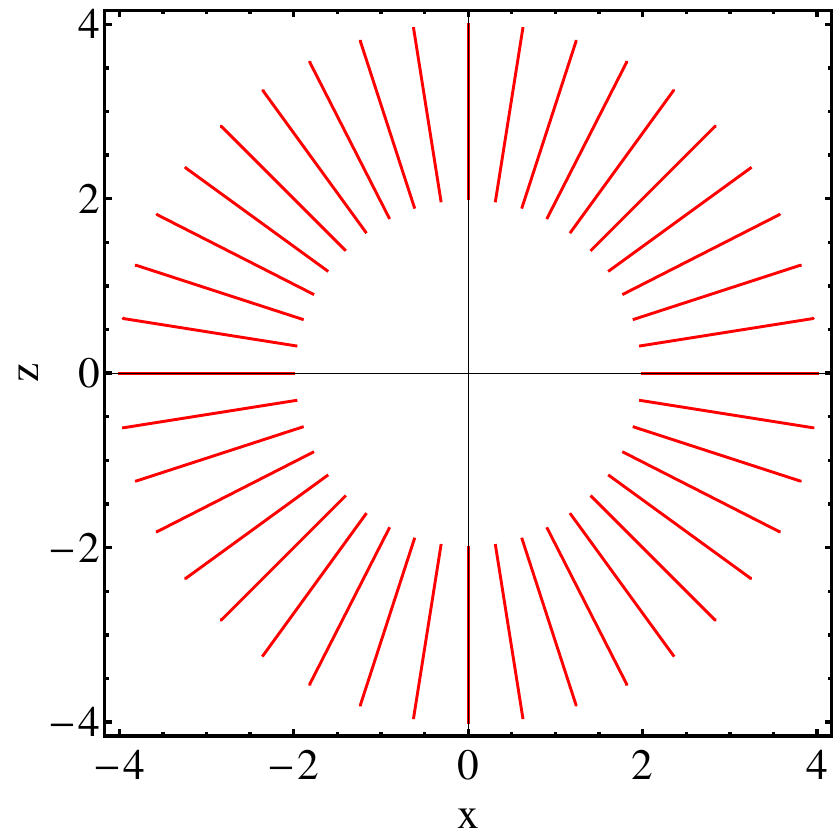}\hfill
\includegraphics[width=0.28\hsize,clip]{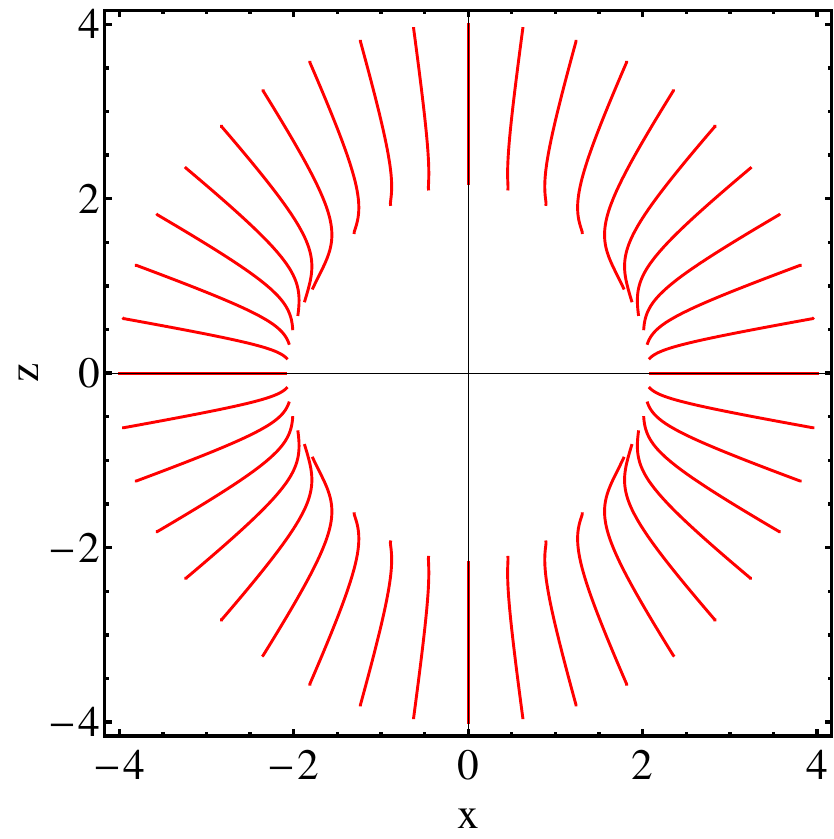}\hfill}

{\hfill
\includegraphics[width=0.28\hsize,clip]{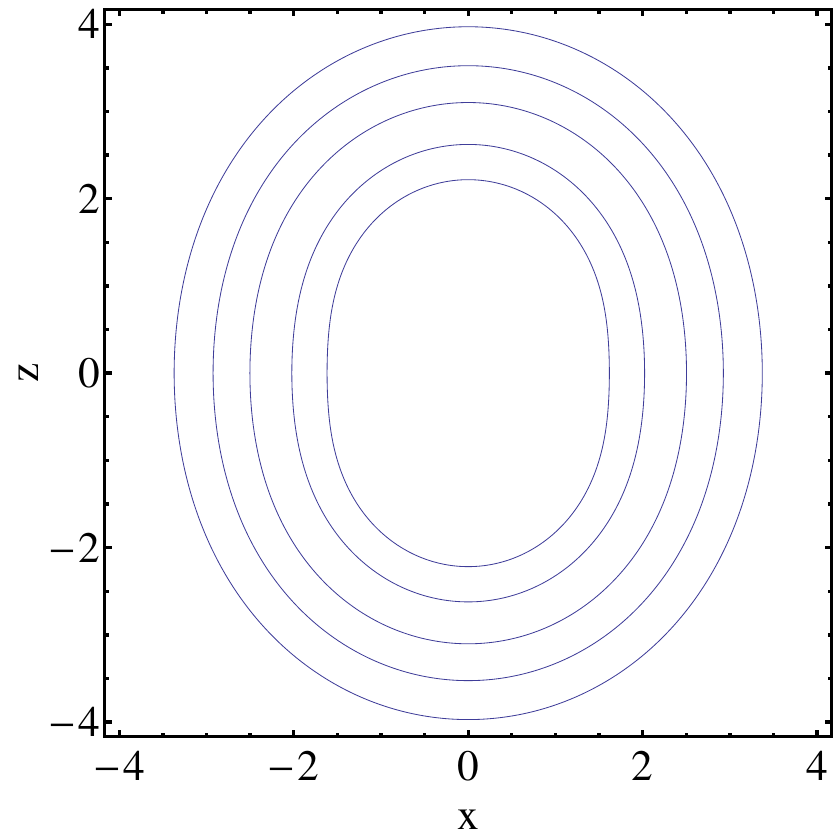}\hfill
\includegraphics[width=0.28\hsize,clip]{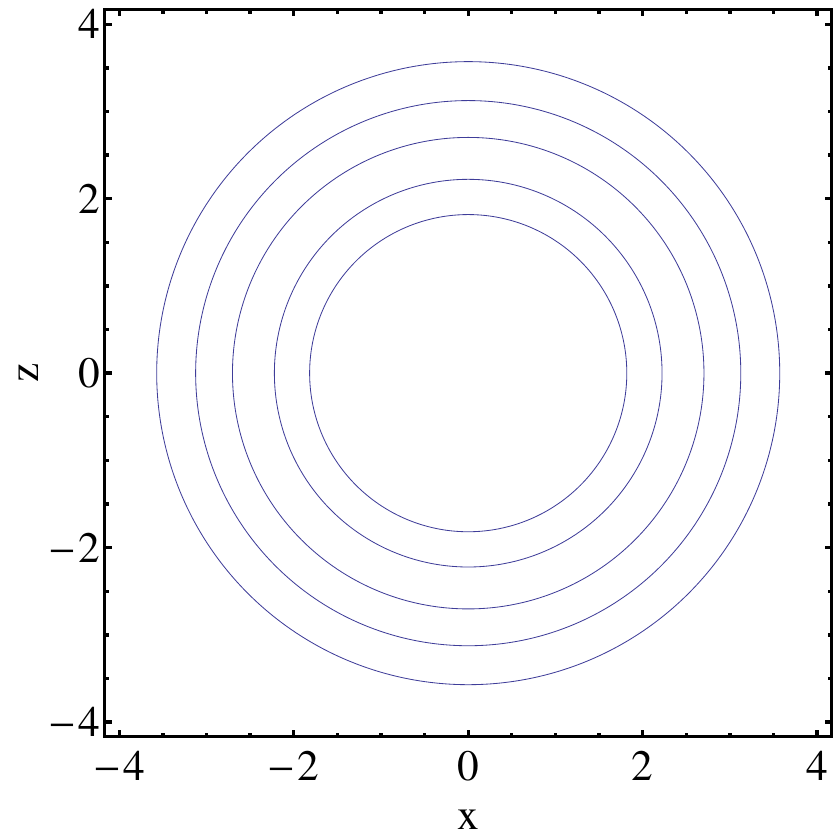}\hfill
\includegraphics[width=0.28\hsize,clip]{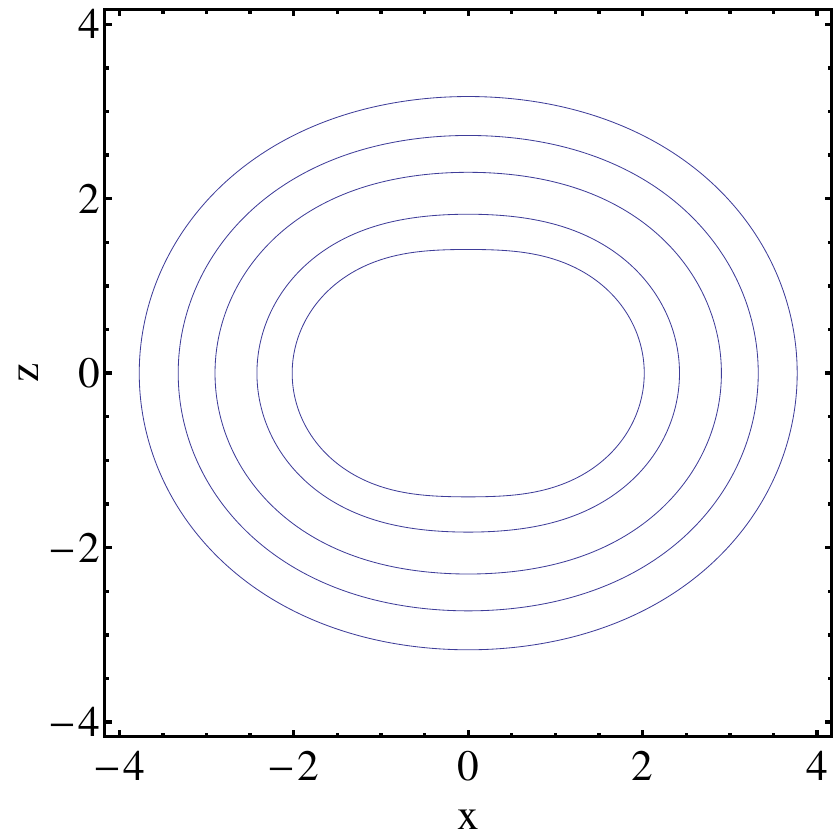}\hfill}

\caption{Top panel: Radial free fall paths of test particles on the $xz$-plane at different $\theta$ with a step of $\pi/20$ in the field of prolate ($q=-0.6$, left), spherical ($q=0$, middle) and oblate ($q=0.6$, right) sources. The initial radial coordinates of test particles are $\dot{r(0)}=4M_0$. The total mass of the source is $M_0=1$ km, $x=\rho\sin\vartheta$ and $z=\rho\cos\vartheta$. Bottom panel: Equipotential contours corresponding to the prolate, spherical and oblate source of gravity. The values of the potential $\Phi_{N}$ from outer to inner part are [0.28, 0.32, 0.37, 0.45, 0.55].}
\label{fig:rfall_newton}
\end{figure*}

The weak field post-Newtonian approximation of the $q$-metric is obtained and discussed in \cite{2019PhRvD..99d4005A}. Here, we give a concise summary of this limit. We assume a small deformation of the source; hence we may expand the $q$-metric to the linear order in $q$. In addition, retaining terms up to $\sim1/r^3$ allows one to consider a weak field regime. Furthermore, we used the following coordinate transformations
\begin{eqnarray}
    r &=& \rho\left\{1-q\frac{m}{\rho}-q\frac{m^2}{\rho^2}\left(1+\frac{m}{\rho}...\right)\sin^2\vartheta\right\}\\
    \theta &=& \vartheta-q\frac{m^2}{\rho^2}\left(1+2\frac{m}{\rho}...\right)\sin\vartheta\cos\vartheta
\end{eqnarray}
one finds 
\begin{eqnarray}\label{eq:ale}
    ds^2&=&(1+2\Phi_N)dt^2-(1+2\Phi_N)^{-1}d\rho^2\\ \nonumber
    &\textcolor{white}{-}&-U(\rho,\vartheta)\rho^2(d\vartheta^2+\sin^2\vartheta d\phi^2),
\end{eqnarray}
where
\begin{eqnarray}
    \Phi_N&=&-\frac{M_0}{\rho}+\frac{Q}{\rho^3}P_2(\cos\vartheta), \\
    U(\rho,\vartheta)&=&1-2\frac{Q}{\rho^3}P_2(\cos\vartheta)
\end{eqnarray}
with $P_2(\cos\vartheta)=(3\cos^2\vartheta-1)/2$ which is a Legendre polynomial and 
\begin{equation}
    M_0=m(1+q), \qquad Q=-\frac{2}{3}m^3q.
\end{equation}
are the total mass and quadrupole moment of the source. We note that for an oblate source $Q<0$. As may be noticed, $\Phi_N$ contains both radial $\rho$ and angular $\vartheta$ coordinates.

Using the approximate line element in Eq.~\eqref{eq:ale} and the Newtonian potential $\Phi_N$ one can study the radial free fall of test particles and equipotential contours on the $xz$ plane.

In Fig.~\ref{fig:rfall_newton} (top panel), radial free fall paths are shown in the Newtonian limit of the $q$-metric. Here, one can explicitly see that for the prolate source ($q<0$) particles curve towards the $z$ axis and for the oblate source ($q>0$) particle curve towards the equatorial plane ($z$=0) as expected. These effects can be explained by considering that a prolate source is extended along the symmetry axis ($z$), resulting in a higher mass density in that region and thus generating an additional gravitational pull, despite the total mass remaining the same for prolate, spherical and oblate sources. Similarly, for oblate sources, the analogous reasoning holds for the $z=0$ plane. This example validates once more our argument that the parameter $q$ is directly related to the mass quadruple moment of the source.

In the bottom panel, we plot equipotential contours for given sources. The values of $q$ and $\Phi_{N}$ are exaggerated to have a good representation of the results. By comparing \ref{fig:rfall_kretsch} and Fig.~\ref{fig:rfall_newton} one can observe that the behavior of the Kretschmann scalar and equipotential contours is similar, though both physical quantities are conceptually different. Regarding the radial free fall, in Fig.~\ref{fig:rfall_kretsch} we have contr-intuitive results due to particular coordinates of the $q$-metric. However, the weak field post-Newtonian limit gives us the exact results as we anticipated.

\section{Conclusions}
\label{sec:conclusions}

In this paper, we investigate tidal effects in the static axially symmetric spacetime with the deformation parameter $q$, known as $q$-metric. We derived the equations governing the deviation of geodesics for particles undergoing radial free-fall and subsequently solved them using numerical computations. Our analysis revealed that the deformation parameter $q$ exerts significant effects on the magnitudes of the physical quantities associated with the tidal effects. It is well known that in Schwarzschild spacetime, a freely falling particle undergoes consistent stretching in the radial direction as it follows its trajectory towards the singularity. In the deformed spacetime, influenced by the parameter $q$, the particle is constrained to finite stretching (compression) in the radial (angular) direction.

In addition, we have analyzed the radial geodesics of free-falling particles in the $q$-metric and found that its coordinates produce counterintuitive results. To investigate further, we constructed contours of the constant Kretschmann scalar, which illustrate the intensity of the gravitational field. The shape of these contours reflects the deformation of the source. In order to further verify our findings, we examined the weak-field post-Newtonian regime of the $q$-metric for small deformations of the source. Using new coordinates, the approximate $q$-metric produced radial free-fall geodesics consistent with our expectations. The plots of the equipotential contours were similar in form to those of the Kretschmann scalar, although they differed in scale and shape close to the central object.

These results can help us better understand the intricate relationship among spacetime parameters, tidal forces, and the dynamics of test particles. For future work, investigating areas of intense gravitational forces within the context of tidal effects in geodesic deviation equations, particularly in proximity to deformed compact objects, can yield valuable insights into the study of gravitational collapse. The use of the quadrupole parameter $q$ and the polar angle $\theta$ is emphasized because these factors play a crucial role in shedding light on a more comprehensive and idealized model for understanding gravitational phenomena in such environments. In essence, examining the interplay of these parameters improves our understanding of the intricate dynamics associated with gravitational collapse near deformed compact objects.

\section*{Acknowledgment}
The authors extend their sincere gratitude to Daniele Malafarina for suggesting this research topic and providing valuable insights into key ideas. KB also expresses his appreciation to Professor Hernando Quevedo for discussions on several conceptual aspects of this work.
KB and DU are supported by Grant No. AP22682939; AU is supported by Grant No. BR21881941, AI is supported by Grant No. AP14870501, all from the Science Committee of the Ministry of Science and Higher Education of the Republic of Kazakhstan. AI expresses gratitude to Oxford University for the opportunity to complete a three-month internship, during which this article was finalized, and extends appreciation to the Yessenov Foundation for its generous financial support. The work was supported by the PNRR-III-C9-2022–I9 call, with project number 760016/27.01.2023. This paper is based on work from COST Action CA21136 {\it Addressing observational tensions in cosmology with systematics and fundamental physics} (CosmoVerse) supported by COST (European Cooperation in Science and Technology).

\bibliography{0refs}

\begin{thebibliography}{73}%
\makeatletter
\providecommand \@ifxundefined [1]{%
 \@ifx{#1\undefined}
}%
\providecommand \@ifnum [1]{%
 \ifnum #1\expandafter \@firstoftwo
 \else \expandafter \@secondoftwo
 \fi
}%
\providecommand \@ifx [1]{%
 \ifx #1\expandafter \@firstoftwo
 \else \expandafter \@secondoftwo
 \fi
}%
\providecommand \natexlab [1]{#1}%
\providecommand \enquote  [1]{``#1''}%
\providecommand \bibnamefont  [1]{#1}%
\providecommand \bibfnamefont [1]{#1}%
\providecommand \citenamefont [1]{#1}%
\providecommand \href@noop [0]{\@secondoftwo}%
\providecommand \href [0]{\begingroup \@sanitize@url \@href}%
\providecommand \@href[1]{\@@startlink{#1}\@@href}%
\providecommand \@@href[1]{\endgroup#1\@@endlink}%
\providecommand \@sanitize@url [0]{\catcode `\\12\catcode `\$12\catcode
  `\&12\catcode `\#12\catcode `\^12\catcode `\_12\catcode `\%12\relax}%
\providecommand \@@startlink[1]{}%
\providecommand \@@endlink[0]{}%
\providecommand \url  [0]{\begingroup\@sanitize@url \@url }%
\providecommand \@url [1]{\endgroup\@href {#1}{\urlprefix }}%
\providecommand \urlprefix  [0]{URL }%
\providecommand \Eprint [0]{\href }%
\providecommand \doibase [0]{https://doi.org/}%
\providecommand \selectlanguage [0]{\@gobble}%
\providecommand \bibinfo  [0]{\@secondoftwo}%
\providecommand \bibfield  [0]{\@secondoftwo}%
\providecommand \translation [1]{[#1]}%
\providecommand \BibitemOpen [0]{}%
\providecommand \bibitemStop [0]{}%
\providecommand \bibitemNoStop [0]{.\EOS\space}%
\providecommand \EOS [0]{\spacefactor3000\relax}%
\providecommand \BibitemShut  [1]{\csname bibitem#1\endcsname}%
\let\auto@bib@innerbib\@empty
\bibitem [{\citenamefont {Misner}\ \emph {et~al.}(1973)\citenamefont {Misner},
  \citenamefont {Thorne},\ and\ \citenamefont
  {Wheeler}}]{misner1973gravitation}%
  \BibitemOpen
  \bibfield  {author} {\bibinfo {author} {\bibfnamefont {C.~W.}\ \bibnamefont
  {Misner}}, \bibinfo {author} {\bibfnamefont {K.~S.}\ \bibnamefont {Thorne}},\
  and\ \bibinfo {author} {\bibfnamefont {J.~A.}\ \bibnamefont {Wheeler}},\
  }\href@noop {} {\emph {\bibinfo {title} {Gravitation}}}\ (\bibinfo
  {publisher} {Macmillan},\ \bibinfo {year} {1973})\BibitemShut {NoStop}%
\bibitem [{\citenamefont {Poisson}(2004)}]{poisson2004relativist}%
  \BibitemOpen
  \bibfield  {author} {\bibinfo {author} {\bibfnamefont {E.}~\bibnamefont
  {Poisson}},\ }\href@noop {} {\emph {\bibinfo {title} {A relativist's toolkit:
  the mathematics of black-hole mechanics}}}\ (\bibinfo  {publisher} {Cambridge
  university press},\ \bibinfo {year} {2004})\BibitemShut {NoStop}%
\bibitem [{\citenamefont {d’Invemo}(1992)}]{d1992introducing}%
  \BibitemOpen
  \bibfield  {author} {\bibinfo {author} {\bibfnamefont {R.}~\bibnamefont
  {d’Invemo}},\ }\href@noop {} {\emph {\bibinfo {title} {Introducing
  Einstein’s relativity}}}\ (\bibinfo  {publisher} {Oxford University
  Press},\ \bibinfo {year} {1992})\BibitemShut {NoStop}%
\bibitem [{\citenamefont {Hobson}\ \emph {et~al.}(2006)\citenamefont {Hobson},
  \citenamefont {Efstathiou},\ and\ \citenamefont
  {Lasenby}}]{hobson2006general}%
  \BibitemOpen
  \bibfield  {author} {\bibinfo {author} {\bibfnamefont {M.~P.}\ \bibnamefont
  {Hobson}}, \bibinfo {author} {\bibfnamefont {G.~P.}\ \bibnamefont
  {Efstathiou}},\ and\ \bibinfo {author} {\bibfnamefont {A.~N.}\ \bibnamefont
  {Lasenby}},\ }\href@noop {} {\emph {\bibinfo {title} {General relativity: an
  introduction for physicists}}}\ (\bibinfo  {publisher} {Cambridge University
  Press},\ \bibinfo {year} {2006})\BibitemShut {NoStop}%
\bibitem [{\citenamefont {Schutz}(2022)}]{schutz2022first}%
  \BibitemOpen
  \bibfield  {author} {\bibinfo {author} {\bibfnamefont {B.}~\bibnamefont
  {Schutz}},\ }\href@noop {} {\emph {\bibinfo {title} {A first course in
  general relativity}}}\ (\bibinfo  {publisher} {Cambridge university press},\
  \bibinfo {year} {2022})\BibitemShut {NoStop}%
\bibitem [{\citenamefont {Abbott}\ \emph {et~al.}(2016)\citenamefont {Abbott},
  \citenamefont {Abbott}, \citenamefont {Abbott}, \citenamefont {Abernathy},
  \citenamefont {Acernese}, \citenamefont {Ackley}, \citenamefont {Adams},
  \citenamefont {Adams}, \citenamefont {Addesso}, \citenamefont {Adhikari}
  \emph {et~al.}}]{abbott2016observation}%
  \BibitemOpen
  \bibfield  {author} {\bibinfo {author} {\bibfnamefont {B.~P.}\ \bibnamefont
  {Abbott}}, \bibinfo {author} {\bibfnamefont {R.}~\bibnamefont {Abbott}},
  \bibinfo {author} {\bibfnamefont {T.}~\bibnamefont {Abbott}}, \bibinfo
  {author} {\bibfnamefont {M.}~\bibnamefont {Abernathy}}, \bibinfo {author}
  {\bibfnamefont {F.}~\bibnamefont {Acernese}}, \bibinfo {author}
  {\bibfnamefont {K.}~\bibnamefont {Ackley}}, \bibinfo {author} {\bibfnamefont
  {C.}~\bibnamefont {Adams}}, \bibinfo {author} {\bibfnamefont
  {T.}~\bibnamefont {Adams}}, \bibinfo {author} {\bibfnamefont
  {P.}~\bibnamefont {Addesso}}, \bibinfo {author} {\bibfnamefont
  {R.}~\bibnamefont {Adhikari}}, \emph {et~al.},\ }\bibfield  {title} {\bibinfo
  {title} {Observation of gravitational waves from a binary black hole
  merger},\ }\href@noop {} {\bibfield  {journal} {\bibinfo  {journal} {Physical
  review letters}\ }\textbf {\bibinfo {volume} {116}},\ \bibinfo {pages}
  {061102} (\bibinfo {year} {2016})}\BibitemShut {NoStop}%
\bibitem [{\citenamefont {Scientific}\ \emph {et~al.}(2017)\citenamefont
  {Scientific}, \citenamefont {Abbott}, \citenamefont {Abbott}, \citenamefont
  {Abbott}, \citenamefont {Acernese}, \citenamefont {Ackley}, \citenamefont
  {Adams}, \citenamefont {Adams}, \citenamefont {Addesso}, \citenamefont
  {Adhikari} \emph {et~al.}}]{scientific2017gw170104}%
  \BibitemOpen
  \bibfield  {author} {\bibinfo {author} {\bibfnamefont {L.}~\bibnamefont
  {Scientific}}, \bibinfo {author} {\bibfnamefont {B.~P.}\ \bibnamefont
  {Abbott}}, \bibinfo {author} {\bibfnamefont {R.}~\bibnamefont {Abbott}},
  \bibinfo {author} {\bibfnamefont {T.}~\bibnamefont {Abbott}}, \bibinfo
  {author} {\bibfnamefont {F.}~\bibnamefont {Acernese}}, \bibinfo {author}
  {\bibfnamefont {K.}~\bibnamefont {Ackley}}, \bibinfo {author} {\bibfnamefont
  {C.}~\bibnamefont {Adams}}, \bibinfo {author} {\bibfnamefont
  {T.}~\bibnamefont {Adams}}, \bibinfo {author} {\bibfnamefont
  {P.}~\bibnamefont {Addesso}}, \bibinfo {author} {\bibfnamefont
  {R.}~\bibnamefont {Adhikari}}, \emph {et~al.},\ }\bibfield  {title} {\bibinfo
  {title} {Gw170104: observation of a 50-solar-mass binary black hole
  coalescence at redshift 0.2},\ }\href@noop {} {\bibfield  {journal} {\bibinfo
   {journal} {Physical review letters}\ }\textbf {\bibinfo {volume} {118}},\
  \bibinfo {pages} {221101} (\bibinfo {year} {2017})}\BibitemShut {NoStop}%
\bibitem [{\citenamefont {Goswami}\ and\ \citenamefont
  {Ellis}(2021)}]{goswami2021tidal}%
  \BibitemOpen
  \bibfield  {author} {\bibinfo {author} {\bibfnamefont {R.}~\bibnamefont
  {Goswami}}\ and\ \bibinfo {author} {\bibfnamefont {G.~F.}\ \bibnamefont
  {Ellis}},\ }\bibfield  {title} {\bibinfo {title} {Tidal forces are
  gravitational waves},\ }\href@noop {} {\bibfield  {journal} {\bibinfo
  {journal} {Classical and Quantum Gravity}\ }\textbf {\bibinfo {volume}
  {38}},\ \bibinfo {pages} {085023} (\bibinfo {year} {2021})}\BibitemShut
  {NoStop}%
\bibitem [{\citenamefont {Pirani}(1956)}]{Pirani:1956tn}%
  \BibitemOpen
  \bibfield  {author} {\bibinfo {author} {\bibfnamefont {F.~A.~E.}\
  \bibnamefont {Pirani}},\ }\bibfield  {title} {\bibinfo {title} {{On the
  Physical significance of the Riemann tensor}},\ }\href
  {https://doi.org/10.1007/s10714-009-0787-9} {\bibfield  {journal} {\bibinfo
  {journal} {Acta Phys. Polon.}\ }\textbf {\bibinfo {volume} {15}},\ \bibinfo
  {pages} {389} (\bibinfo {year} {1956})}\BibitemShut {NoStop}%
\bibitem [{\citenamefont {{Fuchs}}(1990)}]{1990AN....311..219F}%
  \BibitemOpen
  \bibfield  {author} {\bibinfo {author} {\bibfnamefont {H.}~\bibnamefont
  {{Fuchs}}},\ }\bibfield  {title} {\bibinfo {title} {{Parallel Transport and
  Geodesic Deviation in Static Spherically Symmetric Spacetimes}},\ }\href
  {https://doi.org/10.1002/asna.2113110405} {\bibfield  {journal} {\bibinfo
  {journal} {Astronomische Nachrichten}\ }\textbf {\bibinfo {volume} {311}},\
  \bibinfo {pages} {219} (\bibinfo {year} {1990})}\BibitemShut {NoStop}%
\bibitem [{\citenamefont {{Philipp}}\ \emph {et~al.}(2015)\citenamefont
  {{Philipp}}, \citenamefont {{Perlick}}, \citenamefont {{Laemmerzahl}},\ and\
  \citenamefont {{Deshpande}}}]{2015arXiv150806457P}%
  \BibitemOpen
  \bibfield  {author} {\bibinfo {author} {\bibfnamefont {D.}~\bibnamefont
  {{Philipp}}}, \bibinfo {author} {\bibfnamefont {V.}~\bibnamefont
  {{Perlick}}}, \bibinfo {author} {\bibfnamefont {C.}~\bibnamefont
  {{Laemmerzahl}}},\ and\ \bibinfo {author} {\bibfnamefont {K.}~\bibnamefont
  {{Deshpande}}},\ }\bibfield  {title} {\bibinfo {title} {{On geodesic
  deviation in Schwarzschild spacetime}},\ }\href
  {https://doi.org/10.48550/arXiv.1508.06457} {\bibfield  {journal} {\bibinfo
  {journal} {arXiv e-prints}\ ,\ \bibinfo {eid} {arXiv:1508.06457}} (\bibinfo
  {year} {2015})},\ \Eprint {https://arxiv.org/abs/1508.06457}
  {arXiv:1508.06457 [gr-qc]} \BibitemShut {NoStop}%
\bibitem [{\citenamefont {{Ba{\.Z}a{\'n}ski}}\ and\ \citenamefont
  {{Jaranowski}}(1989)}]{1989JMP....30.1794B}%
  \BibitemOpen
  \bibfield  {author} {\bibinfo {author} {\bibfnamefont {S.~L.}\ \bibnamefont
  {{Ba{\.Z}a{\'n}ski}}}\ and\ \bibinfo {author} {\bibfnamefont
  {P.}~\bibnamefont {{Jaranowski}}},\ }\bibfield  {title} {\bibinfo {title}
  {{Geodesic deviation in the Schwarzschild space-time}},\ }\href
  {https://doi.org/10.1063/1.528266} {\bibfield  {journal} {\bibinfo  {journal}
  {Journal of Mathematical Physics}\ }\textbf {\bibinfo {volume} {30}},\
  \bibinfo {pages} {1794} (\bibinfo {year} {1989})}\BibitemShut {NoStop}%
\bibitem [{\citenamefont {Vandeev}\ and\ \citenamefont
  {Semenova}(2022)}]{Vandeev:2022gbi}%
  \BibitemOpen
  \bibfield  {author} {\bibinfo {author} {\bibfnamefont {V.~P.}\ \bibnamefont
  {Vandeev}}\ and\ \bibinfo {author} {\bibfnamefont {A.~N.}\ \bibnamefont
  {Semenova}},\ }\bibfield  {title} {\bibinfo {title} {{Deviation of non-radial
  geodesics in a static spherically symmetric spacetime}},\ }\href
  {https://doi.org/10.1140/epjp/s13360-022-02408-0} {\bibfield  {journal}
  {\bibinfo  {journal} {Eur. Phys. J. Plus}\ }\textbf {\bibinfo {volume}
  {137}},\ \bibinfo {pages} {185} (\bibinfo {year} {2022})},\ \Eprint
  {https://arxiv.org/abs/2204.13200} {arXiv:2204.13200 [gr-qc]} \BibitemShut
  {NoStop}%
\bibitem [{\citenamefont {{Crispino}}\ \emph {et~al.}(2016)\citenamefont
  {{Crispino}}, \citenamefont {{Higuchi}}, \citenamefont {{Oliveira}},\ and\
  \citenamefont {{de Oliveira}}}]{2016EPJC...76..168C}%
  \BibitemOpen
  \bibfield  {author} {\bibinfo {author} {\bibfnamefont {L.~C.~B.}\
  \bibnamefont {{Crispino}}}, \bibinfo {author} {\bibfnamefont
  {A.}~\bibnamefont {{Higuchi}}}, \bibinfo {author} {\bibfnamefont {L.~A.}\
  \bibnamefont {{Oliveira}}},\ and\ \bibinfo {author} {\bibfnamefont {E.~S.}\
  \bibnamefont {{de Oliveira}}},\ }\bibfield  {title} {\bibinfo {title} {{Tidal
  forces in Reissner-Nordstr{\"o}m spacetimes}},\ }\href
  {https://doi.org/10.1140/epjc/s10052-016-3972-5} {\bibfield  {journal}
  {\bibinfo  {journal} {European Physical Journal C}\ }\textbf {\bibinfo
  {volume} {76}},\ \bibinfo {eid} {168} (\bibinfo {year} {2016})}\BibitemShut
  {NoStop}%
\bibitem [{\citenamefont {{Abdel-Megied}}\ and\ \citenamefont
  {{Gad}}(2005)}]{2005CSF....23..313A}%
  \BibitemOpen
  \bibfield  {author} {\bibinfo {author} {\bibfnamefont {M.}~\bibnamefont
  {{Abdel-Megied}}}\ and\ \bibinfo {author} {\bibfnamefont {R.~M.}\
  \bibnamefont {{Gad}}},\ }\bibfield  {title} {\bibinfo {title} {{On the
  singularities of Reissner-Nordstr{\"o}m space-time}},\ }\href
  {https://doi.org/10.1016/j.chaos.2004.04.035} {\bibfield  {journal} {\bibinfo
   {journal} {Chaos Solitons and Fractals}\ }\textbf {\bibinfo {volume} {23}},\
  \bibinfo {pages} {313} (\bibinfo {year} {2005})},\ \Eprint
  {https://arxiv.org/abs/gr-qc/0402077} {arXiv:gr-qc/0402077 [gr-qc]}
  \BibitemShut {NoStop}%
\bibitem [{\citenamefont {{Gad}}(2010)}]{2010Ap&SS.330..107G}%
  \BibitemOpen
  \bibfield  {author} {\bibinfo {author} {\bibfnamefont {R.~M.}\ \bibnamefont
  {{Gad}}},\ }\bibfield  {title} {\bibinfo {title} {{Geodesics and geodesic
  deviation in static charged black holes}},\ }\href
  {https://doi.org/10.1007/s10509-010-0359-1} {\bibfield  {journal} {\bibinfo
  {journal} {\apss}\ }\textbf {\bibinfo {volume} {330}},\ \bibinfo {pages}
  {107} (\bibinfo {year} {2010})},\ \Eprint {https://arxiv.org/abs/0708.2841}
  {arXiv:0708.2841 [math-ph]} \BibitemShut {NoStop}%
\bibitem [{\citenamefont {Sharif}\ and\ \citenamefont
  {Kousar}(2018)}]{sharif2018tidal}%
  \BibitemOpen
  \bibfield  {author} {\bibinfo {author} {\bibfnamefont {M.}~\bibnamefont
  {Sharif}}\ and\ \bibinfo {author} {\bibfnamefont {L.}~\bibnamefont
  {Kousar}},\ }\bibfield  {title} {\bibinfo {title} {Tidal forces in dyonic
  reissner-n{\"o}rdstrom black hole},\ }\href@noop {} {\bibfield  {journal}
  {\bibinfo  {journal} {Communications in Theoretical Physics}\ }\textbf
  {\bibinfo {volume} {69}},\ \bibinfo {pages} {257} (\bibinfo {year}
  {2018})}\BibitemShut {NoStop}%
\bibitem [{\citenamefont {{Lima}}\ and\ \citenamefont
  {{Crispino}}(2020)}]{2020IJMPD..2941014L}%
  \BibitemOpen
  \bibfield  {author} {\bibinfo {author} {\bibfnamefont {H.~C.~D.}\
  \bibnamefont {{Lima}}}\ and\ \bibinfo {author} {\bibfnamefont {L.~C.~B.}\
  \bibnamefont {{Crispino}}},\ }\bibfield  {title} {\bibinfo {title} {{Tidal
  forces in the charged Hayward black hole spacetime}},\ }\href
  {https://doi.org/10.1142/S021827182041014X} {\bibfield  {journal} {\bibinfo
  {journal} {International Journal of Modern Physics D}\ }\textbf {\bibinfo
  {volume} {29}},\ \bibinfo {eid} {2041014} (\bibinfo {year}
  {2020})}\BibitemShut {NoStop}%
\bibitem [{\citenamefont {Shahzad}\ and\ \citenamefont
  {Jawad}(2017)}]{shahzad2017tidal}%
  \BibitemOpen
  \bibfield  {author} {\bibinfo {author} {\bibfnamefont {M.~U.}\ \bibnamefont
  {Shahzad}}\ and\ \bibinfo {author} {\bibfnamefont {A.}~\bibnamefont
  {Jawad}},\ }\bibfield  {title} {\bibinfo {title} {Tidal forces in kiselev
  black hole},\ }\href@noop {} {\bibfield  {journal} {\bibinfo  {journal} {The
  European Physical Journal C}\ }\textbf {\bibinfo {volume} {77}},\ \bibinfo
  {pages} {372} (\bibinfo {year} {2017})}\BibitemShut {NoStop}%
\bibitem [{\citenamefont {Nandi}\ \emph {et~al.}(2001)\citenamefont {Nandi},
  \citenamefont {Nayak}, \citenamefont {Bhadra},\ and\ \citenamefont
  {Alsing}}]{nandi2001tidal}%
  \BibitemOpen
  \bibfield  {author} {\bibinfo {author} {\bibfnamefont {K.}~\bibnamefont
  {Nandi}}, \bibinfo {author} {\bibfnamefont {T.}~\bibnamefont {Nayak}},
  \bibinfo {author} {\bibfnamefont {A.}~\bibnamefont {Bhadra}},\ and\ \bibinfo
  {author} {\bibfnamefont {P.}~\bibnamefont {Alsing}},\ }\bibfield  {title}
  {\bibinfo {title} {Tidal forces in cold black hole spacetimes},\ }\href@noop
  {} {\bibfield  {journal} {\bibinfo  {journal} {International Journal of
  Modern Physics D}\ }\textbf {\bibinfo {volume} {10}},\ \bibinfo {pages} {529}
  (\bibinfo {year} {2001})}\BibitemShut {NoStop}%
\bibitem [{\citenamefont {Cardoso}\ and\ \citenamefont
  {Pani}(2013)}]{cardoso2013tidal}%
  \BibitemOpen
  \bibfield  {author} {\bibinfo {author} {\bibfnamefont {V.}~\bibnamefont
  {Cardoso}}\ and\ \bibinfo {author} {\bibfnamefont {P.}~\bibnamefont {Pani}},\
  }\bibfield  {title} {\bibinfo {title} {Tidal acceleration of black holes and
  superradiance},\ }\href@noop {} {\bibfield  {journal} {\bibinfo  {journal}
  {Classical and Quantum Gravity}\ }\textbf {\bibinfo {volume} {30}},\ \bibinfo
  {pages} {045011} (\bibinfo {year} {2013})}\BibitemShut {NoStop}%
\bibitem [{\citenamefont {Hong}\ \emph {et~al.}(2020)\citenamefont {Hong},
  \citenamefont {Kim},\ and\ \citenamefont {Park}}]{hong2020tidal}%
  \BibitemOpen
  \bibfield  {author} {\bibinfo {author} {\bibfnamefont {S.-T.}\ \bibnamefont
  {Hong}}, \bibinfo {author} {\bibfnamefont {Y.-W.}\ \bibnamefont {Kim}},\ and\
  \bibinfo {author} {\bibfnamefont {Y.-J.}\ \bibnamefont {Park}},\ }\bibfield
  {title} {\bibinfo {title} {Tidal effects in schwarzschild black hole in
  holographic massive gravity},\ }\href@noop {} {\bibfield  {journal} {\bibinfo
   {journal} {Physics Letters B}\ }\textbf {\bibinfo {volume} {811}},\ \bibinfo
  {pages} {135967} (\bibinfo {year} {2020})}\BibitemShut {NoStop}%
\bibitem [{\citenamefont {Toshmatov}\ and\ \citenamefont
  {Ahmedov}(2023)}]{toshmatov2023tidal}%
  \BibitemOpen
  \bibfield  {author} {\bibinfo {author} {\bibfnamefont {B.}~\bibnamefont
  {Toshmatov}}\ and\ \bibinfo {author} {\bibfnamefont {B.}~\bibnamefont
  {Ahmedov}},\ }\bibfield  {title} {\bibinfo {title} {Tidal forces in
  parametrized spacetime: Rezzolla-zhidenko parametrization},\ }\href@noop {}
  {\bibfield  {journal} {\bibinfo  {journal} {Physical Review D}\ }\textbf
  {\bibinfo {volume} {108}},\ \bibinfo {pages} {084035} (\bibinfo {year}
  {2023})}\BibitemShut {NoStop}%
\bibitem [{\citenamefont {Li}\ \emph {et~al.}(2021)\citenamefont {Li},
  \citenamefont {Chen},\ and\ \citenamefont {Jing}}]{li2021tidal}%
  \BibitemOpen
  \bibfield  {author} {\bibinfo {author} {\bibfnamefont {J.}~\bibnamefont
  {Li}}, \bibinfo {author} {\bibfnamefont {S.}~\bibnamefont {Chen}},\ and\
  \bibinfo {author} {\bibfnamefont {J.}~\bibnamefont {Jing}},\ }\bibfield
  {title} {\bibinfo {title} {Tidal effects in 4d einstein--gauss--bonnet black
  hole spacetime},\ }\href@noop {} {\bibfield  {journal} {\bibinfo  {journal}
  {The European Physical Journal C}\ }\textbf {\bibinfo {volume} {81}},\
  \bibinfo {pages} {590} (\bibinfo {year} {2021})}\BibitemShut {NoStop}%
\bibitem [{\citenamefont {{Lima Junior}}\ \emph {et~al.}(2020)\citenamefont
  {{Lima Junior}}, \citenamefont {{Crispino}},\ and\ \citenamefont
  {{Higuchi}}}]{2020EPJP..135..334L}%
  \BibitemOpen
  \bibfield  {author} {\bibinfo {author} {\bibfnamefont {H.~C.~D.}\
  \bibnamefont {{Lima Junior}}}, \bibinfo {author} {\bibfnamefont {L.~C.~B.}\
  \bibnamefont {{Crispino}}},\ and\ \bibinfo {author} {\bibfnamefont
  {A.}~\bibnamefont {{Higuchi}}},\ }\bibfield  {title} {\bibinfo {title}
  {{On-axis tidal forces in Kerr spacetime}},\ }\href
  {https://doi.org/10.1140/epjp/s13360-020-00342-7} {\bibfield  {journal}
  {\bibinfo  {journal} {European Physical Journal Plus}\ }\textbf {\bibinfo
  {volume} {135}},\ \bibinfo {eid} {334} (\bibinfo {year} {2020})},\ \Eprint
  {https://arxiv.org/abs/2003.09506} {arXiv:2003.09506 [gr-qc]} \BibitemShut
  {NoStop}%
\bibitem [{\citenamefont {J.-a.}(1083)}]{Marck1983}%
  \BibitemOpen
  \bibfield  {author} {\bibinfo {author} {\bibfnamefont {M.}~\bibnamefont
  {J.-a.}},\ }\bibfield  {title} {\bibinfo {title} {{Solution to the equations
  of parallel transport in Kerr geometry; tidal tensor}},\ }\href
  {https://doi.org/10.1098/rspa.1983.0021} {\bibfield  {journal} {\bibinfo
  {journal} {Proc. R. Soc. Lond. A}\ }\textbf {\bibinfo {volume} {385}},\
  \bibinfo {pages} {431} (\bibinfo {year} {1083})}\BibitemShut {NoStop}%
\bibitem [{\citenamefont {{Mashhoon}}\ and\ \citenamefont
  {{McClune}}(1993)}]{1993MNRAS.262..881M}%
  \BibitemOpen
  \bibfield  {author} {\bibinfo {author} {\bibfnamefont {B.}~\bibnamefont
  {{Mashhoon}}}\ and\ \bibinfo {author} {\bibfnamefont {J.~C.}\ \bibnamefont
  {{McClune}}},\ }\bibfield  {title} {\bibinfo {title} {{Relativistic tidal
  impulse}},\ }\href {https://doi.org/10.1093/mnras/262.4.881} {\bibfield
  {journal} {\bibinfo  {journal} {\mnras}\ }\textbf {\bibinfo {volume} {262}},\
  \bibinfo {pages} {881} (\bibinfo {year} {1993})}\BibitemShut {NoStop}%
\bibitem [{\citenamefont {Chicone}\ and\ \citenamefont
  {Mashhoon}(2005)}]{Chicone:2004ic}%
  \BibitemOpen
  \bibfield  {author} {\bibinfo {author} {\bibfnamefont {C.}~\bibnamefont
  {Chicone}}\ and\ \bibinfo {author} {\bibfnamefont {B.}~\bibnamefont
  {Mashhoon}},\ }\bibfield  {title} {\bibinfo {title} {{Ultrarelativistic
  motion: Inertial and tidal effects in Fermi coordinates}},\ }\href
  {https://doi.org/10.1088/0264-9381/22/1/013} {\bibfield  {journal} {\bibinfo
  {journal} {Class. Quant. Grav.}\ }\textbf {\bibinfo {volume} {22}},\ \bibinfo
  {pages} {195} (\bibinfo {year} {2005})},\ \Eprint
  {https://arxiv.org/abs/gr-qc/0409017} {arXiv:gr-qc/0409017} \BibitemShut
  {NoStop}%
\bibitem [{\citenamefont {Chicone}\ and\ \citenamefont
  {Mashhoon}(2006)}]{Chicone:2006rm}%
  \BibitemOpen
  \bibfield  {author} {\bibinfo {author} {\bibfnamefont {C.}~\bibnamefont
  {Chicone}}\ and\ \bibinfo {author} {\bibfnamefont {B.}~\bibnamefont
  {Mashhoon}},\ }\bibfield  {title} {\bibinfo {title} {{Tidal dynamics in Kerr
  spacetime}},\ }\href {https://doi.org/10.1088/0264-9381/23/12/002} {\bibfield
   {journal} {\bibinfo  {journal} {Class. Quant. Grav.}\ }\textbf {\bibinfo
  {volume} {23}},\ \bibinfo {pages} {4021} (\bibinfo {year} {2006})},\ \Eprint
  {https://arxiv.org/abs/gr-qc/0602071} {arXiv:gr-qc/0602071} \BibitemShut
  {NoStop}%
\bibitem [{\citenamefont {Bini}\ \emph {et~al.}(2012)\citenamefont {Bini},
  \citenamefont {Boshkayev},\ and\ \citenamefont {Geralico}}]{Bini:2012zze}%
  \BibitemOpen
  \bibfield  {author} {\bibinfo {author} {\bibfnamefont {D.}~\bibnamefont
  {Bini}}, \bibinfo {author} {\bibfnamefont {K.}~\bibnamefont {Boshkayev}},\
  and\ \bibinfo {author} {\bibfnamefont {A.}~\bibnamefont {Geralico}},\
  }\bibfield  {title} {\bibinfo {title} {{Tidal indicators in the spacetime of
  a rotating deformed mass}},\ }\href
  {https://doi.org/10.1088/0264-9381/29/14/145003} {\bibfield  {journal}
  {\bibinfo  {journal} {Class. Quant. Grav.}\ }\textbf {\bibinfo {volume}
  {29}},\ \bibinfo {pages} {145003} (\bibinfo {year} {2012})},\ \Eprint
  {https://arxiv.org/abs/1306.4803} {arXiv:1306.4803 [gr-qc]} \BibitemShut
  {NoStop}%
\bibitem [{\citenamefont {Zipoy}(1966)}]{zipoy1966topology}%
  \BibitemOpen
  \bibfield  {author} {\bibinfo {author} {\bibfnamefont {D.~M.}\ \bibnamefont
  {Zipoy}},\ }\bibfield  {title} {\bibinfo {title} {Topology of some spheroidal
  metrics},\ }\href@noop {} {\bibfield  {journal} {\bibinfo  {journal} {Journal
  of Mathematical Physics}\ }\textbf {\bibinfo {volume} {7}},\ \bibinfo {pages}
  {1137} (\bibinfo {year} {1966})}\BibitemShut {NoStop}%
\bibitem [{\citenamefont {Voorhees}(1970)}]{voorhees1970static}%
  \BibitemOpen
  \bibfield  {author} {\bibinfo {author} {\bibfnamefont {B.}~\bibnamefont
  {Voorhees}},\ }\bibfield  {title} {\bibinfo {title} {Static axially symmetric
  gravitational fields},\ }\href@noop {} {\bibfield  {journal} {\bibinfo
  {journal} {Physical Review D}\ }\textbf {\bibinfo {volume} {2}},\ \bibinfo
  {pages} {2119} (\bibinfo {year} {1970})}\BibitemShut {NoStop}%
\bibitem [{\citenamefont {Papadopoulos}\ \emph {et~al.}(1981)\citenamefont
  {Papadopoulos}, \citenamefont {Stewart},\ and\ \citenamefont
  {Witten}}]{Papadopoulos:1981wr}%
  \BibitemOpen
  \bibfield  {author} {\bibinfo {author} {\bibfnamefont {D.}~\bibnamefont
  {Papadopoulos}}, \bibinfo {author} {\bibfnamefont {B.}~\bibnamefont
  {Stewart}},\ and\ \bibinfo {author} {\bibfnamefont {L.}~\bibnamefont
  {Witten}},\ }\bibfield  {title} {\bibinfo {title} {{Some Properties of a
  Particular Static, Axially Symmetric Space-time}},\ }\href
  {https://doi.org/10.1103/PhysRevD.24.320} {\bibfield  {journal} {\bibinfo
  {journal} {Phys. Rev. D}\ }\textbf {\bibinfo {volume} {24}},\ \bibinfo
  {pages} {320} (\bibinfo {year} {1981})}\BibitemShut {NoStop}%
\bibitem [{\citenamefont {Quevedo}(2011{\natexlab{a}})}]{quevedo2011exterior}%
  \BibitemOpen
  \bibfield  {author} {\bibinfo {author} {\bibfnamefont {H.}~\bibnamefont
  {Quevedo}},\ }\bibfield  {title} {\bibinfo {title} {Exterior and interior
  metrics with quadrupole moment},\ }\href@noop {} {\bibfield  {journal}
  {\bibinfo  {journal} {General Relativity and Gravitation}\ }\textbf {\bibinfo
  {volume} {43}},\ \bibinfo {pages} {1141} (\bibinfo {year}
  {2011}{\natexlab{a}})}\BibitemShut {NoStop}%
\bibitem [{\citenamefont {Quevedo}(2011{\natexlab{b}})}]{Quevedo:2010mn}%
  \BibitemOpen
  \bibfield  {author} {\bibinfo {author} {\bibfnamefont {H.}~\bibnamefont
  {Quevedo}},\ }\bibfield  {title} {\bibinfo {title} {{Mass Quadrupole as a
  Source of Naked Singularities}},\ }\href
  {https://doi.org/10.1142/S0218271811019852} {\bibfield  {journal} {\bibinfo
  {journal} {Int. J. Mod. Phys. D}\ }\textbf {\bibinfo {volume} {20}},\
  \bibinfo {pages} {1779} (\bibinfo {year} {2011}{\natexlab{b}})},\ \Eprint
  {https://arxiv.org/abs/1012.4030} {arXiv:1012.4030 [gr-qc]} \BibitemShut
  {NoStop}%
\bibitem [{\citenamefont {Herrera}\ \emph {et~al.}(2000)\citenamefont
  {Herrera}, \citenamefont {Paiva},\ and\ \citenamefont
  {Santos}}]{Herrera:1998rj}%
  \BibitemOpen
  \bibfield  {author} {\bibinfo {author} {\bibfnamefont {L.}~\bibnamefont
  {Herrera}}, \bibinfo {author} {\bibfnamefont {F.~M.}\ \bibnamefont {Paiva}},\
  and\ \bibinfo {author} {\bibfnamefont {N.~O.}\ \bibnamefont {Santos}},\
  }\bibfield  {title} {\bibinfo {title} {{Geodesics in the gamma space-time}},\
  }\href {https://doi.org/10.1142/S021827180000061X} {\bibfield  {journal}
  {\bibinfo  {journal} {Int. J. Mod. Phys. D}\ }\textbf {\bibinfo {volume}
  {9}},\ \bibinfo {pages} {649} (\bibinfo {year} {2000})},\ \Eprint
  {https://arxiv.org/abs/gr-qc/9812023} {arXiv:gr-qc/9812023} \BibitemShut
  {NoStop}%
\bibitem [{\citenamefont {Chowdhury}\ \emph {et~al.}(2012)\citenamefont
  {Chowdhury}, \citenamefont {Patil}, \citenamefont {Malafarina},\ and\
  \citenamefont {Joshi}}]{Chowdhury:2011aa}%
  \BibitemOpen
  \bibfield  {author} {\bibinfo {author} {\bibfnamefont {A.~N.}\ \bibnamefont
  {Chowdhury}}, \bibinfo {author} {\bibfnamefont {M.}~\bibnamefont {Patil}},
  \bibinfo {author} {\bibfnamefont {D.}~\bibnamefont {Malafarina}},\ and\
  \bibinfo {author} {\bibfnamefont {P.~S.}\ \bibnamefont {Joshi}},\ }\bibfield
  {title} {\bibinfo {title} {{Circular geodesics and accretion disks in
  Janis-Newman-Winicour and Gamma metric}},\ }\href
  {https://doi.org/10.1103/PhysRevD.85.104031} {\bibfield  {journal} {\bibinfo
  {journal} {Phys. Rev. D}\ }\textbf {\bibinfo {volume} {85}},\ \bibinfo
  {pages} {104031} (\bibinfo {year} {2012})},\ \Eprint
  {https://arxiv.org/abs/1112.2522} {arXiv:1112.2522 [gr-qc]} \BibitemShut
  {NoStop}%
\bibitem [{\citenamefont {Boshkayev}\ \emph {et~al.}(2016)\citenamefont
  {Boshkayev}, \citenamefont {Gasperin}, \citenamefont {Gutierrez-Pineres},
  \citenamefont {Quevedo},\ and\ \citenamefont
  {Toktarbay}}]{Boshkayev:2015jaa}%
  \BibitemOpen
  \bibfield  {author} {\bibinfo {author} {\bibfnamefont {K.}~\bibnamefont
  {Boshkayev}}, \bibinfo {author} {\bibfnamefont {E.}~\bibnamefont {Gasperin}},
  \bibinfo {author} {\bibfnamefont {A.~C.}\ \bibnamefont {Gutierrez-Pineres}},
  \bibinfo {author} {\bibfnamefont {H.}~\bibnamefont {Quevedo}},\ and\ \bibinfo
  {author} {\bibfnamefont {S.}~\bibnamefont {Toktarbay}},\ }\bibfield  {title}
  {\bibinfo {title} {{Motion of test particles in the field of a naked
  singularity}},\ }\href {https://doi.org/10.1103/PhysRevD.93.024024}
  {\bibfield  {journal} {\bibinfo  {journal} {Phys. Rev. D}\ }\textbf {\bibinfo
  {volume} {93}},\ \bibinfo {pages} {024024} (\bibinfo {year} {2016})},\
  \Eprint {https://arxiv.org/abs/1509.03827} {arXiv:1509.03827 [gr-qc]}
  \BibitemShut {NoStop}%
\bibitem [{\citenamefont {Capistrano}\ \emph {et~al.}(2019)\citenamefont
  {Capistrano}, \citenamefont {Seidel},\ and\ \citenamefont
  {Cabral}}]{Capistrano:2019qdv}%
  \BibitemOpen
  \bibfield  {author} {\bibinfo {author} {\bibfnamefont {A.~a. J.~S.}\
  \bibnamefont {Capistrano}}, \bibinfo {author} {\bibfnamefont {P.~T.~Z.}\
  \bibnamefont {Seidel}},\ and\ \bibinfo {author} {\bibfnamefont {L.~A.}\
  \bibnamefont {Cabral}},\ }\bibfield  {title} {\bibinfo {title} {{Effective
  apsidal precession from a monopole solution in a Zipoy spacetime}},\ }\href
  {https://doi.org/10.1140/epjc/s10052-019-7238-x} {\bibfield  {journal}
  {\bibinfo  {journal} {Eur. Phys. J. C}\ }\textbf {\bibinfo {volume} {79}},\
  \bibinfo {pages} {730} (\bibinfo {year} {2019})}\BibitemShut {NoStop}%
\bibitem [{\citenamefont {{Stewart, B., Papadopoulos, D., Witten, L.,
  Berezdivin, R., Herrera, L.}}(1982)}]{Stewart:1982}%
  \BibitemOpen
  \bibfield  {author} {\bibinfo {author} {\bibnamefont {{Stewart, B.,
  Papadopoulos, D., Witten, L., Berezdivin, R., Herrera, L.}}},\ }\bibfield
  {title} {\bibinfo {title} {{An interior solution for the gamma metric}},\
  }\href {https://doi.org/10.1007/BF00756201} {\bibfield  {journal} {\bibinfo
  {journal} {Gen. Relat. Gravit.}\ }\textbf {\bibinfo {volume} {14}},\ \bibinfo
  {pages} {97–} (\bibinfo {year} {1982})}\BibitemShut {NoStop}%
\bibitem [{\citenamefont {Herrera}\ \emph {et~al.}(2005)\citenamefont
  {Herrera}, \citenamefont {Magli},\ and\ \citenamefont
  {Malafarina}}]{Herrera:2004ra}%
  \BibitemOpen
  \bibfield  {author} {\bibinfo {author} {\bibfnamefont {L.}~\bibnamefont
  {Herrera}}, \bibinfo {author} {\bibfnamefont {G.}~\bibnamefont {Magli}},\
  and\ \bibinfo {author} {\bibfnamefont {D.}~\bibnamefont {Malafarina}},\
  }\bibfield  {title} {\bibinfo {title} {{Non-spherical sources of static
  gravitational fields: Investigating the boundaries of the no-hair theorem}},\
  }\href {https://doi.org/10.1007/s10714-005-0120-1} {\bibfield  {journal}
  {\bibinfo  {journal} {Gen. Rel. Grav.}\ }\textbf {\bibinfo {volume} {37}},\
  \bibinfo {pages} {1371} (\bibinfo {year} {2005})},\ \Eprint
  {https://arxiv.org/abs/gr-qc/0407037} {arXiv:gr-qc/0407037} \BibitemShut
  {NoStop}%
\bibitem [{\citenamefont {Abdikamalov}\ \emph {et~al.}(2019)\citenamefont
  {Abdikamalov}, \citenamefont {Abdujabbarov}, \citenamefont {Ayzenberg},
  \citenamefont {Malafarina}, \citenamefont {Bambi},\ and\ \citenamefont
  {Ahmedov}}]{Abdikamalov:2019ztb}%
  \BibitemOpen
  \bibfield  {author} {\bibinfo {author} {\bibfnamefont {A.~B.}\ \bibnamefont
  {Abdikamalov}}, \bibinfo {author} {\bibfnamefont {A.~A.}\ \bibnamefont
  {Abdujabbarov}}, \bibinfo {author} {\bibfnamefont {D.}~\bibnamefont
  {Ayzenberg}}, \bibinfo {author} {\bibfnamefont {D.}~\bibnamefont
  {Malafarina}}, \bibinfo {author} {\bibfnamefont {C.}~\bibnamefont {Bambi}},\
  and\ \bibinfo {author} {\bibfnamefont {B.}~\bibnamefont {Ahmedov}},\
  }\bibfield  {title} {\bibinfo {title} {{Black hole mimicker hiding in the
  shadow: Optical properties of the $\gamma$ metric}},\ }\href
  {https://doi.org/10.1103/PhysRevD.100.024014} {\bibfield  {journal} {\bibinfo
   {journal} {Phys. Rev. D}\ }\textbf {\bibinfo {volume} {100}},\ \bibinfo
  {pages} {024014} (\bibinfo {year} {2019})},\ \Eprint
  {https://arxiv.org/abs/1904.06207} {arXiv:1904.06207 [gr-qc]} \BibitemShut
  {NoStop}%
\bibitem [{\citenamefont {Arrieta-Villamizar}\ \emph
  {et~al.}(2021)\citenamefont {Arrieta-Villamizar}, \citenamefont
  {Vel\'asquez-Cadavid}, \citenamefont {Pimentel}, \citenamefont
  {Lora-Clavijo},\ and\ \citenamefont
  {Guti\'errez-Pi\~neres}}]{Arrieta-Villamizar:2020brc}%
  \BibitemOpen
  \bibfield  {author} {\bibinfo {author} {\bibfnamefont {J.~A.}\ \bibnamefont
  {Arrieta-Villamizar}}, \bibinfo {author} {\bibfnamefont {J.~M.}\ \bibnamefont
  {Vel\'asquez-Cadavid}}, \bibinfo {author} {\bibfnamefont {O.~M.}\
  \bibnamefont {Pimentel}}, \bibinfo {author} {\bibfnamefont {F.~D.}\
  \bibnamefont {Lora-Clavijo}},\ and\ \bibinfo {author} {\bibfnamefont {A.~C.}\
  \bibnamefont {Guti\'errez-Pi\~neres}},\ }\bibfield  {title} {\bibinfo {title}
  {{Shadows around the q-metric}},\ }\href
  {https://doi.org/10.1088/1361-6382/abc223} {\bibfield  {journal} {\bibinfo
  {journal} {Class. Quant. Grav.}\ }\textbf {\bibinfo {volume} {38}},\ \bibinfo
  {pages} {015008} (\bibinfo {year} {2021})},\ \Eprint
  {https://arxiv.org/abs/2007.13600} {arXiv:2007.13600 [gr-qc]} \BibitemShut
  {NoStop}%
\bibitem [{\citenamefont {Shaikh}(2023)}]{Shaikh:2022ivr}%
  \BibitemOpen
  \bibfield  {author} {\bibinfo {author} {\bibfnamefont {R.}~\bibnamefont
  {Shaikh}},\ }\bibfield  {title} {\bibinfo {title} {{Testing black hole
  mimickers with the Event Horizon Telescope image of Sagittarius A*}},\ }\href
  {https://doi.org/10.1093/mnras/stad1383} {\bibfield  {journal} {\bibinfo
  {journal} {Mon. Not. Roy. Astron. Soc.}\ }\textbf {\bibinfo {volume} {523}},\
  \bibinfo {pages} {375} (\bibinfo {year} {2023})},\ \Eprint
  {https://arxiv.org/abs/2208.01995} {arXiv:2208.01995 [gr-qc]} \BibitemShut
  {NoStop}%
\bibitem [{\citenamefont {Turimov}\ and\ \citenamefont
  {Ahmedov}(2023)}]{Turimov:2023lbn}%
  \BibitemOpen
  \bibfield  {author} {\bibinfo {author} {\bibfnamefont {B.}~\bibnamefont
  {Turimov}}\ and\ \bibinfo {author} {\bibfnamefont {B.}~\bibnamefont
  {Ahmedov}},\ }\bibfield  {title} {\bibinfo {title} {{Observable Properties of
  Thin Accretion Disk in the \ensuremath{\gamma} Spacetime}},\ }\href
  {https://doi.org/10.3390/sym15101858} {\bibfield  {journal} {\bibinfo
  {journal} {Symmetry}\ }\textbf {\bibinfo {volume} {15}},\ \bibinfo {pages}
  {1858} (\bibinfo {year} {2023})}\BibitemShut {NoStop}%
\bibitem [{\citenamefont {Toshmatov}\ and\ \citenamefont
  {Malafarina}(2019)}]{Toshmatov:2019bda}%
  \BibitemOpen
  \bibfield  {author} {\bibinfo {author} {\bibfnamefont {B.}~\bibnamefont
  {Toshmatov}}\ and\ \bibinfo {author} {\bibfnamefont {D.}~\bibnamefont
  {Malafarina}},\ }\bibfield  {title} {\bibinfo {title} {{Spinning test
  particles in the $\gamma$ spacetime}},\ }\href
  {https://doi.org/10.1103/PhysRevD.100.104052} {\bibfield  {journal} {\bibinfo
   {journal} {Phys. Rev. D}\ }\textbf {\bibinfo {volume} {100}},\ \bibinfo
  {pages} {104052} (\bibinfo {year} {2019})},\ \Eprint
  {https://arxiv.org/abs/1910.11565} {arXiv:1910.11565 [gr-qc]} \BibitemShut
  {NoStop}%
\bibitem [{\citenamefont {Benavides-Gallego}\ \emph {et~al.}(2019)\citenamefont
  {Benavides-Gallego}, \citenamefont {Abdujabbarov}, \citenamefont
  {Malafarina}, \citenamefont {Ahmedov},\ and\ \citenamefont
  {Bambi}}]{Benavides-Gallego:2018htf}%
  \BibitemOpen
  \bibfield  {author} {\bibinfo {author} {\bibfnamefont {C.~A.}\ \bibnamefont
  {Benavides-Gallego}}, \bibinfo {author} {\bibfnamefont {A.}~\bibnamefont
  {Abdujabbarov}}, \bibinfo {author} {\bibfnamefont {D.}~\bibnamefont
  {Malafarina}}, \bibinfo {author} {\bibfnamefont {B.}~\bibnamefont
  {Ahmedov}},\ and\ \bibinfo {author} {\bibfnamefont {C.}~\bibnamefont
  {Bambi}},\ }\bibfield  {title} {\bibinfo {title} {{Charged particle motion
  and electromagnetic field in $\gamma$ spacetime}},\ }\href
  {https://doi.org/10.1103/PhysRevD.99.044012} {\bibfield  {journal} {\bibinfo
  {journal} {Phys. Rev. D}\ }\textbf {\bibinfo {volume} {99}},\ \bibinfo
  {pages} {044012} (\bibinfo {year} {2019})},\ \Eprint
  {https://arxiv.org/abs/1812.04846} {arXiv:1812.04846 [gr-qc]} \BibitemShut
  {NoStop}%
\bibitem [{\citenamefont {Faraji}\ and\ \citenamefont
  {Trova}(2022)}]{Faraji:2021vid}%
  \BibitemOpen
  \bibfield  {author} {\bibinfo {author} {\bibfnamefont {S.}~\bibnamefont
  {Faraji}}\ and\ \bibinfo {author} {\bibfnamefont {A.}~\bibnamefont {Trova}},\
  }\bibfield  {title} {\bibinfo {title} {{Dynamics of charged particles and
  quasi-periodic oscillations in the vicinity of a distorted, deformed compact
  object embedded in a uniform magnetic field}},\ }\href
  {https://doi.org/10.1093/mnras/stac882} {\bibfield  {journal} {\bibinfo
  {journal} {Mon. Not. Roy. Astron. Soc.}\ }\textbf {\bibinfo {volume} {513}},\
  \bibinfo {pages} {3399} (\bibinfo {year} {2022})},\ \Eprint
  {https://arxiv.org/abs/2103.03229} {arXiv:2103.03229 [astro-ph.HE]}
  \BibitemShut {NoStop}%
\bibitem [{\citenamefont {Malafarina}\ and\ \citenamefont
  {Sagynbayeva}(2021)}]{Malafarina:2020kmk}%
  \BibitemOpen
  \bibfield  {author} {\bibinfo {author} {\bibfnamefont {D.}~\bibnamefont
  {Malafarina}}\ and\ \bibinfo {author} {\bibfnamefont {S.}~\bibnamefont
  {Sagynbayeva}},\ }\bibfield  {title} {\bibinfo {title} {{What a difference a
  quadrupole makes?}},\ }\href {https://doi.org/10.1007/s10714-021-02881-5}
  {\bibfield  {journal} {\bibinfo  {journal} {Gen. Rel. Grav.}\ }\textbf
  {\bibinfo {volume} {53}},\ \bibinfo {pages} {112} (\bibinfo {year} {2021})},\
  \Eprint {https://arxiv.org/abs/2009.12839} {arXiv:2009.12839 [gr-qc]}
  \BibitemShut {NoStop}%
\bibitem [{\citenamefont {Toshmatov}\ \emph {et~al.}(2019)\citenamefont
  {Toshmatov}, \citenamefont {Malafarina},\ and\ \citenamefont
  {Dadhich}}]{Toshmatov:2019qih}%
  \BibitemOpen
  \bibfield  {author} {\bibinfo {author} {\bibfnamefont {B.}~\bibnamefont
  {Toshmatov}}, \bibinfo {author} {\bibfnamefont {D.}~\bibnamefont
  {Malafarina}},\ and\ \bibinfo {author} {\bibfnamefont {N.}~\bibnamefont
  {Dadhich}},\ }\bibfield  {title} {\bibinfo {title} {{Harmonic oscillations of
  neutral particles in the $\gamma$-metric}},\ }\href
  {https://doi.org/10.1103/PhysRevD.100.044001} {\bibfield  {journal} {\bibinfo
   {journal} {Phys. Rev. D}\ }\textbf {\bibinfo {volume} {100}},\ \bibinfo
  {pages} {044001} (\bibinfo {year} {2019})},\ \Eprint
  {https://arxiv.org/abs/1905.01088} {arXiv:1905.01088 [gr-qc]} \BibitemShut
  {NoStop}%
\bibitem [{\citenamefont {Benavides-Gallego}\ \emph {et~al.}(2020)\citenamefont
  {Benavides-Gallego}, \citenamefont {Abdujabbarov}, \citenamefont
  {Malafarina},\ and\ \citenamefont {Bambi}}]{Benavides-Gallego:2020fri}%
  \BibitemOpen
  \bibfield  {author} {\bibinfo {author} {\bibfnamefont {C.~A.}\ \bibnamefont
  {Benavides-Gallego}}, \bibinfo {author} {\bibfnamefont {A.}~\bibnamefont
  {Abdujabbarov}}, \bibinfo {author} {\bibfnamefont {D.}~\bibnamefont
  {Malafarina}},\ and\ \bibinfo {author} {\bibfnamefont {C.}~\bibnamefont
  {Bambi}},\ }\bibfield  {title} {\bibinfo {title} {{Quasiharmonic oscillations
  of charged particles in static axially symmetric space-times immersed in a
  uniform magnetic field}},\ }\href
  {https://doi.org/10.1103/PhysRevD.101.124024} {\bibfield  {journal} {\bibinfo
   {journal} {Phys. Rev. D}\ }\textbf {\bibinfo {volume} {101}},\ \bibinfo
  {pages} {124024} (\bibinfo {year} {2020})},\ \Eprint
  {https://arxiv.org/abs/2005.07554} {arXiv:2005.07554 [gr-qc]} \BibitemShut
  {NoStop}%
\bibitem [{\citenamefont {{Boshkayev}}\ \emph {et~al.}(2024)\citenamefont
  {{Boshkayev}}, \citenamefont {{Konysbayev}}, \citenamefont {{Kurmanov}},
  \citenamefont {{Muccino}},\ and\ \citenamefont
  {{Quevedo}}}]{2024MNRAS.531.3876B}%
  \BibitemOpen
  \bibfield  {author} {\bibinfo {author} {\bibfnamefont {K.}~\bibnamefont
  {{Boshkayev}}}, \bibinfo {author} {\bibfnamefont {T.}~\bibnamefont
  {{Konysbayev}}}, \bibinfo {author} {\bibfnamefont {Y.}~\bibnamefont
  {{Kurmanov}}}, \bibinfo {author} {\bibfnamefont {M.}~\bibnamefont
  {{Muccino}}},\ and\ \bibinfo {author} {\bibfnamefont {H.}~\bibnamefont
  {{Quevedo}}},\ }\bibfield  {title} {\bibinfo {title} {{Quasi-periodic
  oscillations in rotating and deformed space-times}},\ }\href
  {https://doi.org/10.1093/mnras/stae1388} {\bibfield  {journal} {\bibinfo
  {journal} {\mnras}\ }\textbf {\bibinfo {volume} {531}},\ \bibinfo {pages}
  {3876} (\bibinfo {year} {2024})},\ \Eprint {https://arxiv.org/abs/2312.03630}
  {arXiv:2312.03630 [astro-ph.HE]} \BibitemShut {NoStop}%
\bibitem [{\citenamefont {Boshkayev}\ \emph {et~al.}(2020)\citenamefont
  {Boshkayev}, \citenamefont {Luongo},\ and\ \citenamefont
  {Muccino}}]{Boshkayev:2020igc}%
  \BibitemOpen
  \bibfield  {author} {\bibinfo {author} {\bibfnamefont {K.}~\bibnamefont
  {Boshkayev}}, \bibinfo {author} {\bibfnamefont {O.}~\bibnamefont {Luongo}},\
  and\ \bibinfo {author} {\bibfnamefont {M.}~\bibnamefont {Muccino}},\
  }\bibfield  {title} {\bibinfo {title} {{Neutrino oscillation in the
  $q$-metric}},\ }\href {https://doi.org/10.1140/epjc/s10052-020-08533-3}
  {\bibfield  {journal} {\bibinfo  {journal} {Eur. Phys. J. C}\ }\textbf
  {\bibinfo {volume} {80}},\ \bibinfo {pages} {964} (\bibinfo {year} {2020})},\
  \Eprint {https://arxiv.org/abs/2010.08254} {arXiv:2010.08254 [gr-qc]}
  \BibitemShut {NoStop}%
\bibitem [{\citenamefont {Chakrabarty}\ \emph {et~al.}(2022)\citenamefont
  {Chakrabarty}, \citenamefont {Borah}, \citenamefont {Abdujabbarov},
  \citenamefont {Malafarina},\ and\ \citenamefont
  {Ahmedov}}]{Chakrabarty:2021bpr}%
  \BibitemOpen
  \bibfield  {author} {\bibinfo {author} {\bibfnamefont {H.}~\bibnamefont
  {Chakrabarty}}, \bibinfo {author} {\bibfnamefont {D.}~\bibnamefont {Borah}},
  \bibinfo {author} {\bibfnamefont {A.}~\bibnamefont {Abdujabbarov}}, \bibinfo
  {author} {\bibfnamefont {D.}~\bibnamefont {Malafarina}},\ and\ \bibinfo
  {author} {\bibfnamefont {B.}~\bibnamefont {Ahmedov}},\ }\bibfield  {title}
  {\bibinfo {title} {{Effects of gravitational lensing on neutrino oscillation
  in $ \gamma $-spacetime}},\ }\href
  {https://doi.org/10.1140/epjc/s10052-021-09982-0} {\bibfield  {journal}
  {\bibinfo  {journal} {Eur. Phys. J. C}\ }\textbf {\bibinfo {volume} {82}},\
  \bibinfo {pages} {24} (\bibinfo {year} {2022})},\ \Eprint
  {https://arxiv.org/abs/2109.02395} {arXiv:2109.02395 [gr-qc]} \BibitemShut
  {NoStop}%
\bibitem [{\citenamefont {Chakrabarty}\ and\ \citenamefont
  {Tang}(2023)}]{Chakrabarty:2022fbd}%
  \BibitemOpen
  \bibfield  {author} {\bibinfo {author} {\bibfnamefont {H.}~\bibnamefont
  {Chakrabarty}}\ and\ \bibinfo {author} {\bibfnamefont {Y.}~\bibnamefont
  {Tang}},\ }\bibfield  {title} {\bibinfo {title} {{Constraining deviations
  from spherical symmetry using \ensuremath{\gamma}-metric}},\ }\href
  {https://doi.org/10.1103/PhysRevD.107.084020} {\bibfield  {journal} {\bibinfo
   {journal} {Phys. Rev. D}\ }\textbf {\bibinfo {volume} {107}},\ \bibinfo
  {pages} {084020} (\bibinfo {year} {2023})},\ \Eprint
  {https://arxiv.org/abs/2204.06807} {arXiv:2204.06807 [gr-qc]} \BibitemShut
  {NoStop}%
\bibitem [{\citenamefont {Boshkayev}\ \emph {et~al.}(2021)\citenamefont
  {Boshkayev}, \citenamefont {Konysbayev}, \citenamefont {Kurmanov},
  \citenamefont {Luongo}, \citenamefont {Malafarina},\ and\ \citenamefont
  {Quevedo}}]{Boshkayev:2021chc}%
  \BibitemOpen
  \bibfield  {author} {\bibinfo {author} {\bibfnamefont {K.}~\bibnamefont
  {Boshkayev}}, \bibinfo {author} {\bibfnamefont {T.}~\bibnamefont
  {Konysbayev}}, \bibinfo {author} {\bibfnamefont {E.}~\bibnamefont
  {Kurmanov}}, \bibinfo {author} {\bibfnamefont {O.}~\bibnamefont {Luongo}},
  \bibinfo {author} {\bibfnamefont {D.}~\bibnamefont {Malafarina}},\ and\
  \bibinfo {author} {\bibfnamefont {H.}~\bibnamefont {Quevedo}},\ }\bibfield
  {title} {\bibinfo {title} {{Luminosity of accretion disks in compact objects
  with a quadrupole}},\ }\href {https://doi.org/10.1103/PhysRevD.104.084009}
  {\bibfield  {journal} {\bibinfo  {journal} {Phys. Rev. D}\ }\textbf {\bibinfo
  {volume} {104}},\ \bibinfo {pages} {084009} (\bibinfo {year} {2021})},\
  \Eprint {https://arxiv.org/abs/2106.04932} {arXiv:2106.04932 [gr-qc]}
  \BibitemShut {NoStop}%
\bibitem [{\citenamefont {Shaikh}\ \emph {et~al.}(2022)\citenamefont {Shaikh},
  \citenamefont {Paul}, \citenamefont {Banerjee},\ and\ \citenamefont
  {Sarkar}}]{Shaikh:2021cvl}%
  \BibitemOpen
  \bibfield  {author} {\bibinfo {author} {\bibfnamefont {R.}~\bibnamefont
  {Shaikh}}, \bibinfo {author} {\bibfnamefont {S.}~\bibnamefont {Paul}},
  \bibinfo {author} {\bibfnamefont {P.}~\bibnamefont {Banerjee}},\ and\
  \bibinfo {author} {\bibfnamefont {T.}~\bibnamefont {Sarkar}},\ }\bibfield
  {title} {\bibinfo {title} {{Shadows and thin accretion disk images of the
  $\gamma $-metric}},\ }\href {https://doi.org/10.1140/epjc/s10052-022-10664-8}
  {\bibfield  {journal} {\bibinfo  {journal} {Eur. Phys. J. C}\ }\textbf
  {\bibinfo {volume} {82}},\ \bibinfo {pages} {696} (\bibinfo {year} {2022})},\
  \Eprint {https://arxiv.org/abs/2105.12057} {arXiv:2105.12057 [gr-qc]}
  \BibitemShut {NoStop}%
\bibitem [{\citenamefont {Quevedo}\ and\ \citenamefont
  {Mashhoon}(1991)}]{Quevedo:1991zz}%
  \BibitemOpen
  \bibfield  {author} {\bibinfo {author} {\bibfnamefont {H.}~\bibnamefont
  {Quevedo}}\ and\ \bibinfo {author} {\bibfnamefont {B.}~\bibnamefont
  {Mashhoon}},\ }\bibfield  {title} {\bibinfo {title} {{Generalization of Kerr
  spacetime}},\ }\href {https://doi.org/10.1103/PhysRevD.43.3902} {\bibfield
  {journal} {\bibinfo  {journal} {Phys. Rev. D}\ }\textbf {\bibinfo {volume}
  {43}},\ \bibinfo {pages} {3902} (\bibinfo {year} {1991})}\BibitemShut
  {NoStop}%
\bibitem [{\citenamefont {Toktarbay}\ and\ \citenamefont
  {Quevedo}(2014)}]{Toktarbay:2014yru}%
  \BibitemOpen
  \bibfield  {author} {\bibinfo {author} {\bibfnamefont {S.}~\bibnamefont
  {Toktarbay}}\ and\ \bibinfo {author} {\bibfnamefont {H.}~\bibnamefont
  {Quevedo}},\ }\bibfield  {title} {\bibinfo {title} {{A stationary
  q-metric}},\ }\href {https://doi.org/10.1134/S0202289314040136} {\bibfield
  {journal} {\bibinfo  {journal} {Grav. Cosmol.}\ }\textbf {\bibinfo {volume}
  {20}},\ \bibinfo {pages} {252} (\bibinfo {year} {2014})},\ \Eprint
  {https://arxiv.org/abs/1510.04155} {arXiv:1510.04155 [gr-qc]} \BibitemShut
  {NoStop}%
\bibitem [{\citenamefont {Allahyari}\ \emph {et~al.}(2020)\citenamefont
  {Allahyari}, \citenamefont {Firouzjahi},\ and\ \citenamefont
  {Mashhoon}}]{Allahyari:2019umx}%
  \BibitemOpen
  \bibfield  {author} {\bibinfo {author} {\bibfnamefont {A.}~\bibnamefont
  {Allahyari}}, \bibinfo {author} {\bibfnamefont {H.}~\bibnamefont
  {Firouzjahi}},\ and\ \bibinfo {author} {\bibfnamefont {B.}~\bibnamefont
  {Mashhoon}},\ }\bibfield  {title} {\bibinfo {title} {{Quasinormal modes of
  generalized black holes: $\delta$-Kerr spacetime}},\ }\href
  {https://doi.org/10.1088/1361-6382/ab6860} {\bibfield  {journal} {\bibinfo
  {journal} {Class. Quant. Grav.}\ }\textbf {\bibinfo {volume} {37}},\ \bibinfo
  {pages} {055006} (\bibinfo {year} {2020})},\ \Eprint
  {https://arxiv.org/abs/1908.10813} {arXiv:1908.10813 [gr-qc]} \BibitemShut
  {NoStop}%
\bibitem [{\citenamefont {Li}\ \emph {et~al.}(2022)\citenamefont {Li},
  \citenamefont {Mirzaev}, \citenamefont {Abdujabbarov}, \citenamefont
  {Malafarina}, \citenamefont {Ahmedov},\ and\ \citenamefont
  {Han}}]{Li:2022eue}%
  \BibitemOpen
  \bibfield  {author} {\bibinfo {author} {\bibfnamefont {S.}~\bibnamefont
  {Li}}, \bibinfo {author} {\bibfnamefont {T.}~\bibnamefont {Mirzaev}},
  \bibinfo {author} {\bibfnamefont {A.~A.}\ \bibnamefont {Abdujabbarov}},
  \bibinfo {author} {\bibfnamefont {D.}~\bibnamefont {Malafarina}}, \bibinfo
  {author} {\bibfnamefont {B.}~\bibnamefont {Ahmedov}},\ and\ \bibinfo {author}
  {\bibfnamefont {W.-B.}\ \bibnamefont {Han}},\ }\bibfield  {title} {\bibinfo
  {title} {{Constraining the deformation of a rotating black hole mimicker from
  its shadow}},\ }\href {https://doi.org/10.1103/PhysRevD.106.084041}
  {\bibfield  {journal} {\bibinfo  {journal} {Phys. Rev. D}\ }\textbf {\bibinfo
  {volume} {106}},\ \bibinfo {pages} {084041} (\bibinfo {year} {2022})},\
  \Eprint {https://arxiv.org/abs/2207.10933} {arXiv:2207.10933 [gr-qc]}
  \BibitemShut {NoStop}%
\bibitem [{\citenamefont {Gurtug}\ \emph {et~al.}(2022)\citenamefont {Gurtug},
  \citenamefont {Halilsoy},\ and\ \citenamefont {Mangut}}]{Gurtug:2021noy}%
  \BibitemOpen
  \bibfield  {author} {\bibinfo {author} {\bibfnamefont {O.}~\bibnamefont
  {Gurtug}}, \bibinfo {author} {\bibfnamefont {M.}~\bibnamefont {Halilsoy}},\
  and\ \bibinfo {author} {\bibfnamefont {M.}~\bibnamefont {Mangut}},\
  }\bibfield  {title} {\bibinfo {title} {{The charged
  Zipoy\textendash{}Voorhees metric with astrophysical applications}},\ }\href
  {https://doi.org/10.1140/epjc/s10052-022-10626-0} {\bibfield  {journal}
  {\bibinfo  {journal} {Eur. Phys. J. C}\ }\textbf {\bibinfo {volume} {82}},\
  \bibinfo {pages} {671} (\bibinfo {year} {2022})},\ \Eprint
  {https://arxiv.org/abs/2110.12188} {arXiv:2110.12188 [gr-qc]} \BibitemShut
  {NoStop}%
\bibitem [{\citenamefont {Halilsoy}(1992)}]{Halilsoy:1992zz}%
  \BibitemOpen
  \bibfield  {author} {\bibinfo {author} {\bibfnamefont {M.}~\bibnamefont
  {Halilsoy}},\ }\bibfield  {title} {\bibinfo {title} {{New metrics for
  spinning spheroids}},\ }\href {https://doi.org/10.1063/1.529822} {\bibfield
  {journal} {\bibinfo  {journal} {J. Math. Phys.}\ }\textbf {\bibinfo {volume}
  {33}},\ \bibinfo {pages} {4225} (\bibinfo {year} {1992})}\BibitemShut
  {NoStop}%
\bibitem [{\citenamefont {Narzilloev}\ \emph {et~al.}(2020)\citenamefont
  {Narzilloev}, \citenamefont {Malafarina}, \citenamefont {Abdujabbarov},\ and\
  \citenamefont {Bambi}}]{Narzilloev:2020qdc}%
  \BibitemOpen
  \bibfield  {author} {\bibinfo {author} {\bibfnamefont {B.}~\bibnamefont
  {Narzilloev}}, \bibinfo {author} {\bibfnamefont {D.}~\bibnamefont
  {Malafarina}}, \bibinfo {author} {\bibfnamefont {A.}~\bibnamefont
  {Abdujabbarov}},\ and\ \bibinfo {author} {\bibfnamefont {C.}~\bibnamefont
  {Bambi}},\ }\bibfield  {title} {\bibinfo {title} {{On the properties of a
  deformed extension of the NUT space-time}},\ }\href
  {https://doi.org/10.1140/epjc/s10052-020-8370-3} {\bibfield  {journal}
  {\bibinfo  {journal} {Eur. Phys. J. C}\ }\textbf {\bibinfo {volume} {80}},\
  \bibinfo {pages} {784} (\bibinfo {year} {2020})},\ \Eprint
  {https://arxiv.org/abs/2003.11828} {arXiv:2003.11828 [gr-qc]} \BibitemShut
  {NoStop}%
\bibitem [{\citenamefont {Gibbons}\ and\ \citenamefont
  {Volkov}(2017)}]{Gibbons:2017jzk}%
  \BibitemOpen
  \bibfield  {author} {\bibinfo {author} {\bibfnamefont {G.~W.}\ \bibnamefont
  {Gibbons}}\ and\ \bibinfo {author} {\bibfnamefont {M.~S.}\ \bibnamefont
  {Volkov}},\ }\bibfield  {title} {\bibinfo {title} {{Weyl metrics and
  wormholes}},\ }\href {https://doi.org/10.1088/1475-7516/2017/05/039}
  {\bibfield  {journal} {\bibinfo  {journal} {JCAP}\ }\textbf {\bibinfo
  {volume} {05}},\ \bibinfo {pages} {039}},\ \Eprint
  {https://arxiv.org/abs/1701.05533} {arXiv:1701.05533 [hep-th]} \BibitemShut
  {NoStop}%
\bibitem [{\citenamefont {Narzilloev}\ \emph {et~al.}(2021)\citenamefont
  {Narzilloev}, \citenamefont {Malafarina}, \citenamefont {Abdujabbarov},
  \citenamefont {Ahmedov},\ and\ \citenamefont {Bambi}}]{Narzilloev:2021ygl}%
  \BibitemOpen
  \bibfield  {author} {\bibinfo {author} {\bibfnamefont {B.}~\bibnamefont
  {Narzilloev}}, \bibinfo {author} {\bibfnamefont {D.}~\bibnamefont
  {Malafarina}}, \bibinfo {author} {\bibfnamefont {A.}~\bibnamefont
  {Abdujabbarov}}, \bibinfo {author} {\bibfnamefont {B.}~\bibnamefont
  {Ahmedov}},\ and\ \bibinfo {author} {\bibfnamefont {C.}~\bibnamefont
  {Bambi}},\ }\bibfield  {title} {\bibinfo {title} {{Particle motion around a
  static axially symmetric wormhole}},\ }\href
  {https://doi.org/10.1103/PhysRevD.104.064016} {\bibfield  {journal} {\bibinfo
   {journal} {Phys. Rev. D}\ }\textbf {\bibinfo {volume} {104}},\ \bibinfo
  {pages} {064016} (\bibinfo {year} {2021})},\ \Eprint
  {https://arxiv.org/abs/2105.09174} {arXiv:2105.09174 [gr-qc]} \BibitemShut
  {NoStop}%
\bibitem [{\citenamefont {Stephani}\ \emph {et~al.}(2009)\citenamefont
  {Stephani}, \citenamefont {Kramer}, \citenamefont {MacCallum}, \citenamefont
  {Hoenselaers},\ and\ \citenamefont {Herlt}}]{stephani2009exact}%
  \BibitemOpen
  \bibfield  {author} {\bibinfo {author} {\bibfnamefont {H.}~\bibnamefont
  {Stephani}}, \bibinfo {author} {\bibfnamefont {D.}~\bibnamefont {Kramer}},
  \bibinfo {author} {\bibfnamefont {M.}~\bibnamefont {MacCallum}}, \bibinfo
  {author} {\bibfnamefont {C.}~\bibnamefont {Hoenselaers}},\ and\ \bibinfo
  {author} {\bibfnamefont {E.}~\bibnamefont {Herlt}},\ }\href@noop {} {\emph
  {\bibinfo {title} {Exact solutions of Einstein's field equations}}}\
  (\bibinfo  {publisher} {Cambridge university press},\ \bibinfo {year}
  {2009})\BibitemShut {NoStop}%
\bibitem [{\citenamefont {Weyl}(1917)}]{weyl1917theory}%
  \BibitemOpen
  \bibfield  {author} {\bibinfo {author} {\bibfnamefont {H.}~\bibnamefont
  {Weyl}},\ }\bibfield  {title} {\bibinfo {title} {The theory of gravitation},\
  }\href@noop {} {\bibfield  {journal} {\bibinfo  {journal} {Annalen Phys}\
  }\textbf {\bibinfo {volume} {54}},\ \bibinfo {pages} {117} (\bibinfo {year}
  {1917})}\BibitemShut {NoStop}%
\bibitem [{\citenamefont {Quevedo}(2011{\natexlab{c}})}]{quevedo2011mass}%
  \BibitemOpen
  \bibfield  {author} {\bibinfo {author} {\bibfnamefont {H.}~\bibnamefont
  {Quevedo}},\ }\bibfield  {title} {\bibinfo {title} {Mass quadrupole as a
  source of naked singularities},\ }\href@noop {} {\bibfield  {journal}
  {\bibinfo  {journal} {International Journal of Modern Physics D}\ }\textbf
  {\bibinfo {volume} {20}},\ \bibinfo {pages} {1779} (\bibinfo {year}
  {2011}{\natexlab{c}})}\BibitemShut {NoStop}%
\bibitem [{\citenamefont {{Shirokov}}(1973)}]{1973GReGr...4..131S}%
  \BibitemOpen
  \bibfield  {author} {\bibinfo {author} {\bibfnamefont {M.~F.}\ \bibnamefont
  {{Shirokov}}},\ }\bibfield  {title} {\bibinfo {title} {{On one new effect of
  the Einsteinian theory of gravitation}},\ }\href
  {https://doi.org/10.1007/BF00762799} {\bibfield  {journal} {\bibinfo
  {journal} {General Relativity and Gravitation}\ }\textbf {\bibinfo {volume}
  {4}},\ \bibinfo {pages} {131} (\bibinfo {year} {1973})}\BibitemShut {NoStop}%
\bibitem [{Note1()}]{Note1}%
  \BibitemOpen
  \bibinfo {note} {For clarity, under angular tidal force we consider polar
  tidal component, $\theta $, of axially symmetric spacetime, since contrary to
  spherical symmetric spacetime azimuthal, $\phi $, and polar, $\theta $, angle
  are not the same values.}\BibitemShut {Stop}%
\bibitem [{\citenamefont {{Allahyari}}\ \emph {et~al.}(2019)\citenamefont
  {{Allahyari}}, \citenamefont {{Firouzjahi}},\ and\ \citenamefont
  {{Mashhoon}}}]{2019PhRvD..99d4005A}%
  \BibitemOpen
  \bibfield  {author} {\bibinfo {author} {\bibfnamefont {A.}~\bibnamefont
  {{Allahyari}}}, \bibinfo {author} {\bibfnamefont {H.}~\bibnamefont
  {{Firouzjahi}}},\ and\ \bibinfo {author} {\bibfnamefont {B.}~\bibnamefont
  {{Mashhoon}}},\ }\bibfield  {title} {\bibinfo {title} {{Quasinormal modes of
  a black hole with quadrupole moment}},\ }\href
  {https://doi.org/10.1103/PhysRevD.99.044005} {\bibfield  {journal} {\bibinfo
  {journal} {\prd}\ }\textbf {\bibinfo {volume} {99}},\ \bibinfo {eid} {044005}
  (\bibinfo {year} {2019})},\ \Eprint {https://arxiv.org/abs/1812.03376}
  {arXiv:1812.03376 [gr-qc]} \BibitemShut {NoStop}%
\bibitem [{\citenamefont {{Semer{\'a}k}}\ and\ \citenamefont
  {{Basovn{\'\i}k}}(2016)}]{2016PhRvD..94d4006S}%
  \BibitemOpen
  \bibfield  {author} {\bibinfo {author} {\bibfnamefont {O.}~\bibnamefont
  {{Semer{\'a}k}}}\ and\ \bibinfo {author} {\bibfnamefont {M.}~\bibnamefont
  {{Basovn{\'\i}k}}},\ }\bibfield  {title} {\bibinfo {title} {{Geometry of
  deformed black holes. I. Majumdar-Papapetrou binary}},\ }\href
  {https://doi.org/10.1103/PhysRevD.94.044006} {\bibfield  {journal} {\bibinfo
  {journal} {\prd}\ }\textbf {\bibinfo {volume} {94}},\ \bibinfo {eid} {044006}
  (\bibinfo {year} {2016})},\ \Eprint {https://arxiv.org/abs/1608.05948}
  {arXiv:1608.05948 [gr-qc]} \BibitemShut {NoStop}%
\end{thebibliography}%

\end{document}